\documentclass[12pt,a4paper]{article}

\usepackage{ifthen} 
\usepackage{xspace}
\newboolean{pdflatex}
\setboolean{pdflatex}{true} 

\newboolean{articletitles}
\setboolean{articletitles}{true} 

\newboolean{uprightparticles}
\setboolean{uprightparticles}{false} 

\usepackage{bigints}

\def\paperauthors{LHCb collaboration} 
\def\paperasciititle{Long-lived particle reconstruction 
downstream of the LHCb
magnet} 
\def\papertitle{Long-lived particle reconstruction 
downstream of the \lhcb 
magnet} 
\def\paperkeywords{{High Energy Physics}, {LHCb}} 
\def\papercopyright{\the\year\ CERN for the benefit of the LHCb collaboration} 
\def\paperlicence{CC BY 4.0 licence}
\def\paperlicenceurl{https://creativecommons.org/licenses/by/4.0/}

\usepackage{xcolor}

\usepackage[top=1in, bottom=1.25in, left=1in, right=1in]{geometry}

\usepackage{microtype}
\usepackage{lineno}  
\usepackage{xspace} 
\usepackage{caption} 
\usepackage{subcaption}

\usepackage{graphicx}  
\usepackage{color}
\usepackage{colortbl}
\graphicspath{{./figs/}} 
\DeclareGraphicsExtensions{.pdf,.PDF,png,.PNG}

\usepackage{amsmath} 
\usepackage{amssymb}
\usepackage{amsfonts}
\usepackage{upgreek} 

\newcommand*\patchAmsMathEnvironmentForLineno[1]{%
\expandafter\let\csname old#1\expandafter\endcsname\csname #1\endcsname
\expandafter\let\csname oldend#1\expandafter\endcsname\csname
end#1\endcsname
 \renewenvironment{#1}%
   {\linenomath\csname old#1\endcsname}%
   {\csname oldend#1\endcsname\endlinenomath}%
}
\newcommand*\patchBothAmsMathEnvironmentsForLineno[1]{%
  \patchAmsMathEnvironmentForLineno{#1}%
  \patchAmsMathEnvironmentForLineno{#1*}%
}
\AtBeginDocument{%
\patchBothAmsMathEnvironmentsForLineno{equation}%
\patchBothAmsMathEnvironmentsForLineno{align}%
\patchBothAmsMathEnvironmentsForLineno{flalign}%
\patchBothAmsMathEnvironmentsForLineno{alignat}%
\patchBothAmsMathEnvironmentsForLineno{gather}%
\patchBothAmsMathEnvironmentsForLineno{multline}%
\patchBothAmsMathEnvironmentsForLineno{eqnarray}%
}

\usepackage{hyperxmp}

\usepackage[unicode=true, 
            pdfinfo={
                Author={\paperauthors},
                Title={\paperasciititle},
                Keywords={\paperkeywords},
                Copyright={Copyright (C) \papercopyright},
                LicenseUrl={\paperlicenceurl}}]{hyperref}

\usepackage[colorinlistoftodos,textsize=scriptsize]{todonotes}

\usepackage[all]{hypcap} 

\usepackage{xspace} 
\usepackage{upgreek}

\def\lhcb   {\mbox{LHCb}\xspace}

\def\lhc    {\mbox{LHC}\xspace}

\def\MagUp {\mbox{\em Mag\kern -0.05em Up}\xspace}

\ifthenelse{\boolean{uprightparticles}}%
{

 \def\Pmu         {\ensuremath{\upmu}\xspace}

 \def\Ppi         {\ensuremath{\uppi}\xspace}

 \def\Ppsi        {\ensuremath{\uppsi}\xspace}

 \def\PDelta      {\ensuremath{\Delta}\xspace}                 
 \def\PXi         {\ensuremath{\Xi}\xspace}                 
 \def\PLambda     {\ensuremath{\Lambda}\xspace}                 
 \def\PSigma      {\ensuremath{\Sigma}\xspace}                 
 \def\POmega      {\ensuremath{\Omega}\xspace}                 
 \def\PUpsilon    {\ensuremath{\Upsilon}\xspace}
 \let\oldPi\Pi
 \def\PPi         {\ensuremath{\oldPi}\xspace}

 \def\PB      {\ensuremath{\mathrm{B}}\xspace}                 
                  
 \def\PD      {\ensuremath{\mathrm{D}}\xspace}

 \def\PJ      {\ensuremath{\mathrm{J}}\xspace}                 
 \def\PK      {\ensuremath{\mathrm{K}}\xspace}

 \def\Pb      {\ensuremath{\mathrm{b}}\xspace}                 
 \def\Pc      {\ensuremath{\mathrm{c}}\xspace}

 \def\Pi      {\ensuremath{\mathrm{i}}\xspace}

 \def\Pp      {\ensuremath{\mathrm{p}}\xspace}

 \def\Ps      {\ensuremath{\mathrm{s}}\xspace}

 \def\thebaroffset{0.0em}
}
{

 \def\Pmu         {\ensuremath{\mu}\xspace}

 \def\Ppi         {\ensuremath{\pi}\xspace}

 \def\Ppsi        {\ensuremath{\psi}\xspace}                 
                  
 \mathchardef\PDelta="7101
 \mathchardef\PXi="7104
 \mathchardef\PLambda="7103
 \mathchardef\PSigma="7106
 \mathchardef\POmega="710A
 \mathchardef\PUpsilon="7107
 \mathchardef\PPi="7105
                  
 \def\PB      {\ensuremath{B}\xspace}                 
                  
 \def\PD      {\ensuremath{D}\xspace}

 \def\PJ      {\ensuremath{J}\xspace}                 
 \def\PK      {\ensuremath{K}\xspace}

 \def\Pb      {\ensuremath{b}\xspace}                 
 \def\Pc      {\ensuremath{c}\xspace}

 \def\Pi      {\ensuremath{i}\xspace}

 \def\Pp      {\ensuremath{p}\xspace}

 \def\Ps      {\ensuremath{s}\xspace}

 \def\thebaroffset{0.18em}
}
\newcommand{\offsetoverline}[2][\thebaroffset]{\kern #1\overline{\kern -#1 #2}}%

\makeatletter
\ifcase \@ptsize \relax
  \newcommand{\miniscule}{\@setfontsize\miniscule{4}{5}}
\or
  \newcommand{\miniscule}{\@setfontsize\miniscule{5}{6}}
\or
  \newcommand{\miniscule}{\@setfontsize\miniscule{5}{6}}
\fi
\makeatother

\DeclareRobustCommand{\optbar}[1]{\shortstack{{\miniscule (\rule[.5ex]{1.25em}{.18mm})}
  \\ [-.7ex] $#1$}}

\def\mup        {{\ensuremath{\Pmu^+}}\xspace}
\def\mun        {{\ensuremath{\Pmu^-}}\xspace} 

\def\squark    {{\ensuremath{\Ps}}\xspace}

\def\cquark    {{\ensuremath{\Pc}}\xspace}

\def\bquark    {{\ensuremath{\Pb}}\xspace}

\def\pion   {{\ensuremath{\Ppi}}\xspace}

\def\pip    {{\ensuremath{\pion^+}}\xspace}
\def\pim    {{\ensuremath{\pion^-}}\xspace}

\def\kaon    {{\ensuremath{\PK}}\xspace}

\def\KorKbar {\kern \thebaroffset\optbar{\kern -\thebaroffset \PK}{}\xspace}

\def\KS      {{\ensuremath{\kaon^0_{\mathrm{S}}}}\xspace}

\def\D       {{\ensuremath{\PD}}\xspace}

\def\DorDbar {\kern \thebaroffset\optbar{\kern -\thebaroffset \PD}\xspace}

\def\Dp      {{\ensuremath{\D^+}}\xspace}
\def\Dm      {{\ensuremath{\D^-}}\xspace}

\def\DpDm    {\ensuremath{\Dp {\kern -0.16em \Dm}}\xspace}

\def\B       {{\ensuremath{\PB}}\xspace}

\def\BorBbar {\kern \thebaroffset\optbar{\kern -\thebaroffset \PB}\xspace}
\def\Bz      {{\ensuremath{\B^0}}\xspace}

\def\Bd      {{\ensuremath{\B^0}}\xspace}

\def\BdorBdbar {\kern \thebaroffset\optbar{\kern -\thebaroffset \Bd}\xspace}

\def\Bs      {{\ensuremath{\B^0_\squark}}\xspace}

\def\BsorBsbar {\kern \thebaroffset\optbar{\kern -\thebaroffset \Bs}\xspace}

\def\jpsi     {{\ensuremath{{\PJ\mskip -3mu/\mskip -2mu\Ppsi}}}\xspace}

\def\Y#1S{\ensuremath{\PUpsilon{(#1S)}}\xspace}

\def\proton      {{\ensuremath{\Pp}}\xspace}

\def\Lz          {{\ensuremath{\PLambda}}\xspace}
\def\Lbar        {{\ensuremath{\offsetoverline{\PLambda}}}\xspace}
\def\LorLbar     {\kern \thebaroffset\optbar{\kern -\thebaroffset \PLambda}\xspace}

\def\Lb           {{\ensuremath{\Lz^0_\bquark}}\xspace}

\def\to                 {\ensuremath{\rightarrow}\xspace}

\def\CPT               {{\ensuremath{C\!PT}}\xspace}

\def\AT#1     {\ensuremath{A_{\mathrm{T}}^{#1}}\xspace}           

\def\C#1      {\ensuremath{\mathcal{C}_{#1}}\xspace}                       
\def\Cp#1     {\ensuremath{\mathcal{C}_{#1}^{'}}\xspace}                    
\def\Ceff#1   {\ensuremath{\mathcal{C}_{#1}^{\mathrm{(eff)}}}\xspace}        
\def\Cpeff#1  {\ensuremath{\mathcal{C}_{#1}^{'\mathrm{(eff)}}}\xspace}       
\def\Ope#1    {\ensuremath{\mathcal{O}_{#1}}\xspace}                       
\def\Opep#1   {\ensuremath{\mathcal{O}_{#1}^{'}}\xspace}                    

\newcommand{\nospaceunit}[1]{\ensuremath{\text{#1}}}       
\newcommand{\aunit}[1]{\ensuremath{\text{\,#1}}}       

\newcommand{\tev}{\aunit{Te\kern -0.1em V}\xspace}
\newcommand{\gev}{\aunit{Ge\kern -0.1em V}\xspace}
\newcommand{\mev}{\aunit{Me\kern -0.1em V}\xspace}
\newcommand{\kev}{\aunit{ke\kern -0.1em V}\xspace}
\newcommand{\ev}{\aunit{e\kern -0.1em V}\xspace}
 
\newcommand{\mevc}{\ensuremath{\aunit{Me\kern -0.1em V\!/}c}\xspace}
\newcommand{\gevc}{\ensuremath{\aunit{Ge\kern -0.1em V\!/}c}\xspace}
\newcommand{\mevcc}{\ensuremath{\aunit{Me\kern -0.1em V\!/}c^2}\xspace}
\newcommand{\gevcc}{\ensuremath{\aunit{Ge\kern -0.1em V\!/}c^2}\xspace}

\def\m    {\aunit{m}\xspace}

\def\cm   {\aunit{cm}\xspace}

\def\mm   {\aunit{mm}\xspace}

\def\mum  {\ensuremath{\,\upmu\nospaceunit{m}}\xspace}

\def\fb   {\ensuremath{\aunit{fb}}\xspace}
\def\invfb   {\ensuremath{\fb^{-1}}\xspace}

\def\sec  {\ensuremath{\aunit{s}}\xspace}

\def\ns   {\ensuremath{\aunit{ns}}\xspace}
\def\ps   {\ensuremath{\aunit{ps}}\xspace}

\def\mhz  {\ensuremath{\aunit{MHz}}\xspace}

\def\gsim{{~\raise.15em\hbox{$>$}\kern-.85em
          \lower.35em\hbox{$\sim$}~}\xspace}
\def\lsim{{~\raise.15em\hbox{$<$}\kern-.85em
          \lower.35em\hbox{$\sim$}~}\xspace}

\def\sPlot{\mbox{\em sPlot}\xspace}

\def\pt         {\ensuremath{p_{\mathrm{T}}}\xspace}

\def\rad{\aunit{rad}\xspace}

\def\evtgen     {\mbox{\textsc{EvtGen}}\xspace}

\def\geant      {\mbox{\textsc{Geant4}}\xspace}

\def\photos     {\mbox{\textsc{Photos}}\xspace}

\def\pythia     {\mbox{\textsc{Pythia}}\xspace}

\def\tell1  {TELL1\xspace}
\def\ukl1   {UKL1\xspace}

\newcommand{\eg}{\mbox{\itshape e.g.}\xspace}
\newcommand{\ie}{\mbox{\itshape i.e.}\xspace}

\newcommand{\lhcborcid}[1]{\href{https://orcid.org/#1}{\hspace*{0.1em}\raisebox{-0.45ex}{\includegraphics[width=1em]{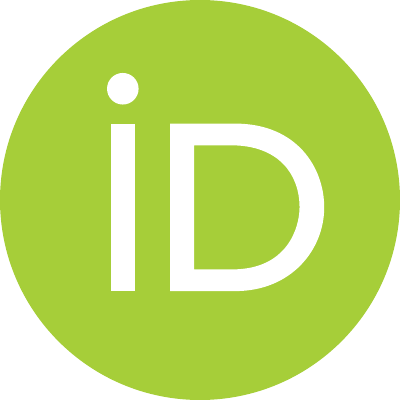}}}}

\usepackage{pstricks}

\usepackage{cite} 
\usepackage{mciteplus}

\usepackage{comment}

\def\runI   {\mbox{Run 1}\xspace}
\def\runII  {\mbox{Run 2}\xspace}
\def\runIII {\mbox{Run 3}\xspace}
\def\runIV {\mbox{Run 4}\xspace}
\def\BdToJpsiKS{$\Bz\rightarrow\jpsi\KS$\xspace}
\def\LbToJpsiLz{$\Lb\rightarrow\jpsi\Lz$\xspace}
\def\bfBdToJpsiKS{$\boldsymbol{\Bz\rightarrow\jpsi\KS}$\xspace}
\def\bfLbToJpsiLz{$\boldsymbol{\Lb\rightarrow\jpsi\Lz}$\xspace}

\def\good{{\em Good}\xspace}
\def\ghost{{\em Ghost}\xspace}

\begin{document}

\renewcommand{\thefootnote}{\fnsymbol{footnote}}
\setcounter{footnote}{1}


\begin{titlepage}
\pagenumbering{roman}

\vspace*{-1.5cm}
\centerline{\large EUROPEAN ORGANIZATION FOR NUCLEAR RESEARCH (CERN)}
\vspace*{1.5cm}
\noindent
\begin{tabular*}{\linewidth}{lc@{\extracolsep{\fill}}r@{\extracolsep{0pt}}}
\ifthenelse{\boolean{pdflatex}}
{\vspace*{-1.5cm}\mbox{\!\!\!\includegraphics[width=.14\textwidth]{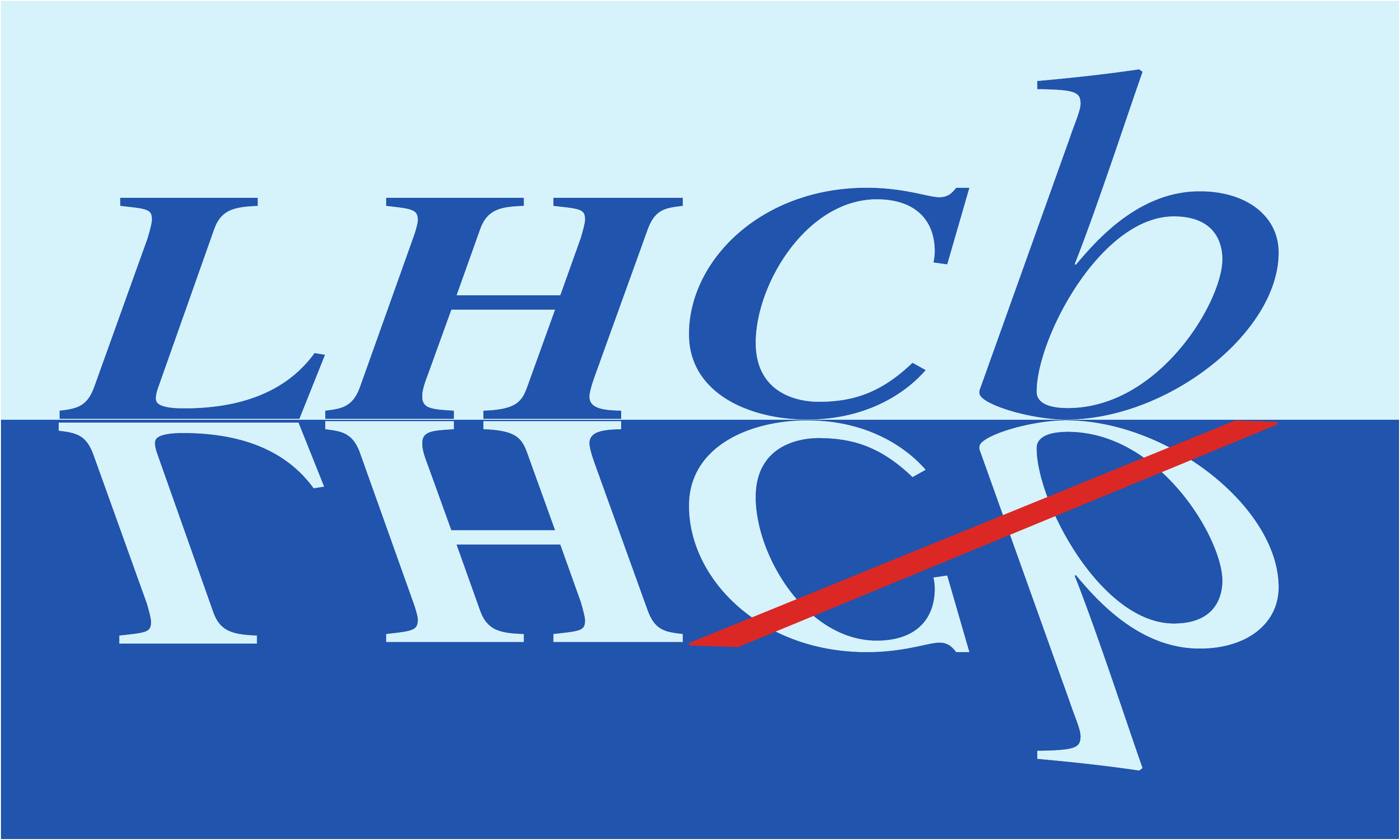}} & &}%
{\vspace*{-1.2cm}\mbox{\!\!\!\includegraphics[width=.12\textwidth]{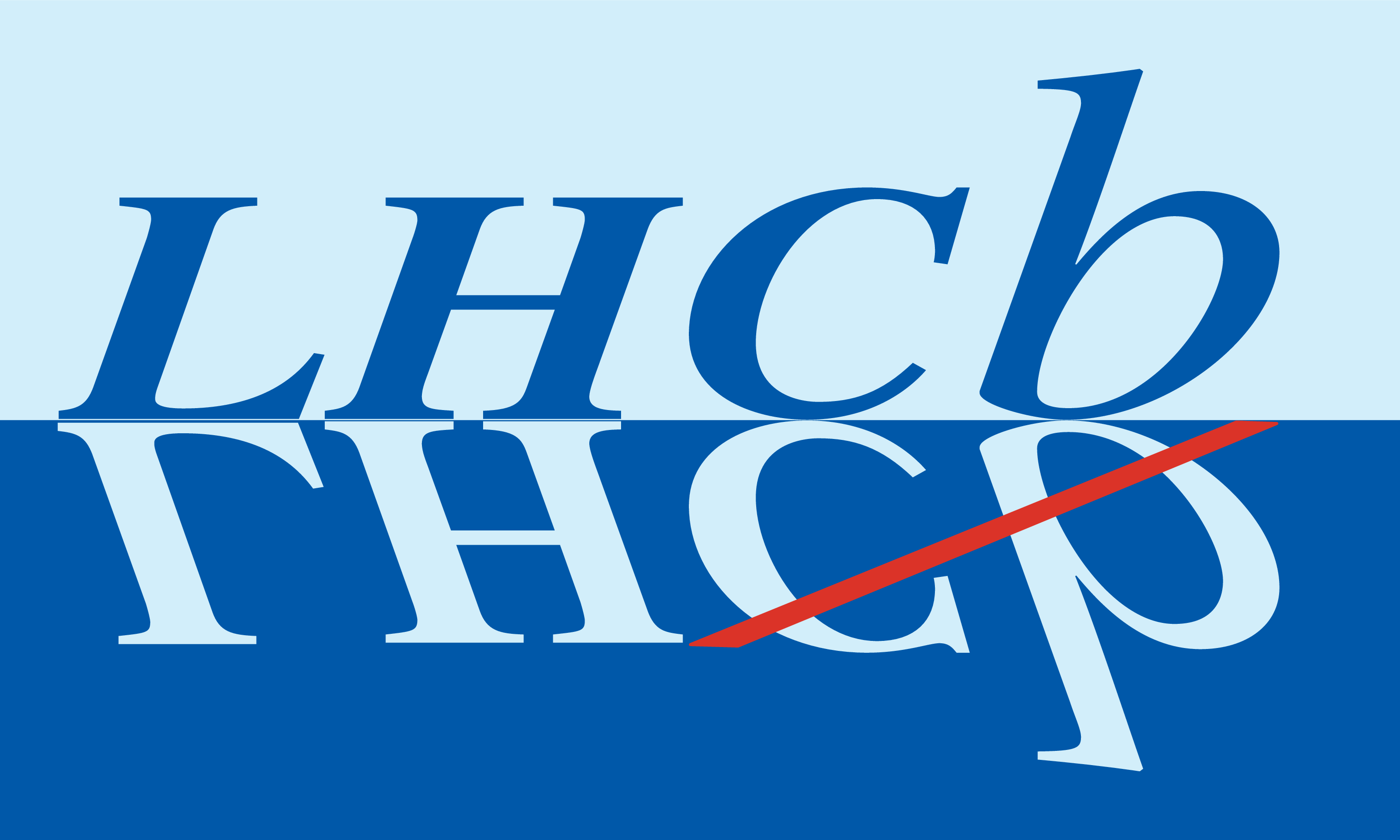}} & &}%
\\
  & & CERN-LHCb-DP-2022-001 \\ 
 & & January 6, 2025 \\ 
 & & \\
\end{tabular*}

\vspace*{3.0cm}

{\normalfont\bfseries\boldmath\huge
\begin{center}
  \papertitle 
\end{center}
}

\vspace*{2.0cm}

\begin{center}
\paperauthors\footnote{Authors are listed at the end of this paper.}
\end{center}

\vspace*{0.5cm}

\begin{abstract}
  \noindent
Charged-particle trajectories are usually reconstructed with the LHCb 
detector using combined information from the tracking devices placed upstream and downstream of the \mbox{4\,T\,m} dipole magnet. Trajectories reconstructed using only information from the tracker downstream of the dipole magnet, which are referred to as \mbox{T tracks}, have not been used for physics analysis to date. 
The challenges of the reconstruction of long-lived particles using \mbox{T tracks} for use in physics analyses are discussed and  
solutions are proposed.
The feasibility and the tracking performance are studied using
samples of long-lived $\varLambda$ 
and $K_S^0$ 
hadrons decaying between 6.0 and 7.6 metres downstream of the proton-proton collision point,
thereby traversing most of the magnetic field region and providing maximal sensitivity to magnetic and electric dipole moments.
The reconstruction can be expanded 
upstream to about 2.5 meters
for use in direct searches of exotic long-lived particles.
The data used in this analysis have been recorded between 2015 and 2018 and correspond to an integrated luminosity of 6\,fb$^{-1}$. 
The results obtained demonstrate the possibility to further extend the 
decay
volume and the physics reach of the LHCb 
experiment.
\end{abstract}

\vspace*{2.0cm}

\begin{center}
Published in The European Physical Journal C (2025) 85:7
\end{center}

\vspace{\fill}

{\footnotesize 
\centerline{\copyright~\papercopyright. \href{\paperlicenceurl}{\paperlicence}.}}
\vspace*{2mm}

\end{titlepage}


\newpage
\setcounter{page}{2}
\mbox{~}
%
%
%
%


\renewcommand{\thefootnote}{\arabic{footnote}}
\setcounter{footnote}{0}



\pagestyle{plain} 
\setcounter{page}{1}
\pagenumbering{arabic}

%


\section{Introduction}
\label{sec:Introduction}

After two successful data-taking campaigns from 2009 to 2013 (\runI),  and from 2015 to 2018 (\runII), the \lhc 
\runIII~started in 2022. Motivated by further exploration and precision studies, the \lhcb~detector~\cite{LHCb-DP-2008-001,LHCb-DP-2014-002} has undergone upgrades to most of its components in order to operate at an instantaneous luminosity of $2\times 10^{33}\cm^{-2}\sec^{-1}$ and integrate an expected total luminosity of \mbox{$\sim$ 50\invfb} by the end of LHC \runIV~\cite{LHCb:2023hlw,LHCb-TDR-012,LHCb-TDR-013,LHCb-TDR-015,LHCb-TDR-014}. 
A triggerless readout followed by a fully software-based trigger will operate at an average proton-proton ($\proton\proton$) bunch-crossing rate of 30\mhz~\cite{LHCb-TDR-016,LHCb-TDR-017}.

This paper demonstrates the feasibility of the~\lhcb detector in reconstructing long-lived particles (LLPs)
using only information from the tracking system located downstream of the dipole magnet, about 8 m away
from the $\proton\proton$ interaction point (IP). 
These decays have not been used for physics analysis to date due to their reconstruction challenges and the limited momentum resolution.
The performance of such a reconstruction is evaluated using samples of long-lived \Lz and \KS hadrons decaying into the $\proton\pim$ and $\pip\pim$ final states, respectively, at distances between $6.0$ and $7.6\m$ from the IP, which are reconstructed and selected from \LbToJpsiLz and \BdToJpsiKS decays,\footnote{Charge conjugation is implied throughout the paper if not otherwise stated.} using data collected during \runII and corresponding to an integrated luminosity of 6\invfb.

Examples of physics use cases that motivate these studies include 
measurements of the magnetic and electric~\cite{POSPELOV2005119,Jungmann2013, PDG2022_TestConservationLaws} dipole
moments of Standard Model (SM) LLPs,
and direct searches for beyond SM LLPs~\cite{Knapen:2022afb,Beacham:2019nyx,Acosta:2021mdu,Alimena:2019zri}.
For instance,
new analyses of the electromagnetic dipole moments of the \Lz baryon~\cite{Schachinger:1978qs,Pondrom:1981gu},
and first measurements of these properties for the $\overline \Lz$ antibaryon, which offer the opportunity to perform a fundamental test of \CPT symmetry~\cite{PDG2022_TestConservationLaws,VanDyck:1987ay, Miller:2007kk,ATRAP:2013vnt}, would be within experimental reach.
The requirements for these analyses include a source of polarised \Lz baryons not aligned with the magnetic field, which can be obtained from weak decays of heavier baryons, \eg \LbToJpsiLz decays where the \Lz 
helicity
is maximal~\cite{LHCb-PAPER-2012-057, LHCb-PAPER-2020-005, ATLAS:2014swk,CMS:2018wjk},
and the ability to reconstruct the \Lz and $\overline \Lz$ decays after their spin precession in the magnetic field region with sufficient invariant-mass, vertex and helicity-angle resolution~\cite{Botella:2016ksl}.
Additionaly, thanks to its forward pseudorapidity coverage and excellent performance in the reconstruction of heavy-hadron decays, the \lhcb detector is especially well suited for searches of beyond SM LLPs, such as
dark scalars, dark photons, Axion Like Particles (ALPs), Heavy Neutral Leptons (HNLs) or other LLPs emerging from photon, beauty- and charm-hadrons produced in the $pp$ interactions~\cite{Borsato:2021aum}.
Using tracks that require hits in the Vertex Locator (VELO), \lhcb has to date excluded LLPs with lifetimes only up to $\sim 10\ps$ (see \eg~\cite{Antel:2023hkf} and references therein), corresponding to average flight distances of about $30\cm$.
A preliminary study of the~\lhcb sensitivity for exotic LLPs with an average flight distance up to about $2.5\m$ has been reported in Ref.~\cite{Calefice:2022} and developments are currently ongoing. 
The use of tracks with hits only in the tracking stations placed above $7.6\m$ downstream of the dipole magnet, is instead brand new and its first study is discussed here. 
Whereas the main focus of this work is the inclusive and exclusive reconstruction of \Lz and \KS hadrons from
\LbToJpsiLz and \BdToJpsiKS decays, the same reconstruction can be used to search for exotic LLPs with lifetimes up to $\sim 30\ns$, further expanding the \lhcb reach for BSM physics. This will be the topic of a subsequent paper.

The paper is organized as follows. In Secs.~\ref{sec:lhcb} and~\ref{sec:datasamples}, the \lhcb detector and its upgrade, the charged-particle reconstruction, and the data and simulation samples used are described. Sections~\ref{sec:reconstructionLb} and~\ref{sec:reconstructionB0} contain thedescription of the decay reconstruction and selection algorithms, and how they have been adapted to enable the reconstruction of particles decaying in the region of the \lhcb dipole magnet, along with the performance obtained. The summary and prospects are discussed in Sec.~\ref{sec:conclusions}.

\section{\lhcb detector}
\label{sec:lhcb}

The \lhcb detector~\cite{LHCb-DP-2008-001,LHCb-DP-2014-002} is a single-arm forward spectrometer covering the pseudorapidity range $2 < \eta < 5$, designed for the study of particles containing \bquark or \cquark quarks. During \lhc \runI and \runII, a silicon-strip detector (VErtex LOcator, VELO) surrounds the $pp$ interaction 
region~\cite{LHCb-DP-2014-001} 
at a radius of $8$\,mm. It consists of $42$ planar modules arranged along the direction of the beam axis ($z$ axis)
covering a total length of about $1$\,m, each providing a measurement of the $r$ ($R$ sensors) and $\phi$ ($\Phi$ sensors) coordinates. The Tracker Turicensis (TT)
is a planar silicon strip detector with a total active area of about $8$\,m$^2$, located about $2$\,m away from the IP, upstream of the dipole magnet
with a bending power of about \mbox{$4$\,T\,m}. Three hybrid planar tracking stations (T1--T3) are placed downstream of the magnet, between about $7.5$ and $10$\,m away from the IP; the inner detector modules (IT) closer to the beam pipe are identical to those used in the TT stations and have a total active area of $4$\,m$^2$, while the outer detector modules 
(OT)~\cite{LHCb-DP-2013-003} 
are gas-tight straw tubes with a total active area of about $360$\,m$^2$. 
Both the TT and T1--T3 stations are composed of four planes arranged in a $xuvx$ geometry, where the $x$ planes contain vertical strips, while the $u$ and $v$ planes contain strips rotated with respect to the $x$ planes by a stereo angle of $+5^{\circ}$ and $-5^{\circ}$, respectively. The tracking system provides a measurement of the momentum, $p$, of charged particles with a relative uncertainty that varies from $0.5$\% at low momentum to $1.0$\% at $200$\gevc. The minimum distance of a track to a primary $pp$ collision vertex (PV), the impact parameter, is measured with a resolution of 
$(15 + 29/\pt) \mum$, where \pt is the component of the momentum transverse to the beam, in \gevc~\cite{LHCb-DP-2014-002}. Muons are identified by a system composed of alternating layers of iron and multiwire proportional chambers~\cite{LHCb-DP-2013-001}. The online event selection is performed by a trigger~\cite{LHCb-DP-2012-004}, which consists of a hardware stage, based on information from the muon system and calorimeters, followed by a software stage, which applies a full event reconstruction.
Particle identification (PID) is performed using two RICH detector stations, RICH1 located between the VELO and the TT and RICH2 located downstream of the last T stations.

During the \lhc \runIII (2022--2025) and the forthcoming \runIV (starting in 2029) the upgraded \lhcb experiment~\cite{LHCb:2023hlw} plans to collect a total integrated luminosity of \mbox{$\sim 50\invfb$} of $pp$ collisions. In order to meet this goal, the detector aims to operate at an instantaneous luminosity of $2\times 10^{33} \cm^{-2}\sec^{-1}$, five times higher than during \runII,
and read out at 40\mhz with a flexible software-based trigger.
The whole read-out electronics and most of the 
detectors (with the exception of the RICH, calorimeters and muon chambers) have been replaced, maintaining the overall geometry unchanged, while improving, where possible, their acceptance and resolution
~\cite{LHCb:2023hlw,LHCb-TDR-013,LHCb-TDR-015,LHCb-TDR-014,LHCb-TDR-016,LHCb-TDR-017}. The upgraded VELO is made of new state-of-the-art silicon pixel sensors~\cite{LHCb-TDR-013}; the Upstream Tracker (UT) detector~\cite{LHCb-TDR-015} is similar to the original TT in shape, but with thinner sensors, finer segmentation and larger coverage; the Scintillating Fiber (SciFi) Tracker~\cite{LHCb-TDR-015} replaces the three hybrid tracking stations downstream of the magnet by scintillating fibers with a diameter of 250\mum, bonded in a matrix structure with a total active area of about $30$\,m$^2$.
At the same time, a new trigger paradigm consisting of two stages of an entirely software-based trigger, the High Level Trigger (HLT) 1 and 2~\cite{LHCb-TDR-016,LHCb-TDR-017}, is becoming operational. In order to cope with the 30\mhz collision rate, the event reconstruction algorithms running on the HLT computing farms must meet strict requirements in terms of computing efficiency. Ensuring excellent physics performance while leaving the required throughput constitutes a significant challenge.

\section{Track reconstruction and data samples}
\label{sec:datasamples}

\begin{figure}[hbt]
\centering
\def\svgwidth{\columnwidth}
\includegraphics[width=0.6\textwidth]{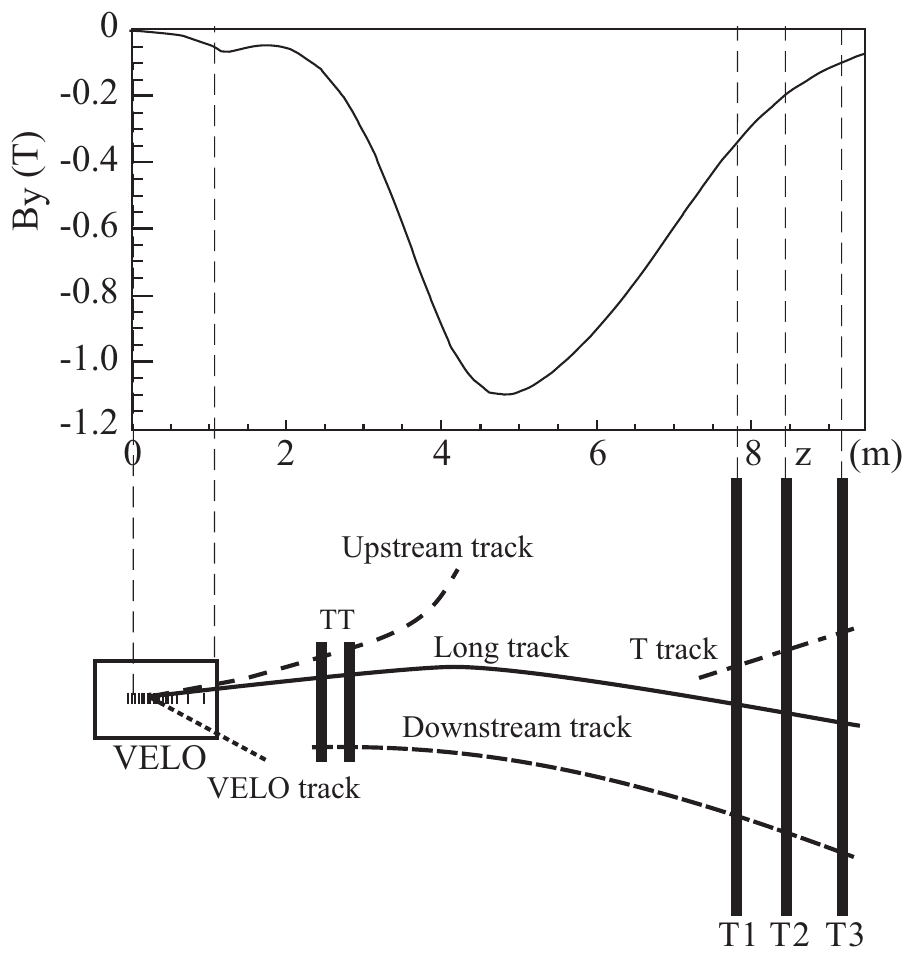}
\caption{Track-type definitions in the LHCb 
tracking system.
The main $B$-field component, along the $y$ direction, is plotted above as a function of the $z$ coordinate, along the beam axis. Figure taken from Ref.~\cite{LHCb-DP-2014-002}.}%
\label{fig:tracktypes}
\end{figure}

Reconstructed tracks of charged-particle trajectories are categorised according to the contributions of hits from the various parts of the tracking system~\cite{LHCb-DP-2014-002,LHCb-2007-006}, as illustrated in Fig.~\ref{fig:tracktypes}:
\begin{description}
\item[Long tracks] traverse the full tracking system. They include hits in both the VELO and the T1--T3 stations,
and optionally in TT (UT in the upgraded detector).
\item[Upstream tracks] traverse only the VELO and the TT (UT) stations. They are typically produced by low momentum particles, which are bent away by the magnetic field, thus failing to reach the T1--T3 stations.
\item[Downstream tracks] traverse both the TT (UT) and T1--T3 stations, but do not leave any hit in the VELO. They typically belong to decay products of long-lived particles decaying beyond the VELO, such as \Lz or \KS hadrons.
\item[VELO tracks] have hits only in the VELO. They include large-angle or backward tracks, useful for the determination of the PV, as well as very low momentum tracks.
\item[T tracks] have hits only in the T1--T3 stations. Similarly to downstream tracks, they include the decay products of long-lived particles decaying far away from the PV, up to several metres. A significant fraction of tracks reconstructed in this category comes from  secondary interactions with the material of the mechanical structures and back-scattering particles coming from the calorimeters and hadron shield behind the muon chambers.
\end{description}

The tracking algorithms used for the reconstruction of 
\runI and \runII data are outlined in Refs.~\cite{LHCb-DP-2008-001,LHCb-DP-2014-002}, and briefly summarised below. 
For \runIII, the tracking algorithms have been updated to take full advantage of the upgraded tracker. However, the main features and especially the track definitions given above remain unchanged.

The search for Long tracks starts with identifying VELO hits that form straight line trajectories, exploiting the negligibly small magnetic field in the VELO~\cite{LHCb-2007-013,LHCb-PUB-2011-001,LHCb-DP-2013-002}. These tracks are reconstructed with a minimum of three hits in the $R$ sensors and three hits in the $\Phi$ sensors.
There are two complementary algorithms to match VELO tracks with the hits in the TT and T1--T3 stations.
In the first algorithm, called forward tracking~\cite{LHCb-2007-015}, VELO tracks are combined with hits in the individual T stations that form clusters along the projected trajectory.
In the second algorithm, a standalone track algorithm is used to reconstruct tracks using only the T1--T3 stations~\cite{LHCb-2007-026,LHCb-2008-042} 
and assuming the tracks originate at the IP, which induces a small variation of the reconstruction efficiency as a function of the LLP decay position.
Only track segments with at least one hit in the $x$ layer and one hit in the stereo layers ($u$ or $v$) for each T station are considered. VELO tracks are then combined with T track segments that provide the best possible match~\cite{LHCb-2007-020,LHCb-2007-129}. 
Finally, TT hits that are compatible with track extrapolations are added for improved momentum resolution.
 
Downstream tracks are reconstructed starting from T tracks, extrapolating them through the magnetic field and matching with TT hits~\cite{LHCb-2007-026,LHCb-PUB-2017-001,LHCb-PUB-2014-007}. Similarly, Upstream tracks are reconstructed by matching VELO tracks with residual hits in the TT stations. In both cases, for Upstream and Downstream tracks, at least three hits in the TT stations are required for the matching.

A fit based on the Kalman filter procedure~\cite{Kalman:1960,Fruhwirth:1987fm,VanTilburg:885750,Fruhwirth:2020zbo} is used on all tracks, using a fifth-order Runge--Kutta method~\cite{Hairer1993,LHCb-2007-140} to describe the track transport through the magnetic field, taking into account the
material of the detector. After the fit, the reconstructed track is represented by a list of
state vectors giving the transverse position, slopes with respect to the $z$ axis, and the charge/momentum ratio, $(x,y, {\rm d}x/{\rm d}z, {\rm d}y/{\rm d}z, q/p)$, specified at given $z$ positions in the experiment~\cite{LHCb-2007-007}.
Duplicate tracks, known as clones, can occur when two or more tracks are reconstructed with the same trajectory and share most of their hits in the T and potentially TT stations. An algorithm, referred to as the clone killer, removes these tracks by only keeping the best track based on the total number of hits of the track and the goodness of the fit.
Tracks involving more subdetectors are always preferred to individual segments.
Therefore, T tracks are those that do not correspond to a Long or Downstream particle trajectory.
No assumptions on the origin vertex of the track are made at this stage.

Fake tracks are those that are not associated with a real charged-particle trajectory, caused by random combinations of hits or mismatch of track segments upstream and downstream of the magnet. The fraction of fake Long tracks was estimated using \runI data to range from typically around $6.5$\% in minimum bias events up to about $20$\% for events with large track multiplicity~\cite{LHCb-DP-2014-002}.
This fake rate is significantly reduced in \runII by about 60\% with a small drop in efficiency through a neural-network algorithm that uses information from the overall $\chi^2$ of the Kalman filter, the $\chi^2$ values of the track segments, the number of hits in the different detectors, and the \pt of the track~\cite{LHCb-DP-2019-001,DeCian:2255039}. 
Studies on simulation estimate that around 5\% of T tracks are fake due to 
detector hits caused by noise~\cite{LHCb-TDR-015}.


The \mbox{$\Lb\rightarrow \jpsi \Lz$} and \mbox{$\Bz\rightarrow \jpsi \KS$}, with \mbox{$\jpsi\rightarrow \mup\mun$}, \mbox{$\Lz \rightarrow \proton\pim$} and \mbox{$\KS \rightarrow \pip\pim$} decays are chosen as benchmark channels to study the capability and the performance of the \lhcb detector in reconstructing LLPs using T tracks. 
The \Lb and \Bz hadrons have a characteristic flight length of around 10\mm, thus decay inside the VELO detector. 
Muons are reconstructed as Long tracks, providing a precise determination of their momenta and of the \jpsi decay vertex, which coincides with the decay vertex of the \Lb or \Bz hadron, 
and allowing for kinematic constraints to be applied on the remaining part of the decay chain.
The \Lz and \KS candidates are reconstructed as combinations of two \mbox{T tracks} with their vertex position along the beam axis between 6.0 and 7.6\,m from the nominal IP.
Hadrons decaying in this region have traversed most of the magnetic field, thereby experiencing maximal spin precession, and their final-state particles reach the T1--T3 stations. 
The selection is based on the inclusive detached \jpsi trigger, which is part of the \lhcb trigger strategy in the \runI and \runII data-taking campaigns~\cite{LHCb-DP-2012-004}. The events used correspond to an integrated luminosity of $6$\invfb collected during \runII.

Samples of simulated events are required to model the detector acceptance, detector response, reconstruction efficiency, and the effect of imposed signal selection requirements. Furthermore, the simulated events are used to train a classifier to discriminate between signal and background. In the simulation, $pp$ collisions are generated using \pythia~\cite{Sjostrand:2007gs} with a specific \lhcb configuration~\cite{Belyaev:2011zza}. Decays of unstable particles are described by \evtgen~\cite{Lange:2001uf}, in which final-state radiation is generated using \photos~\cite{Golonka:2005pn}. The transport of the generated particles and the detector response are implemented using the \geant toolkit~\cite{Agostinelli:2002hh}, as described in Ref.~\cite{Clemencic:2011zza}.

\section{Reconstruction of \texorpdfstring{\bfLbToJpsiLz}{LambdabToJpsiLz} decays}
\label{sec:reconstructionLb}

The reconstruction of the \Lz decay vertex presents two main challenges when it is located downstream of the TT (UT) stations. 
Firstly, the particle trajectories must often be extrapolated over large distances through the strong and inhomogeneous magnetic field, 
accounting for effects induced by the particle interactions with
air and detector material,
using measurements only available downstream of the magnet (the T tracks).
The uncertainty on the extrapolation depends on the measurement precision of the track slopes (${\rm d}x/{\rm d}z$ and ${\rm d}y/{\rm d}z$ parameters introduced in Sec.~\ref{sec:datasamples}), which itself depends on the hit resolutions of the T1--T3 stations, and on the precision and granularity of the magnetic field measurements.
Therefore, the trajectories are influenced in a way that cannot be described by an exact analytical solution. 
Secondly, the measurement of particle momentum ($q/p$ parameter) relies on the relatively low curvature of T tracks, as they are reconstructed using only hits located in a region with a weak magnetic field.
As a consequence, the momentum resolution of T tracks is 
poor compared to Long or Downstream tracks, and the charge can be assigned incorrectly.
In simulation about 0.5\% of all T tracks have a wrong charge assignment.

The \mbox{$\Lb\rightarrow \jpsi \Lz$} with \mbox{$\Lz \rightarrow \proton\pim$} signal candidates are first required to pass the online event selection performed by the detached \mbox{$\jpsi \rightarrow \mup\mun$} trigger~\cite{LHCb-DP-2012-004}. Offline, a loose selection is applied after the decay chain reconstruction, followed by a multivariate classifier based on a Histogrammed Gradient Boosted Decision Tree (HBDT) 
available in the \texttt{scikit-learn} package~\cite{Scikit-learn-paper}.

\subsection{Signal candidate reconstruction and loose selection}
\label{subsec:reconstructionLb}

Signal candidates are reconstructed following a two-stage procedure. First, decay vertices are reconstructed from the final-state particles using an iterative algorithm based on a Kalman filter~\cite{Fruhwirth:1987fm,Fruhwirth:2020zbo}, referred to in the following as vertex fitter. The original track-state vectors are transformed, after transportation to the estimated vertex position of the previous iteration, 
into a convenient representation $(x,y,z,p_x,p_y,p_z,E)$, where $E$ is the particle energy corresponding to momentum vector $(p_x,p_y,p_z)$ at position $(x,y,z)$ for a given mass hypothesis~\cite{LHCb-2007-007}. The prior (seeding) covariance matrix of the vertex position is set to a large diagonal matrix, which removes any dependence on its prior knowledge.
The convergence criterion is that between two consecutive iterations either the vertex position moves by less then 1\mum or the $\chi^2$ changes by less than 0.01. The maximum number of iterations is 10.
In order to find a decay vertex position, tracks must be extrapolated during the Kalman filtering iterative procedure to a common vertex location. This step is 
particularly challenging for T tracks, as outlined above.
For this task an approach based on a fifth-order Runge--Kutta method~\cite{Hairer1993,LHCb-2007-140} is employed instead of the default cubic interpolation used for Long and Downstream tracks.
Since the \LbToJpsiLz with \mbox{$\jpsi \rightarrow \mup\mun$} and \mbox{$\Lz \rightarrow \proton\pim$}
decay chain involves multiple decays in cascade, these are reconstructed one-by-one using a bottom-up approach. In the second stage, the identified cascade decays are fed into a separate fitter, 
referred to as Decay Tree Fitter (DTF)~\cite{Hulsbergen:2005pu}, 
also based on a Kalman filter, where the whole decay chain is fitted simultaneously with
geometric and kinematic constraints 
imposed as appropriate.
The constraints require the 
\mbox{\bquark-hadron} candidate to originate at the PV, 
the \jpsi and \Lz masses to be equal to their known values~\cite{PDG2022},
and the \Lz momentum aligned with its flight direction, defined by the \jpsi and \Lz decay vertices.

The loose selection criteria to identify signal candidates are based on the following variables: $p$ of the proton and the pion, \pt of the proton and \Lz candidates; the invariant masses of the $\mup\mun$, $\proton\pim$ and $\jpsi\Lz$ systems; the $z$ component of the \Lz candidate decay vertex, $z_{\rm vtx}$; the cosine of the angle $\xi_p$ between the proton and the \Lz momenta, and between the proton and the \Lb momenta;  
the $\chi^2_{\rm IP}$ 
and $\chi^2_{\rm vtx}$ of the \Lz and \Lb candidates, where $\chi^2_{\rm IP}$ is the difference in the vertex-fit $\chi^2$ of the PV candidate reconstructed with and without the particle considered, 
and $\chi^2_{\rm vtx}$ is the decay vertex-fit $\chi^2$. In
Table~\ref{tab:preselections} of Appendix~\ref{appendix:LbKS}, the loose selection requirements applied on these variables are summarised. All the subsequent analysis steps are performed using this selected sample.

\begin{figure}[tb]
\centering
{\includegraphics[height=6.5cm]{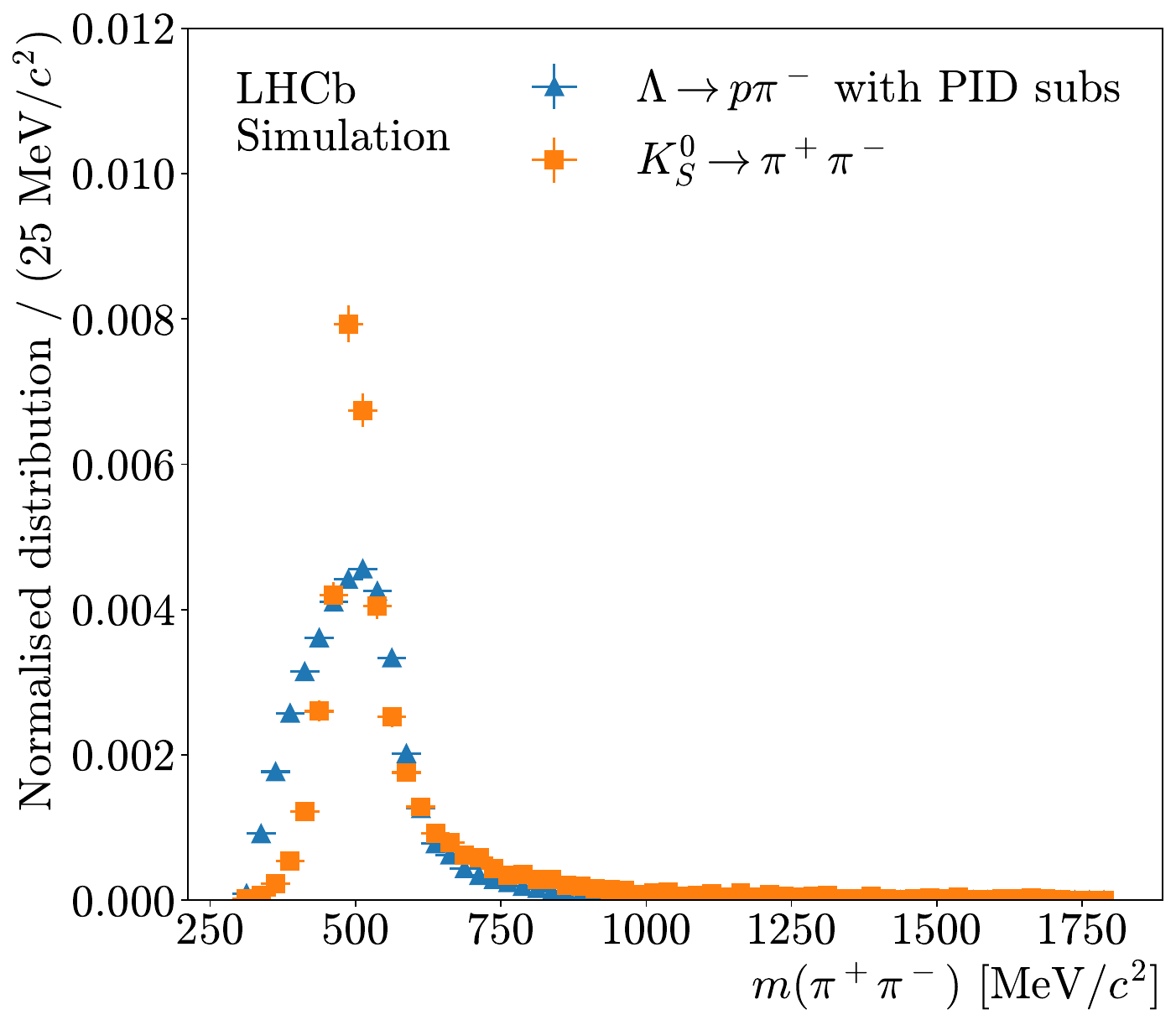}}
\hspace{0.5cm}
{\includegraphics[height=6.5cm]{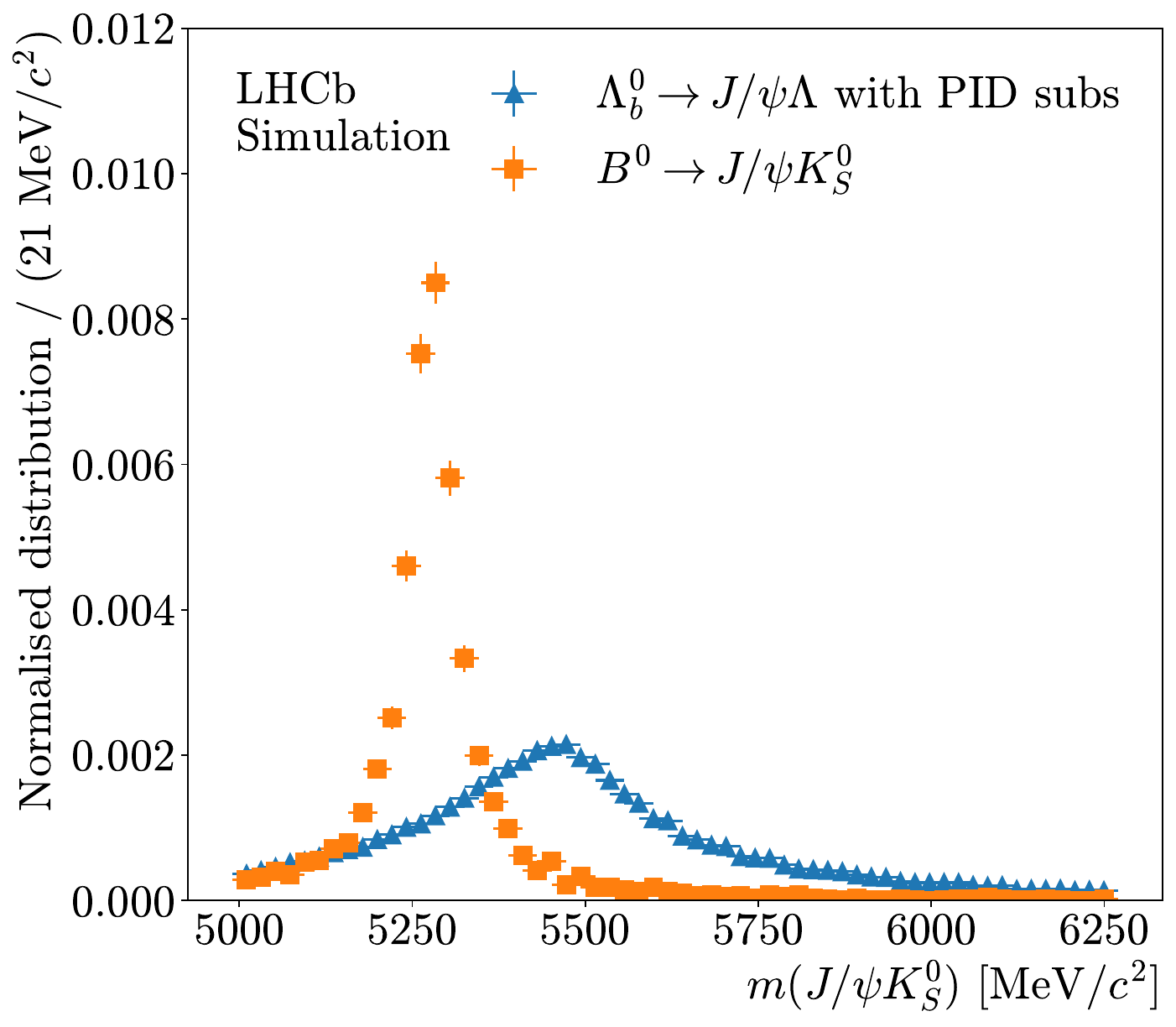}}
\caption{(Left) invariant-mass distribution $m(\pi^+\pi^-)$ of $K_S^0$ 
candidates from simulated \mbox{$B^0 \rightarrow J/\psi K_S^0$} 
decays (orange points) compared with the mass distribution of $\varLambda$ 
candidates from \mbox{$\varLambda_b^0 \rightarrow J/\psi \varLambda$} 
decays, where the proton in the final state is assigned the pion-mass hypothesis (blue points). (Right) invariant-mass distribution $m(J/\psi K_S^0)$ 
from simulated \mbox{$B^0 \rightarrow J/\psi K_S^0$} 
decays (orange points) compared with the corresponding 
distribution from \mbox{$\varLambda_b^0 \rightarrow J/\psi \varLambda$} 
simulation, where the proton in the final state has assigned the pion-mass hypothesis and the intermediate $\varLambda$ 
has assigned the $K_S^0$-mass 
hypothesis (blue points). 
}
\label{fig:B0vsLambdab_inv_mass}
\end{figure}

The \mbox{$\Bz\rightarrow \jpsi \KS$} with \mbox{$\KS \rightarrow \pip\pim$} decay constitutes a large source of 
background since the topology of the decay is almost identical to the signal, with the substitution of a proton by a pion in the final state. Due to the unavailability of PID information for T tracks in the current analysis,
this background cannot be easily distinguished. In 
\mbox{Fig.~\ref{fig:B0vsLambdab_inv_mass} (left)}, the invariant-mass distribution of \KS candidates from simulated \mbox{\BdToJpsiKS} decays is compared with the corresponding distribution of \Lz candidates from \Lb decays where the proton of the final state
has been assigned the pion-mass hypothesis. 
Due to the poor mass resolution, the two distributions are almost completely overlapping, and a veto based on this variable does not help.
Nevertheless, \mbox{Fig.~\ref{fig:B0vsLambdab_inv_mass} (right)} shows that the $m(\jpsi \KS)$ invariant mass discriminates between the two \bquark-hadron decays.
A veto on candidates with $m(\jpsi \KS)$ in the range $\pm 70$\mevcc around the known \Bz mass~\cite{PDG2022} is applied in the following, unless otherwise stated, retaining
$87$\% of the \Lb signal while rejecting $65$\% of the \Bz decays.
The 
PID information from the RICH2, calorimeter and muon systems
will be used for future physics analyses to improve the signal selection and the background rejection.

\subsection{Multivariate analysis for signal discrimination}
\label{sec:mva_signal}

A multivariate analysis is performed to enhance the signal purity.
A Histogram-based Gradient Boosted Decision Tree classifier 
is used~\cite{Scikit-learn-paper}.
The training sample consists of about $43\,000$ simulated \Lb signal decays with about 
$6\times10^6$
background candidates reconstructed in data. The background candidates are selected from the lower and upper sidebands of the $m(\jpsi\Lz)$ distribution, 
chosen as $150$- and $300$\mevcc-wide windows below and above the signal region, respectively, itself defined by a  $\jpsi\Lz$ invariant mass within $\pm 600$\mevcc of the known \Lb mass~\cite{PDG2022}. 
The training is performed using $90$\% of the sample, while the remaining $10$\% is held out for the assessment of the classifier performance.
The classifier includes 
kinematic and topological variables: the longitudinal and transverse momenta of the proton, pion and \jpsi candidates; the coordinates of the \Lz decay vertex,  $(x,y,z)_{\rm vtx}$; the cosine of the \Lz (\Lb) direction angle between its momentum and the flight direction defined by the \Lz (\Lb) and \jpsi decay vertices; the $\chi^2_{\rm vtx}$, $\chi^2_{\rm IP}$ and 
$\chi^2_{\rm dist}$ of the \Lz and \Lb candidates,
where $\chi^2_{\rm dist}$ is the squared distance between the PV and the particle decay vertex normalised to its uncertainty; and
the status flags (converged/failed) of the decay chain vertex fit with and without the \Lz mass constraint. 
The use of the \Lz decay position plays an important role in the rejection of background originating from material interactions.


The following hyperparameters of the HBDT are optimised: the maximum number of leaf nodes for the decision trees; the learning rate, \ie the weight applied to each successive decision tree of the boosting procedure; and the total number of iterations (number of decision trees in the ensemble).
The performance of the trained HBDT is estimated calculating the Area-Under-Curve (AUC) of the curve representing the signal purity versus the signal efficiency, with the threshold applied to the HBDT response varying continuously from zero to one. 

The requirement on the HBDT response is optimised by maximising the figure-of-merit $S/\sqrt{S+B}$,
where $S$ and $B$ are the signal and background yields in the signal region, respectively.
For this purpose, the figure-of-merit is estimated for each HBDT threshold performing a binned fit to the \mbox{\LbToJpsiLz} invariant-mass distribution in data.
The optimal threshold results in a signal efficiency of 72\% and a figure-of-merit value of 54.
Training the classifier with a similar size of the
signal and background 
samples induces a shift of the optimal threshold with no impact on the figure-of-merit.

\subsection{Armenteros-Podolanski plot}
\label{sec:armenteros}

The Armenteros-Podolanski (AP) plot~\cite{Armenteros-Podolanski} is used as a kinematics-based PID technique to reject \KS 
background events
from the \Lz and \mbox{\LbToJpsiLz} samples.
For two-body decays of a particle of mass $M$ into two particles of masses $m_1$ and $m_2$, it is a representation of the transverse momentum versus the asymmetry of the longitudinal momenta of the final-state particles with respect to the parent flight direction. In this diagram, the decays show as a semi-ellipse centred at $((m_1^2-m_2^2)/M^2,0)$ and
with radii $(2p_{\rm cm}/M,p_{\rm cm})$, where $p_{\rm cm}$ is the momentum of the decay products in the centre-of-mass frame of the decaying particle. 

\begin{figure}[htb]
\centering
\includegraphics[height=6cm]{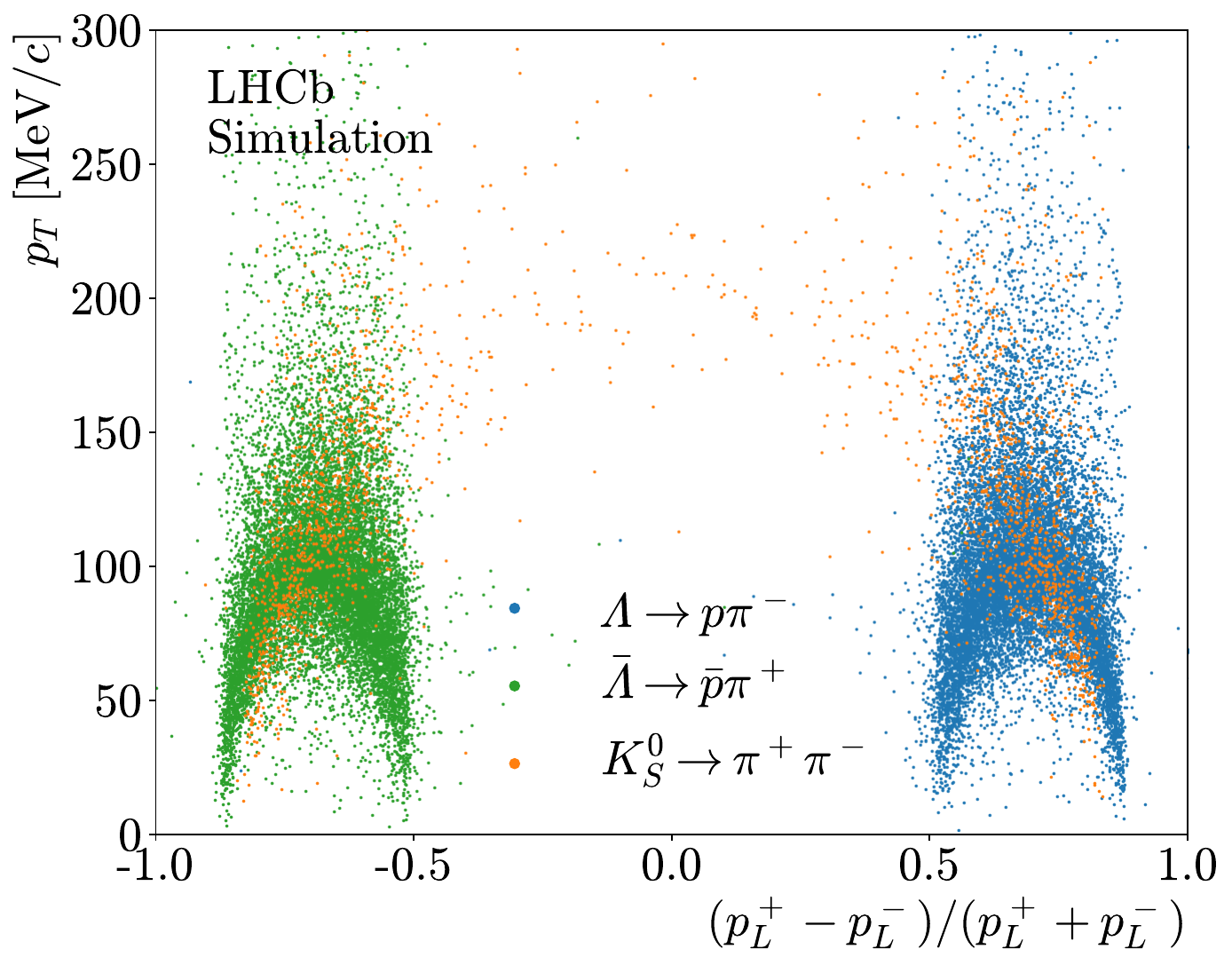}
\includegraphics[height=6cm]{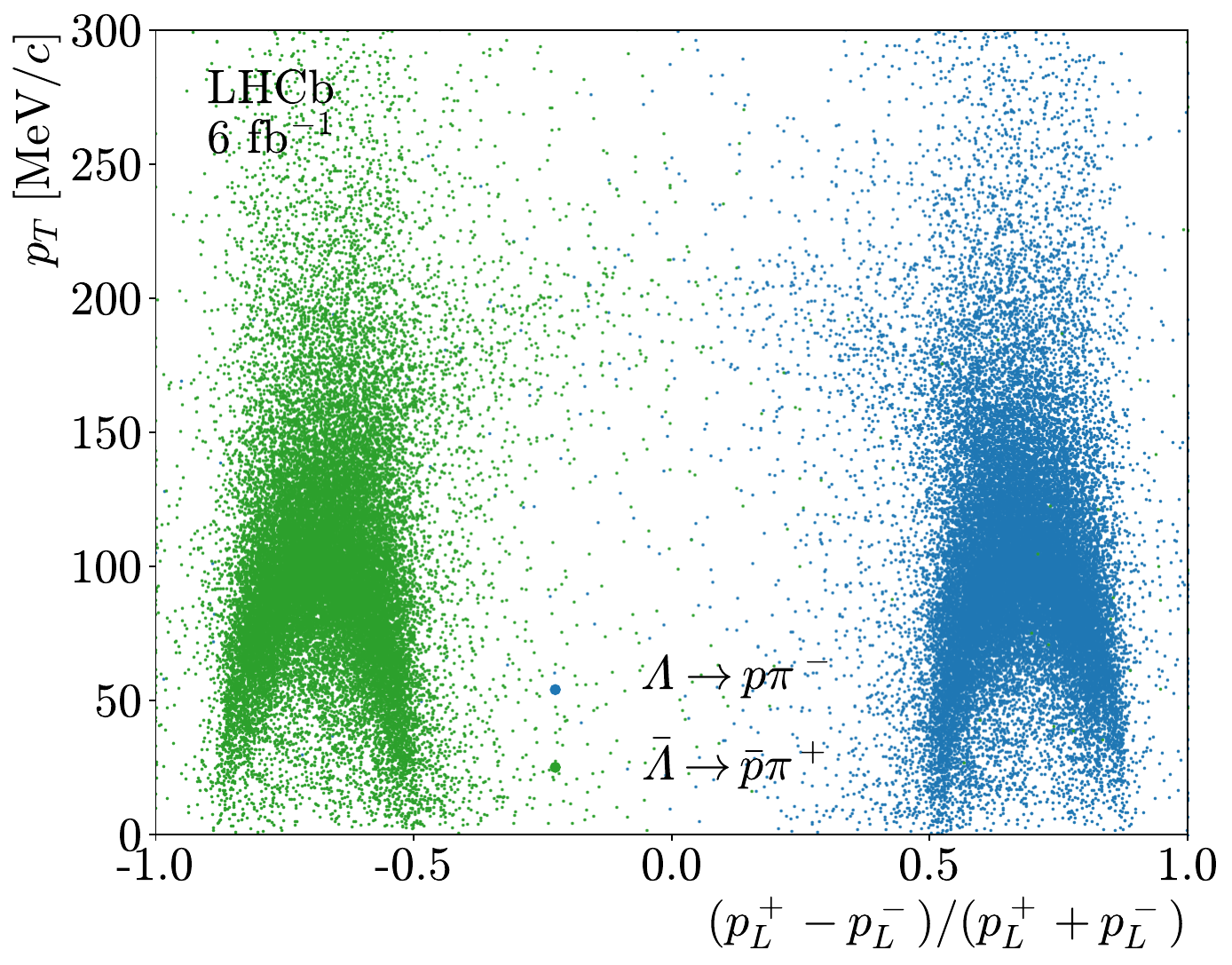}
\caption{The AP plot for \mbox{$\varLambda_b^0 \rightarrow J/\psi \varLambda$} 
candidates passing the loose, HBDT and $B^0$ 
veto selection criteria, for 
(left) simulated \mbox{$\varLambda_b^0 \rightarrow J/\psi \varLambda$} 
and 
\mbox{$B^0 \rightarrow J/\psi K_S^0$} 
signal events and (right) \mbox{Run 2} 
data. 
The symbol $p^+_{\mathrm{L}}$ ($p^-_{\mathrm{L}}$) 
refers to the longitudinal momentum of the positively (negatively) charged final-state particle, \textit{i.e.} 
the $p$ 
and $\pi^+$ 
($\pi^-$ 
and $\overline{p}$) 
in the $\varLambda$ 
and $\offsetoverline{\varLambda}$ 
decay, respectively, and $p_{\mathrm{T}}$ 
their transverse momentum, calculated with respect to the hyperon flight direction. 
}
\label{fig:armenteros_podolanski}
\end{figure}

Figure~\ref{fig:armenteros_podolanski} shows the AP plot for \mbox{\LbToJpsiLz} signal compared to \mbox{\BdToJpsiKS} signal from simulation, and \mbox{\LbToJpsiLz} \runII data, from the decay chain vertex fit with the \jpsi mass constraint and after the loose, HBDT and \Bz veto selection criteria applied.
The depleted central region of the \KS semi-ellipse in the data reveals that the HBDT requirement, combined with the \Bz veto, is effectively removing \KS background events not overlapping with the \Lz and \Lbar semi-ellipses. This is due to the inclusion of the transverse and longitudinal momenta of the proton and pion from the \Lz decay in the classifier. Requiring the magnitude of the longitudinal momentum asymmetry to exceed 0.5 rejects 17\% of the remaining \Bz decays, with 
a signal efficiency of 99\%,
for candidates passing all selection criteria including the \Bz veto. This requirement
removes 7\% of the selected candidates in data.

\subsection{Signal yields and invariant-mass resolution}
\label{subsec:signal_yields_mass_resol}

\begin{figure}[tb]
\centering
{\includegraphics[height=6.5cm]{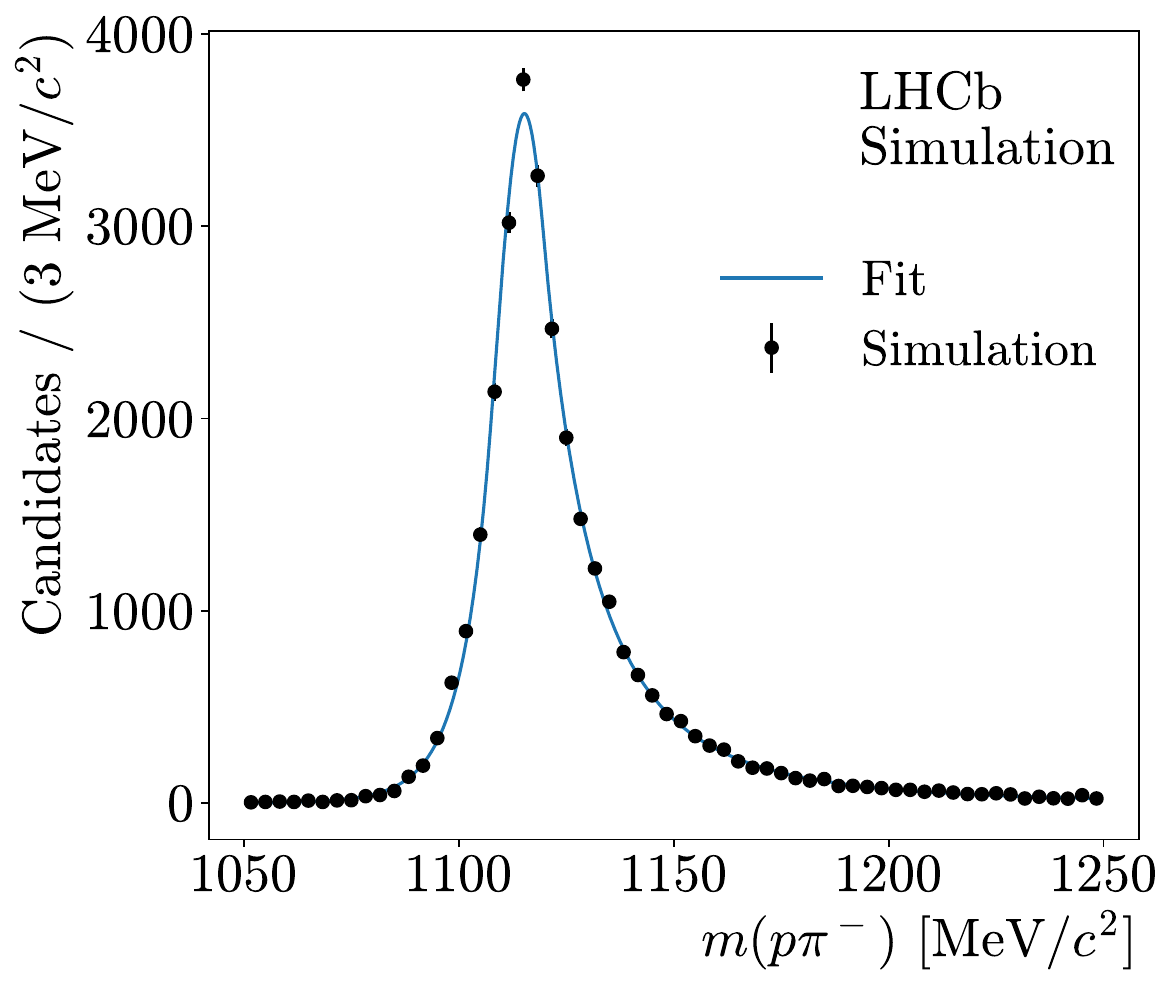}}
{\includegraphics[height=6.42cm]{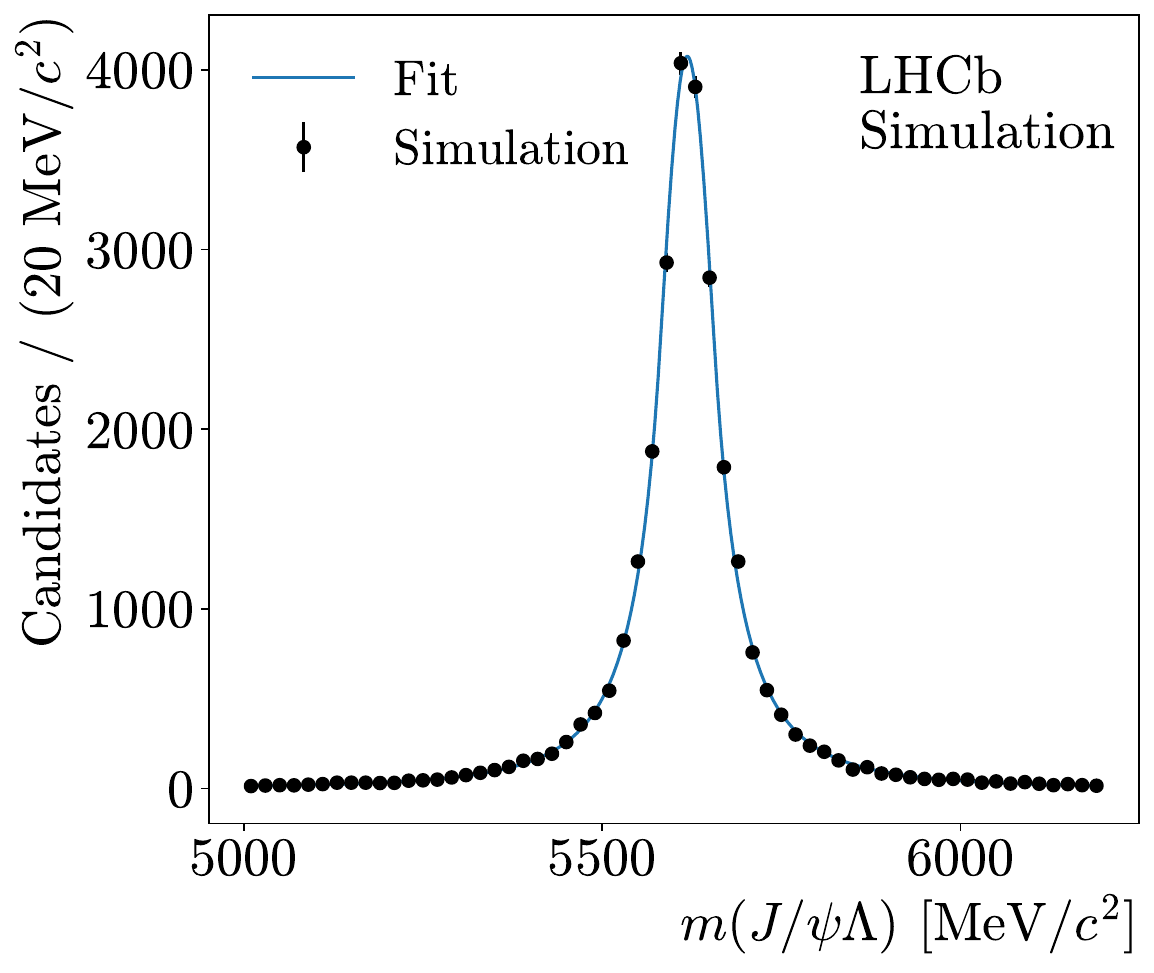}}
\caption{
Invariant-mass distribution (left) $m(p\pi^-)$ 
and (right) $m(J/\psi\varLambda)$ 
for simulated \mbox{$\varLambda_b^0 \rightarrow J/\psi \varLambda$} 
signal decays. 
The black points with error bars represent the simulation, and the overlaid (blue) curves are the 
results of the mass fits. 
}
\label{fig:inv_mass_sim}
\end{figure}

\begin{figure}[tb]
\centering
{\includegraphics[height=6.4cm]{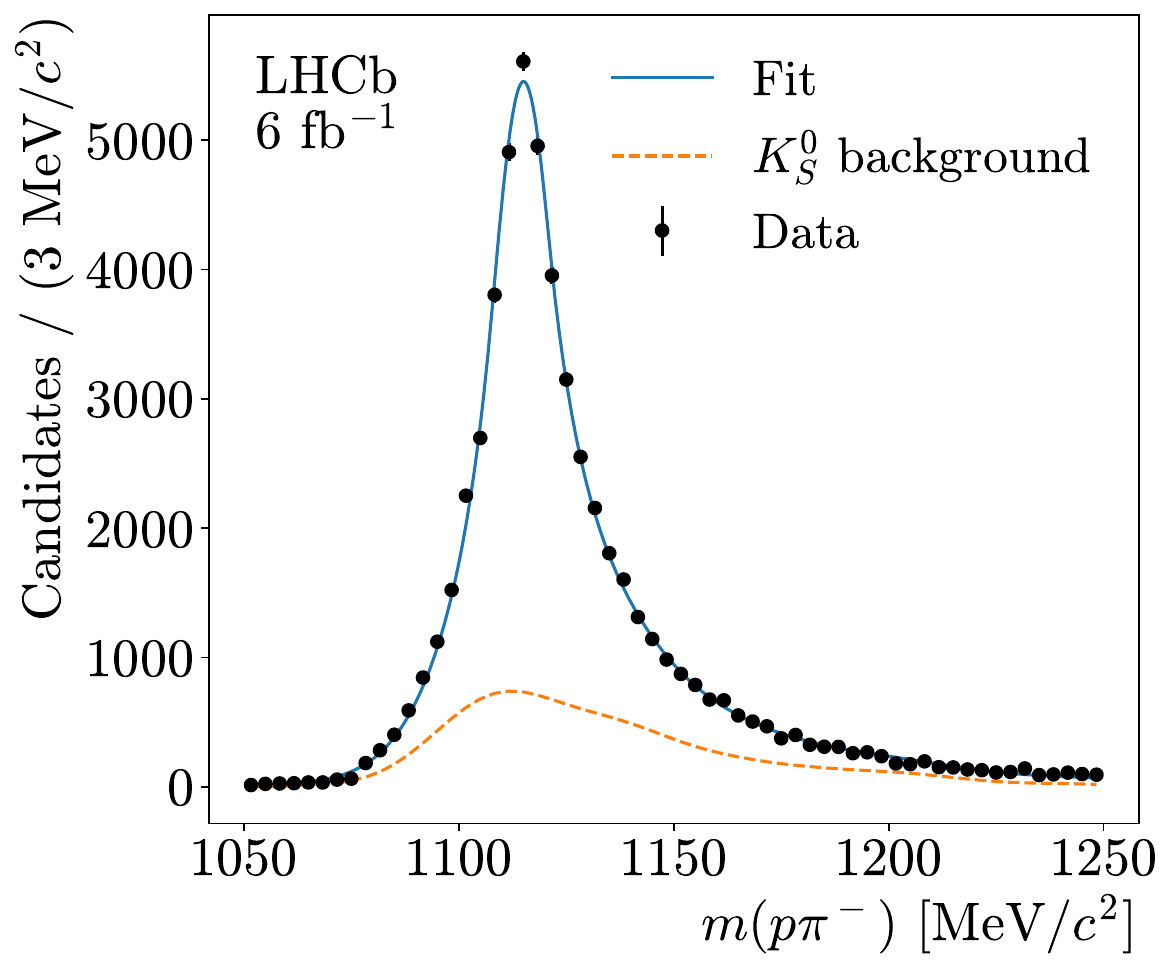}}
\hspace{0.5cm}
{\includegraphics[height=6.38cm]{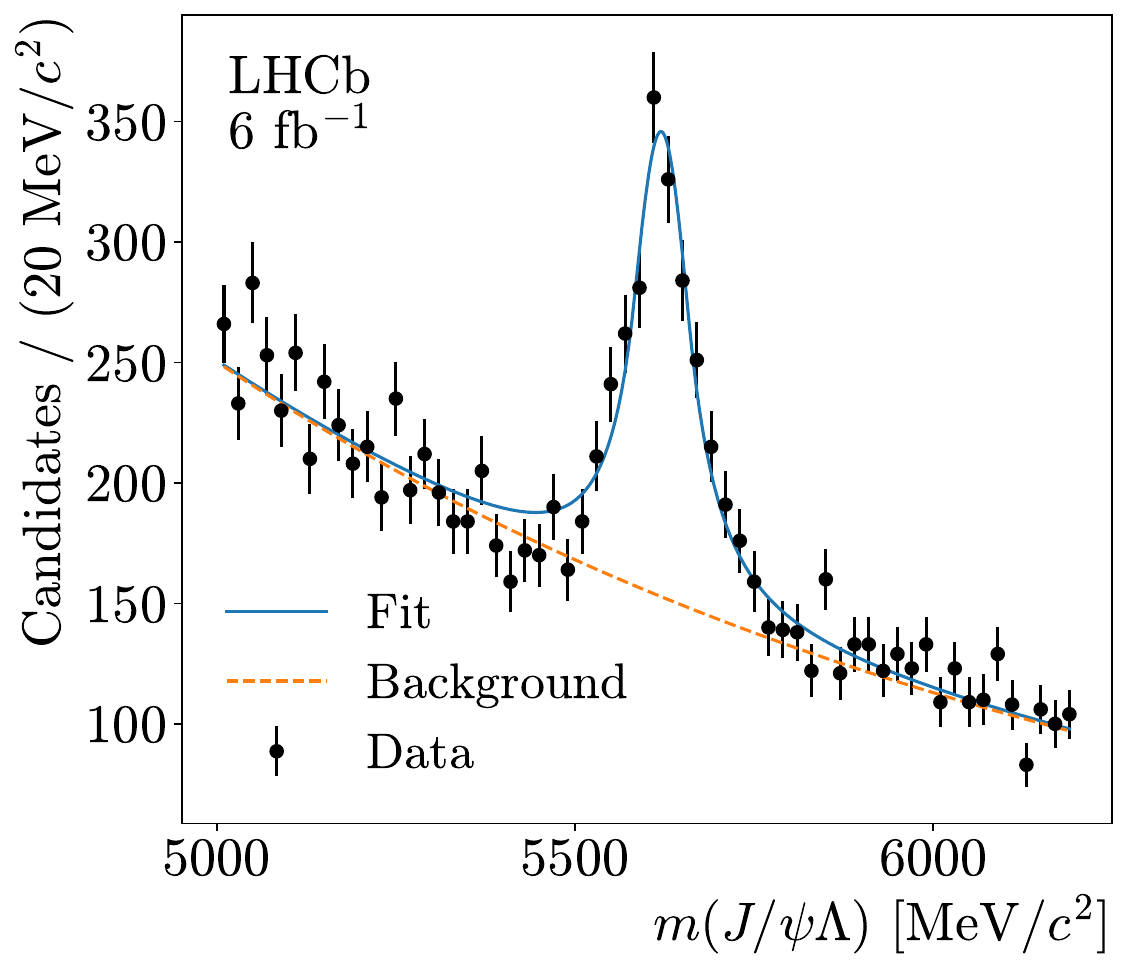}}
\hspace{0.5cm}
\caption{
Invariant-mass distribution (left) $m(p\pi^-)$ 
and (right) $m(J/\psi \varLambda)$ 
for \mbox{$\varLambda_b^0 \rightarrow J/\psi \varLambda$} 
candidates reconstructed using \mbox{Run 2} 
data after all selection criteria. The 
results of the mass fits are superimposed (blue curves) along with the background contribution (dotted orange curves).}
\label{fig:inv_mass_data}
\end{figure}

Figures~\ref{fig:inv_mass_sim} and~\ref{fig:inv_mass_data} show the 
invariant-mass distributions of the $\proton\pim$ and $\jpsi\Lz$ systems, $m(\proton\pim)$ and $m(\jpsi\Lz)$, of the selected \mbox{$\Lb\rightarrow \jpsi\Lz$} candidates from simulation and \runII data, respectively, after all selection criteria. 
The $\jpsi\Lz$ invariant mass is computed 
applying the constraints outlined previously, whereas 
the $\proton\pim$ invariant mass is evaluated without the \Lz mass constraint.
Signal and background yields along with the invariant-mass resolution are obtained by fitting these distributions independently.
The $m(\proton\pim)$ and $m(\jpsi \Lz)$ signals are adequately described with asymmetric and symmetric double-tail Crystal Ball functions~\cite{Skwarnicki:1986xj}, respectively, with tail parameters in data fixed to the values obtained from simulation.
In data, the background in the $m(\proton\pim)$ distribution, which is dominated by \KS decays, is parameterised using a template determined from simulated $\Bz\to\jpsi\KS$ decays.
Instead, the background contribution to the $m(\jpsi \Lz)$ distribution is largely dominated by random combinations of \jpsi and \Lz candidates, with a residual contribution from $\Bz\to\jpsi\KS$ decays, and
is modelled with an exponential function.

The reconstructed sample in simulation amounts to about $31\,000$ signal decays, from which an invariant-mass resolution of $16.4\pm0.2$ and \mbox{$68.0\pm 0.5$\mevcc} is obtained for $m(\proton\pim)$ and $m(\jpsi\Lz)$, respectively. Here and in the following, resolutions are evaluated as the central 68.3\% confidence-level (CL) region of the underlying distribution, with statistical uncertainties based upon its fourth moment~\cite{rao73}.\footnote{See Eq.~(6h.3.1) in the reference.}
%
In data, the fits yield about $43\,500$ $\Lz\rightarrow\proton\pim$ and $6\,140$ $\Lb\to\jpsi\Lz$ signal decays,
with mass resolution of $17.8\pm0.3$ and \mbox{$74.4\pm0.6$\mevcc}, respectively, about 10\% worse than in simulation.
These resolutions are determined 
from the corresponding invariant-mass distributions after statistical background subtraction using the \sPlot technique~\cite{Pivk:2004ty}, 
with $m(p \pim)$ and $m(\jpsi\Lz)$ as discriminant variables, respectively.
The \KS background is determined to be about $15\,000$ candidates in the full $m(\proton\pim)$ range shown in \mbox{Fig.~\ref{fig:inv_mass_data} (left)}, whereas the background yield in the $m(\jpsi\Lz$) distribution, estimated in a region defined by three times the invariant-mass resolution around the mean mass, amounts to $10\,800$.

\subsection{Reconstruction and selection efficiency}
\label{sec:Efficiency}

\begin{figure}[htb]
\centering
\includegraphics[width=1.1\textwidth]{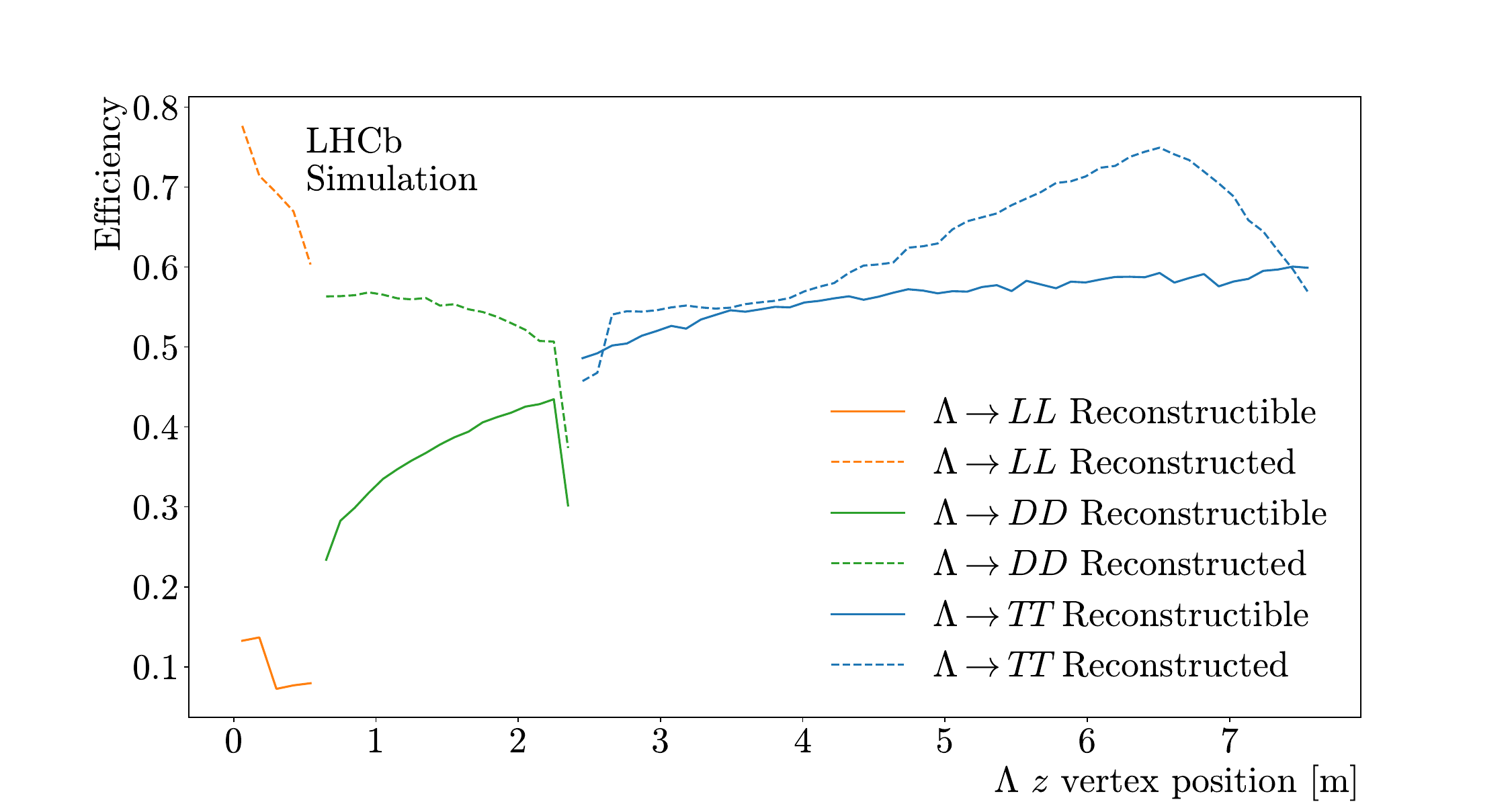}
\caption{Fraction of \mbox{$\varLambda_b^0 \rightarrow J/\psi \varLambda$} 
signal decays that are reconstructible and four-track reconstruction efficiency as a function of the $\varLambda$ 
decay vertex position along the $z$ axis for the different track categories of the final-state particles (Long, Downstream and T).}
\label{fig:efficiency_recocted}
\end{figure}

\begin{figure}[htb]
\centering
\includegraphics[width=1.\textwidth]{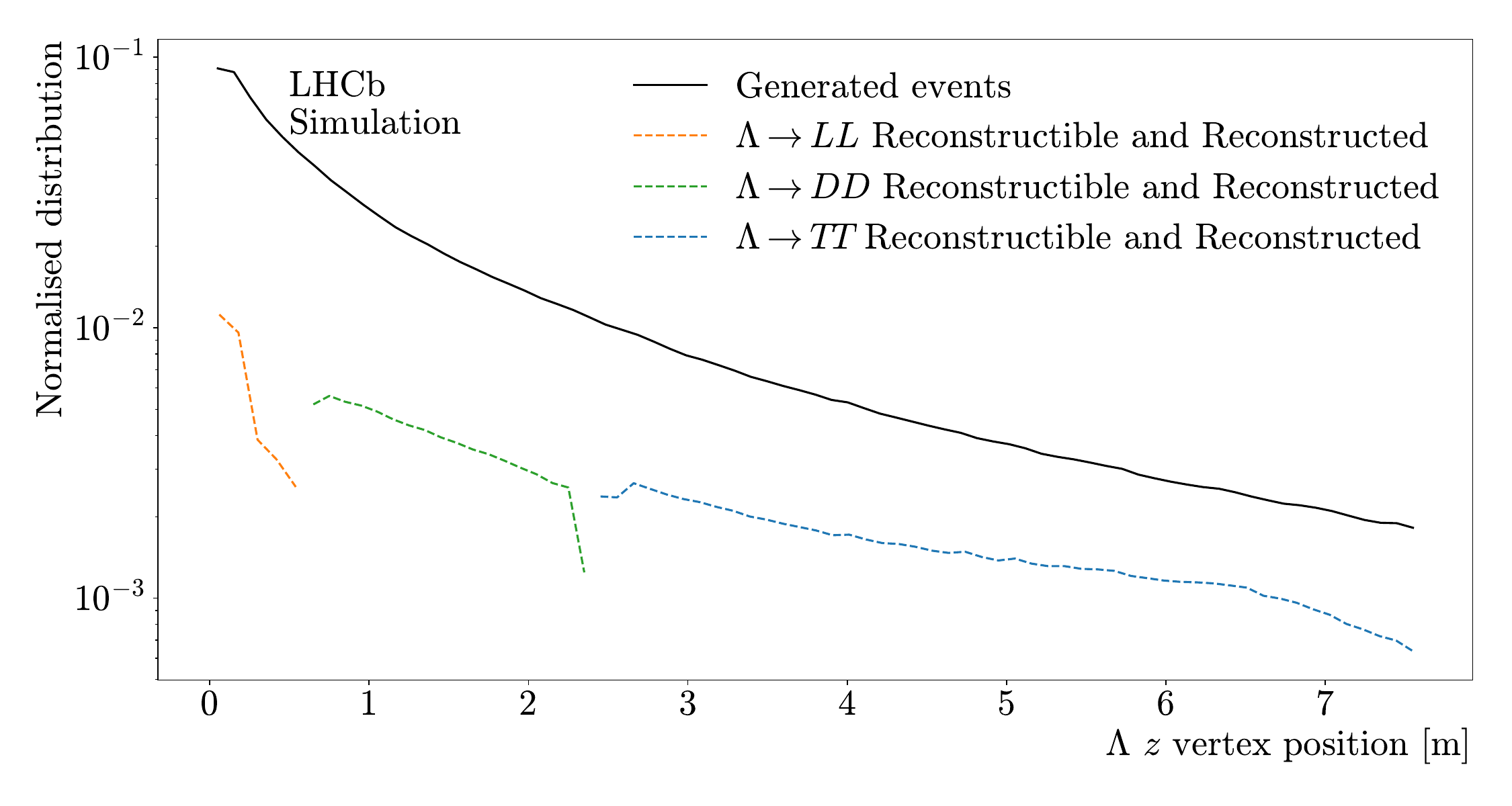}
\caption{Fraction of \mbox{$\varLambda_b^0 \rightarrow J/\psi \varLambda$} 
signal decays that are reconstructible times the four-track reconstruction efficiency as a function of the $\varLambda$ 
decay vertex position along the $z$ axis for the different track categories of the final-state particles. The generated distribution is shown as a solid black line. The two distributions are normalised to the total number of generated events. The generated $\varLambda$ 
lifetime is 0.263\,ns.
}
\label{fig:recoected_distribution}
\end{figure}

The reconstruction and selection efficiency is estimated using the simulation, and
is defined here as the probability that a reconstructible \LbToJpsiLz signal decay 
%
%
is actually reconstructed and passes all the selection requirements.
%
The reconstructible efficiency is itself defined as the ratio between the number of signal events with all their final-state trajectories (\ie the four tracks) within the \lhcb detector acceptance and intersecting the minimum detector elements required for each track, and all the generated signal events with their final-state particles emitted with polar angle between 0.01 and 0.4\rad.
Figure~\ref{fig:efficiency_recocted} compares the reconstructible and four-track reconstruction efficiencies for different decay regions
(track categories) of the \Lz final-state particles, namely Long, Downstream and T,\footnote{These regions are defined with the $z$ position of the \Lz decay vertex in the ranges 0--0.6\,m, 0.6--2.4\,m and 2.4--7.6\,m (see Fig.~\ref{fig:tracktypes}), respectively.}
as a function of the \Lz decay vertex position along the $z$ axis.
Integrating over $z$ in the different track-type regions, the reconstructible fractions (four-track reconstruction efficiencies) amount to about 11\%, 34\% and 55\%  (72\%, 55\% and 60\%) in the Long, Downstream and T volumes, respectively.
It is also illustrative to compare the product of the reconstructible and the four-track reconstruction efficiencies as a function of the \Lz decay vertex $z$ position to the generated distribution, as shown in Fig.~\ref{fig:recoected_distribution}.
Integrated over the different regions, the former accounts for about 17\%, 40\% and 43\% of the events, to be compared to
40\%, 37\% and 23\% of the latter. 

\begin{figure}[tb]
\centering
\includegraphics[width=\textwidth]{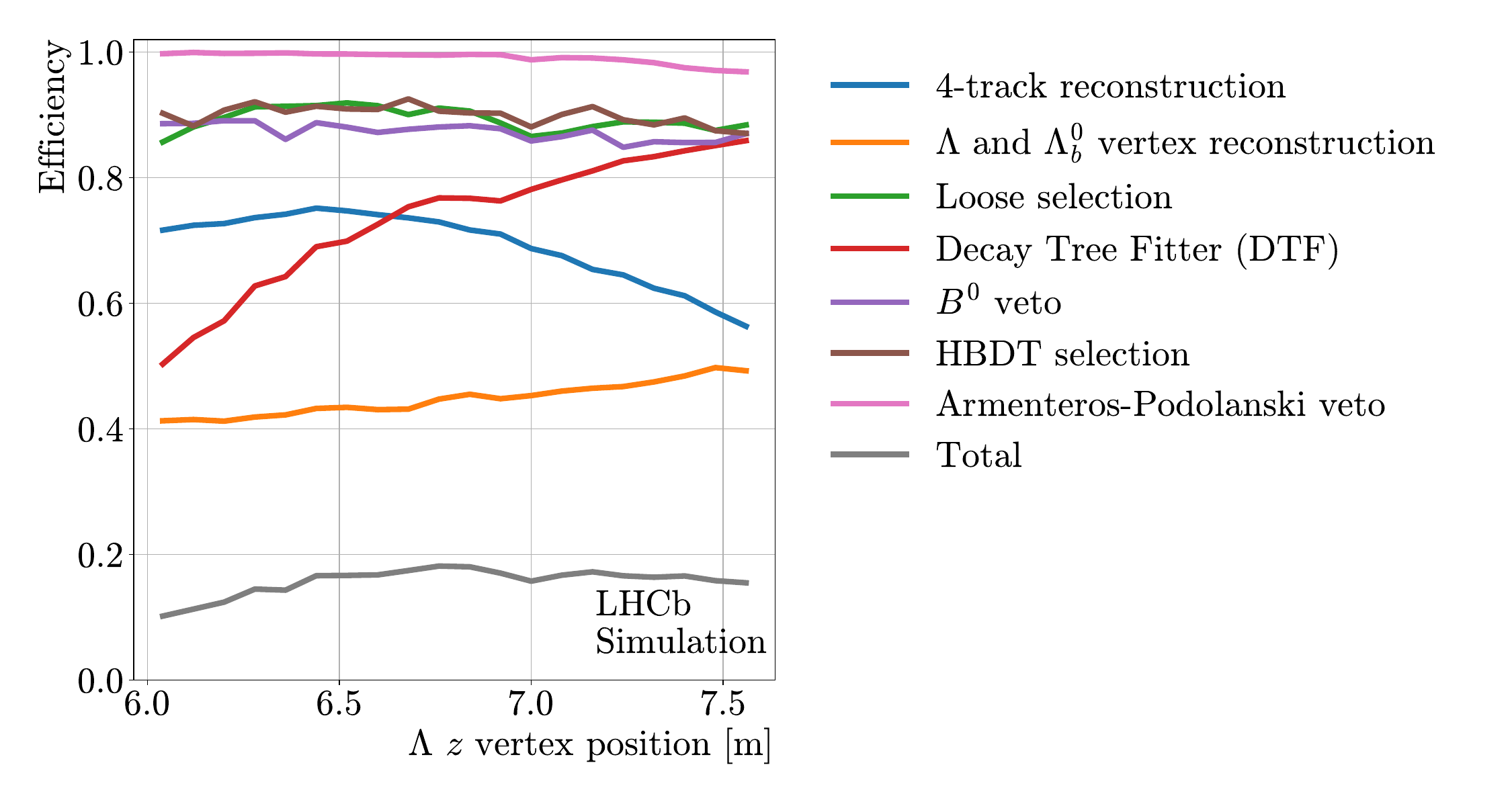}
\caption{Reconstruction and selection efficiency as a function of the true $\varLambda$ 
decay vertex position along the beam axis for simulated \mbox{$\varLambda_b^0 \rightarrow J/\psi \varLambda$} 
signal events (grey) and its breakdown for different reconstruction and selection steps: the reconstruction and quality selections of the four tracks of the final state (blue); the reconstruction of the $\varLambda$ 
and $\varLambda_b^0$ 
decay vertices (orange); the initial selections applied to the final-state and decaying particles as described in Table~\ref{tab:preselections} (Appendix~\ref{appendix:LbKS}, green); the fraction of successful 
decay chain vertex fits (DTF, red); the $B^0$ 
mass veto (violet); the selection efficiency of the HBDT (maroon); and the AP veto (pink). The four-track efficiency contribution and the total efficiency (see text) are evaluated for reconstructible signal decays, while all other contributions are relative to the previous reconstruction step. The inefficiencies of the vertex reconstruction and the decay chain fit are dominated by the reconstruction of the $\varLambda$ candidate.}
\label{fig:efficiency}
\end{figure}

Signal decays with a \Lz baryon decaying in the region between 6.0 and 7.6\,m represent 4\% of all generated events and have a reconstructibility of 59\% and a four-track reconstruction efficiency of 70\%. Overall, this represents about 8\% of all reconstructible and four-track reconstructed decays. 
%
Figure~\ref{fig:efficiency} shows the reconstruction efficiency and its breakdown for the different reconstruction and selection steps as a function of the true \Lz decay-vertex position along the beam axis. The efficiency is about $8$\% at $6.0$\,m  from the IP; it 
rises to about $16$\% at $6.8$\,m; then it decreases slightly staying above $13$\% between $7.0$ and $7.6$\,m.
The reconstruction of the \Lz and \Lb vertices has an efficiency of about $40-50$\%, with a slightly larger success rate when the \Lz baryon decays closer to the T stations (see Fig.~\ref{fig:tracktypes}).
This failure rate is mainly due to the poor track momentum resolution and the presence of non-Gaussian effects not properly accounted for by the $\chi^2$ ansatz of the vertex fitter. A test with a single stage reconstruction based on the decay chain vertex fitter (DTF) provided higher vertex reconstruction efficiency, approximately 60\%, and will be used for future physics analyses.
The efficiency of the loose selection is about $90$\%, mostly independent of the \Lz decay location.
The decay chain vertex fit convergence depends on the \Lz decay vertex position, going from about $50$\% for \Lz baryons decaying farthest from the T stations, to about $86$\% for those closest to them. This highlights the impact of the propagation of T tracks through the magnetic field. The HBDT and AP veto selection efficiencies are about $72$\% and $99\%$, largely independent of the position of the \Lz decay vertex.

\subsection{Momentum, vertex and angular resolutions}
\label{subsec:resol}

\begin{figure}[tb]
\centering
{\includegraphics[height=7cm]{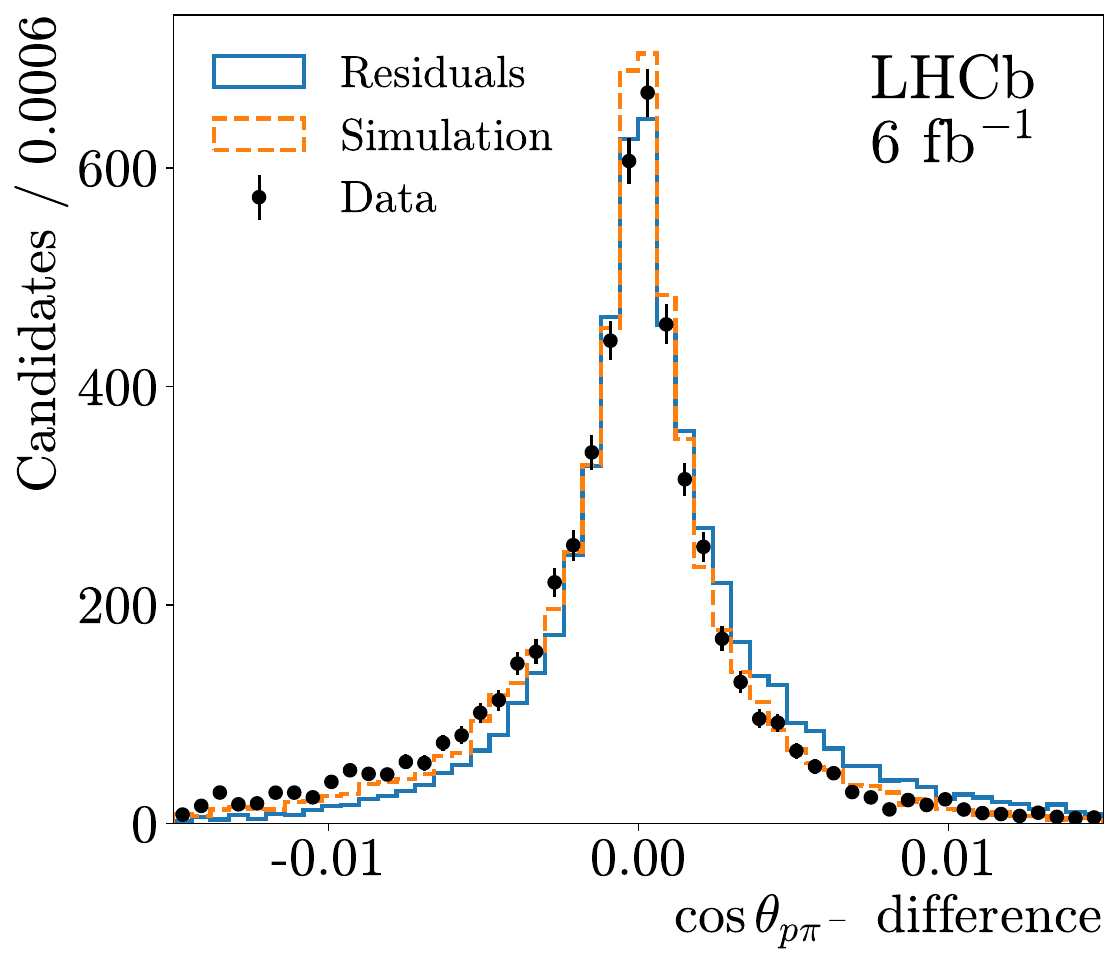}}
\hspace{0.5cm}
\caption{Distributions of the event-by-event differences between the cosine of the proton-pion opening angle at the vertex position reconstructed using the vertex and the decay chain fitters, for \mbox{$\varLambda_b^0 \rightarrow J/\psi \varLambda$} 
signal candidates in simulation (orange histogram) and data (black points). For simulation, the residual distribution using the vertex fitter is overlaid (blue histogram).
}
\label{fig:angular_resolution}
\end{figure}

The momentum resolution of T tracks suffers due to their relatively low curvature, as they are reconstructed exclusively using hits located in a region with weak magnetic field and relatively short lever arm.
%
%
The relative momentum resolution as a function of momentum has been measured in simulation and data using the data procedure described in Appendix ~\ref{appendix:pfromdata}, and is shown in Fig.~\ref{fig:mom_res}. The results in simulation are compared to the resolution determined from residuals, defined as the difference between reconstructed and true momenta.
Candidates satisfying all the selection criteria and requiring in addition the fake-track probability to be below 0.5\%, which has an efficiency of 70\% per track, have been considered. 
The relative resolution from the track fit is about 
20\%, improving to 4--5\% when the constraints of the \Lb decay chain are imposed.
Within uncertainties, data and simulation are consistent,
although the former tend to be 10--20\% above the latter. 
When the \Lz parent particle is reconstructed using the vertex fitter, the procedure relies on the measurement of the $m(p \pim)$ resolution,
along with the resolution of the opening angle between the proton and the pion at the vertex position, both determined as a function of the track momentum. 
The invariant-mass resolution is measured as described previously, 
in regions (bins) of momentum, 
whereas the angular resolution is determined from the distribution of per-event differences between the angle reconstructed using the vertex fitter and DTF, since the resolution for the former is about a factor of four worse.
Figure~\ref{fig:angular_resolution} shows this distribution as obtained from simulation and data signal candidates passing all selection criteria, compared to the distribution of residuals in simulation, \ie replacing the reconstructed angle with the decay chain fitter by its true value, integrated over momenta.
The average resolution on the cosine of this angle is $0.003$ and $0.004$, for simulation and data, respectively.
%
%
%
In contrast, when the \Lz parent particle is reconstructed using the \mbox{\LbToJpsiLz} decay chain with all constraints, the momentum resolution is mostly determined by the $m(\jpsi\Lz)$ invariant mass, since other contributions, including the opening angle between the \jpsi and the \Lz hadrons, are subdominant. 

%
%
\begin{figure}[tb]
\centering
\includegraphics[height=6.7cm]{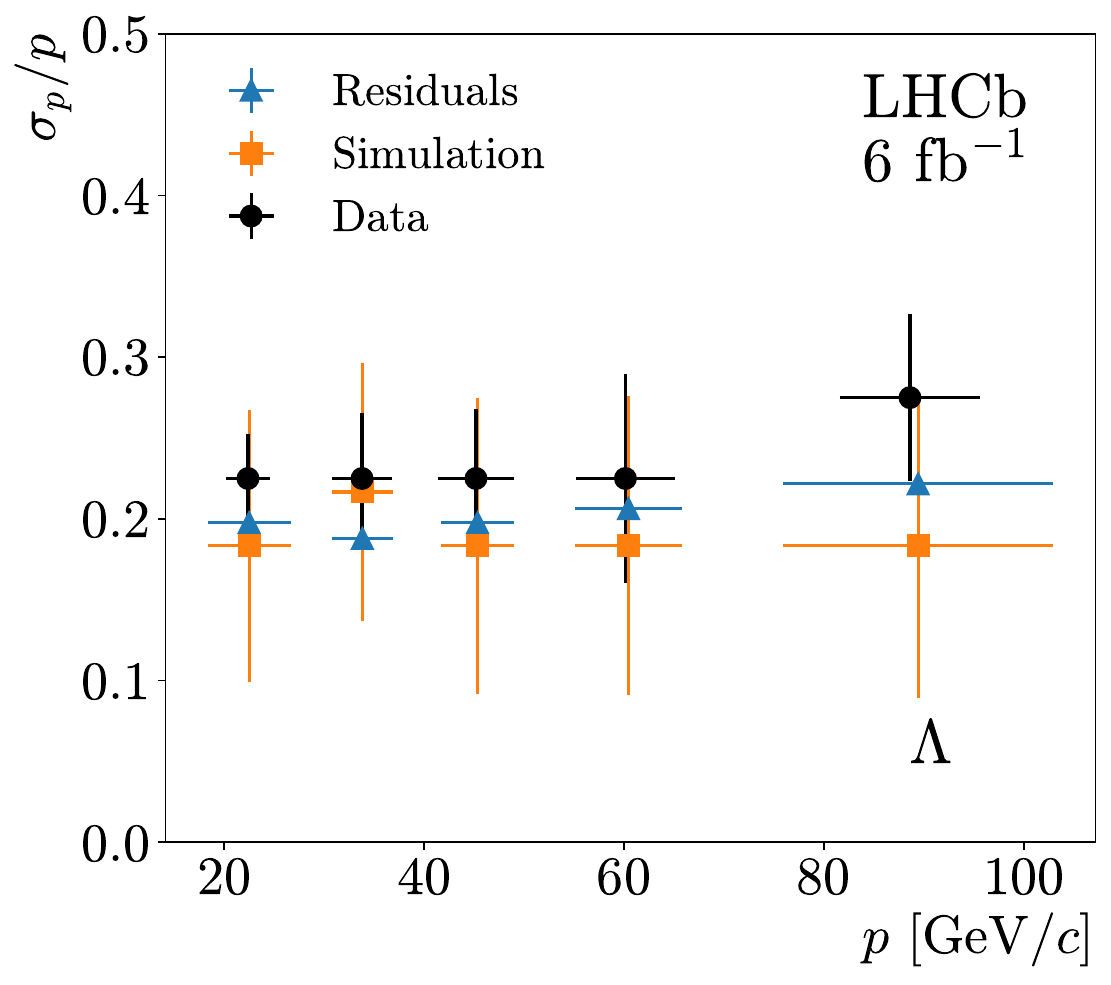}
\includegraphics[height=6.7cm]
{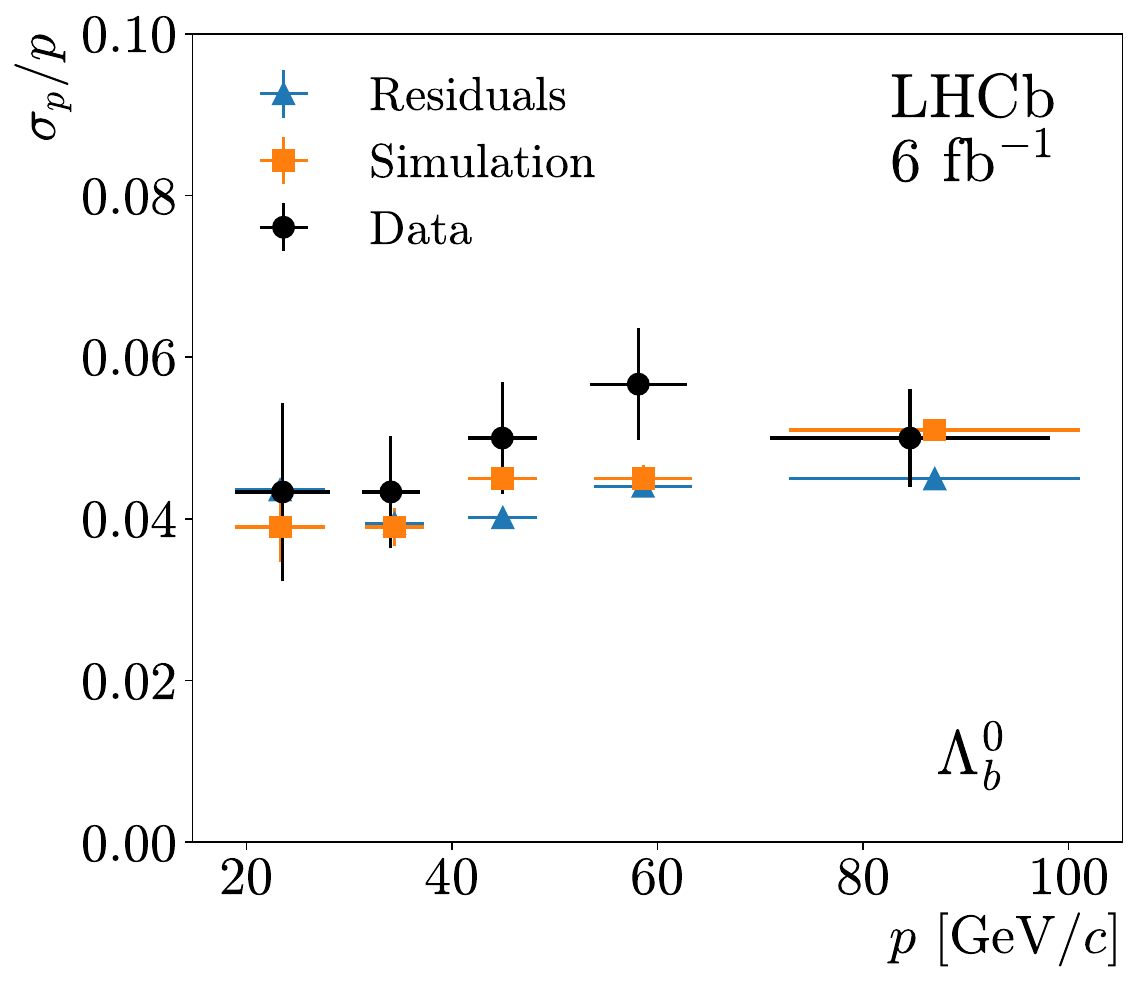}
\caption{Relative momentum resolution as a function of momentum for protons reconstructed as final state of the \mbox{$\varLambda_b^0 \rightarrow J/\psi \varLambda$} 
decay, obtained with the (left) vertex and (right) decay chain fitters. The resolution obtained from residuals in simulation (blue triangles) is compared to that obtained from the data procedure 
applied to both simulation (orange squares) and data (black points). The improved momentum resolution on the right compared to the left is due to the geometric and kinematic constraints imposed when fitting the whole decay chain.}
\label{fig:mom_res}
\end{figure}

\begin{figure}[tb]
\centering
\includegraphics[height=6cm]{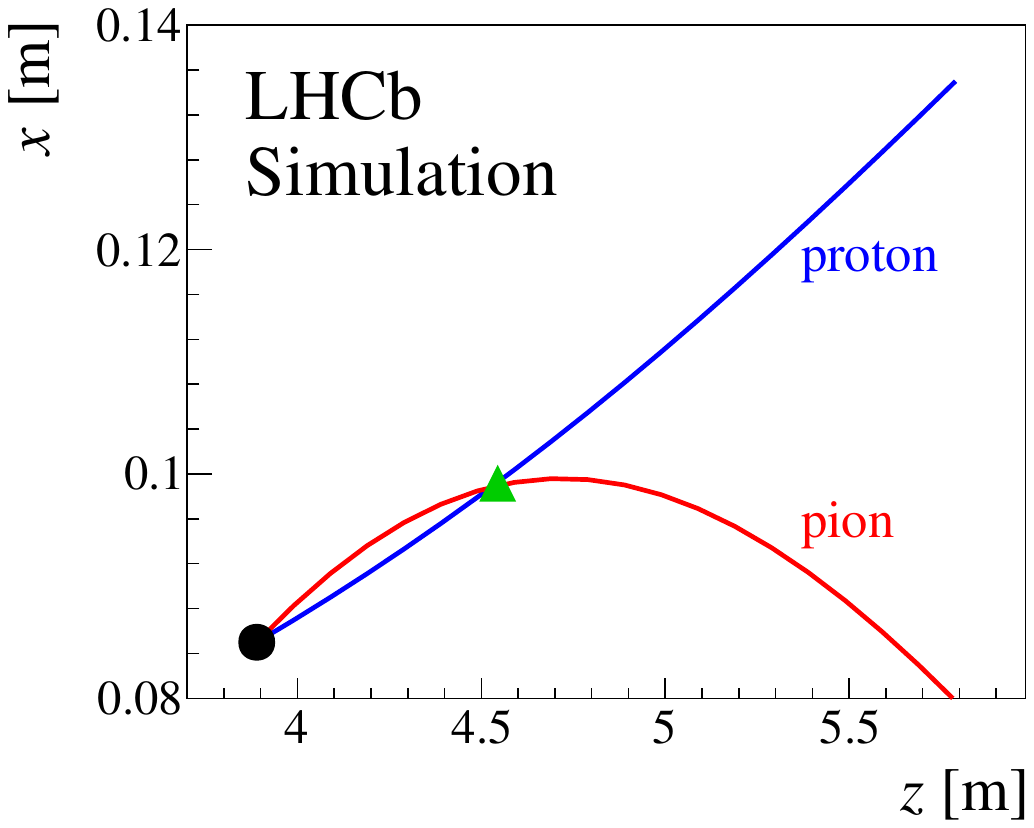}    
\includegraphics[height=6cm]{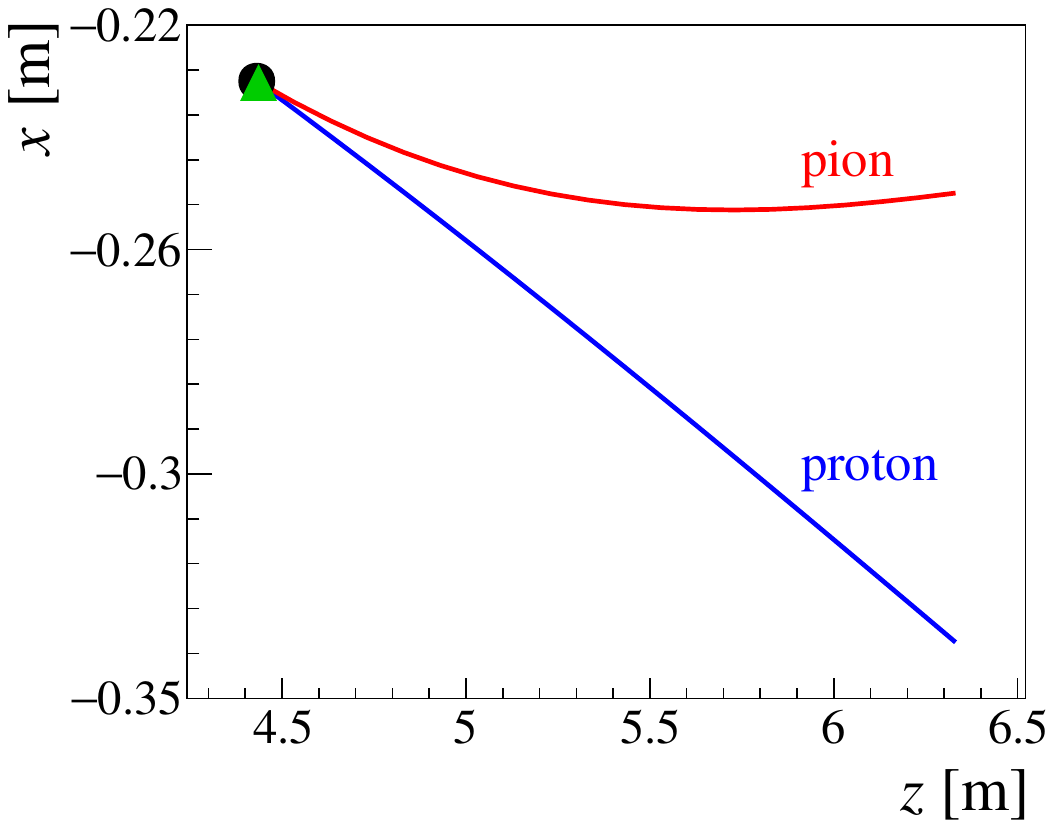}
\caption{Event displays with (blue) proton and (red) pion trajectories together with the true 
(black dot)
and the reconstructed (green triangle)
vertices from simulation. The reconstructed vertex is wrongly assigned to the downstream (larger $z$) crossing point between particle trajectories in most of the decays with closing-track topology (left), while it is found correctly in opening-track topologies (right). }
\label{fig:event_displays}
\end{figure}

\begin{figure}[tb]
\centering
\includegraphics[height=7cm]{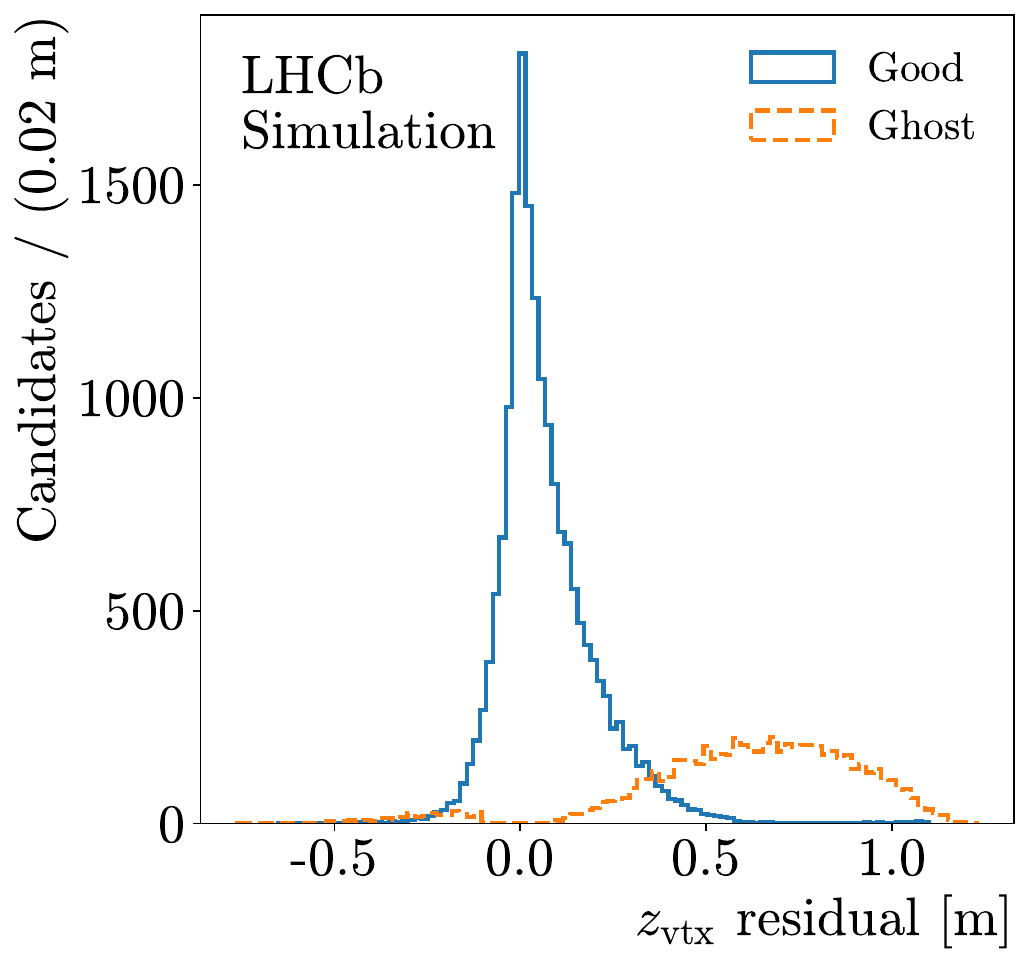}
\caption{Residual distribution of the reconstructed $\varLambda$ 
$z$ position for \mbox{$\varLambda_b^0 \rightarrow J/\psi \varLambda$} 
signal candidates in simulation tagged as {\em Good} and {\em Ghost}, 
with true $\varLambda$ 
vertex along the $z$ axis above 6.0\,m,
before applying the BDT requirement.
}
\label{fig:select_real_ghosts}
\end{figure}

\begin{figure}[tb]
\centering
\includegraphics[width=0.7\textwidth]{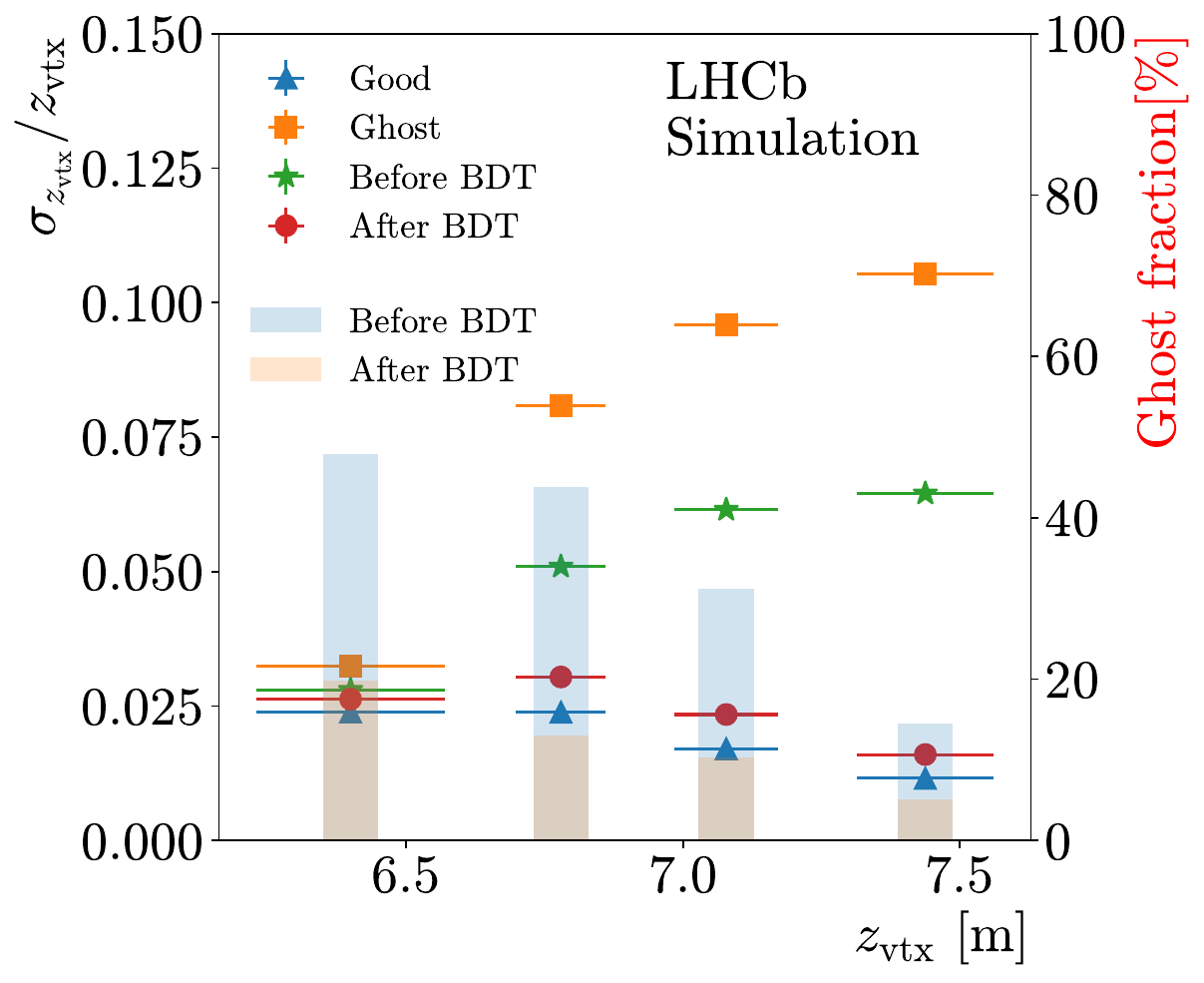}
\caption{Relative precision on the reconstructed $\varLambda$ 
decay vertex position along the beam axis for \mbox{$\varLambda_b^0 \rightarrow J/\psi \varLambda$} 
signal candidates in simulation tagged as {\em Good} 
(blue triangles) and {\em Ghost} 
(orange squares), as well as for events before (green stars) and after (red points) the BDT selection. 
The resolution is evaluated as the central 68.3\% CL region of the underlying residual distribution, adding in quadrature the offset.
The fraction of {\em Ghost} 
events before (light blue shadow) and after (light orange shadow) the BDT selection criteria are also indicated. The resolution achieved after the BDT selection approaches closely that obtained for {\em Good} 
events.}
\label{fig:vertex_resolution}
\end{figure}

The reconstruction of the \Lz decay vertex using T tracks is particularly challenging as a consequence of three concurring effects: the aforementioned low curvature, the long track-transportation over large distances with intense and inhomogeneous field, 
and the presence of decays with a closing-track topology.
These events,
sketched in \mbox{Fig.~\ref{fig:event_displays} (left)}, 
exhibit particle trajectories with two (consistent within track uncertainties) crossing points, inducing the vertex fitter to converge frequently to the wrong vertex (denoted in the following as \ghost). 
In contrast, opening-track decays, shown in \mbox{Fig.~\ref{fig:event_displays} (right)}, do not have a ghost vertex and converge to the correct position (denoted \good).
Figure~\ref{fig:select_real_ghosts}
illustrates the residual distributions of the reconstructed \Lz vertex $z$ position, $z_{\rm vtx}$, for \good and \ghost signal events from simulation. Here, all selection criteria and additional vertex quality requirements, in particular 
$z_{\rm vtx}$ less than $7.6$\,m and its uncertainty, $\sigma_{z_{\rm vtx}}$, estimated event-by-event by the vertexing algorithm, below $0.3$\,m are applied.\footnote{These additional vertex quality requirements are applied in the following to all studies related to vertex resolution.} 
Simulated events are labelled in these categories by propagating through the detector volume the true trajectories and searching for the $z$ positions of closest approach between the two tracks in the $xz$ plane. When there exists a unique minimum, or with two minima and the reconstructed $z$ position closer to the true decay vertex, the event is tagged as \good, and \ghost otherwise. \ghost events, amounting to about 30\% of the reconstructed decays, show a broad residual distribution, strongly biased towards larger $z$ positions. 
Note that \ghost events biased towards low $z$ positions can also occur, but their contribution is largely suppressed, as can be observed in 
Fig.~\ref{fig:select_real_ghosts}.
%
Consequently, there is a significant vertex bias and resolution degradation, as illustrated in Fig.~\ref{fig:vertex_resolution}. 
Indeed, for \good events the \Lz vertex position resolution along the beam axis amounts to about $2.5$\% for vertices reconstructed around $6.0$\,m from the nominal IP, and improves by about a factor of two for vertices located closer to the T stations, around $7.5$\,m.
For \ghost events, the corresponding resolutions at these positions are
3\% and 10\%.
For each region in the reconstructed $z$ position, 
the resolution function for \good events has its most probable value near zero although it has an asymmetric shape towards larger $z$ values, as consequence of the large and non-Gaussian momentum uncertainties and the track transport from higher to lower $z$ positions. These features are also observed when averaging over $z$, as illustrated in Fig.~\ref{fig:select_real_ghosts}.
The resolution function can be well described by the sum of two Gaussian functions with different means and widths, one centred at zero describing the core resolution, and the other with an offset to account for the positive tail.
For \ghost events it is well represented by a single Gaussian function with an offset, neglecting the very small contribution of \ghost events with negative residual.


Closing- and opening-track geometries can be identified through a Boosted Decision Tree (BDT) classifier. Among the variables included in the classifier, the horizontality $h$ plays a preeminent role. This variable is defined as 
the $y$ component of the unit vector normal to the \Lz decay plane multiplied by the dipole magnet polarity, 
\mbox{$\text{sign}(B_y)$}, 
and the proton/antiproton charge, 
\mbox{$\text{sign}( \Lz )=+1~(-1)$} for \Lz (\Lbar) baryons, \ie
\begin{equation}
    h = \text{sign}( \Lz ) \text{sign}(B_y) \frac{\left({\bf p}_\proton \times {\bf p}_{\pim}\right)_y}{|{\bf p}_\proton \times {\bf p}_{\pim}|}, 
\end{equation}
where ${\bf p}_\proton$ and ${\bf p}_{\pim}$ are the three-momenta of the proton and pion. 
Decays with the extreme values $h = \pm 1$ lie exactly in the $xz$ bending plane, with $h = -1$ ($h = +1$) events having a completely closing (opening)-track topology, whereas $h = 0$ events lie in the $yz$ plane. Although having $h<0$ is necessary to identify \ghost events, this 
condition is not sufficient, requiring additional information.  
Other features of the classifier include kinematic and topological variables such as the $z_{\rm vtx}$ coordinate, the vertex uncertainty along the $y$ component and the $\chi^2_{\rm vtx}$ of the \Lz candidate,  
the opening angle between the proton and the pion, the coordinates of the point of closest approach between the proton and pion, the $m(p\pim)$ invariant mass, the \Lz direction angle, and the fake-track probabilities of the proton and pion. The momentum of the final-state particles, the \Lz decay length and the \Lb decay vertex position as obtained from the decay chain fitter are also used.
The BDT is trained with \good and \ghost simulated signal events. The chosen threshold on the BDT response has a signal efficiency of 75\% and removes 75\% of the \ghost events, reducing from 30\% to 6\% the amount of \ghost events in the simulation sample, as illustrated in 
Fig.~\ref{fig:select_real_ghosts}~(right).
As a result, the $z$ vertex resolution becomes close to that of \good events, as it is shown in Fig.~\ref{fig:vertex_resolution}.

\begin{figure}[tb]
\centering
{\includegraphics[height=7cm]{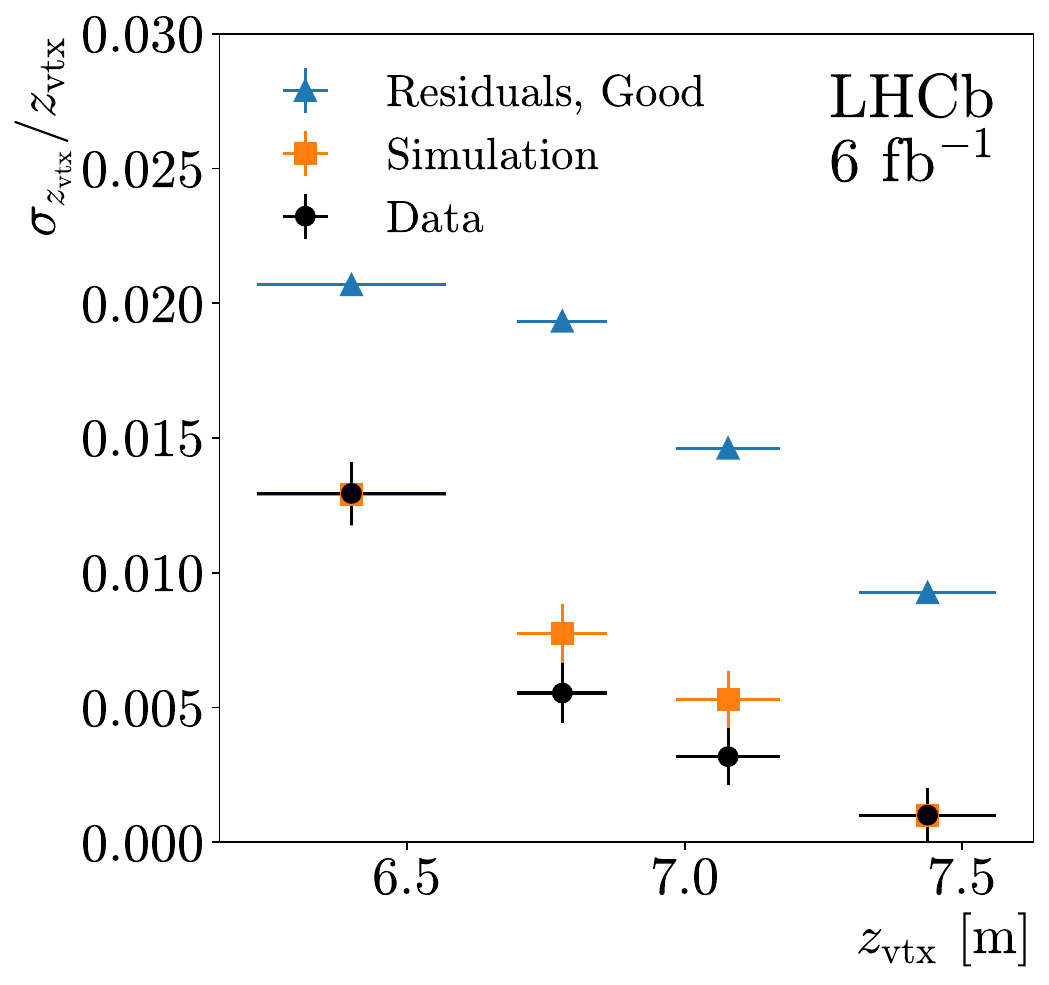}}
{\includegraphics[height=6.9cm]{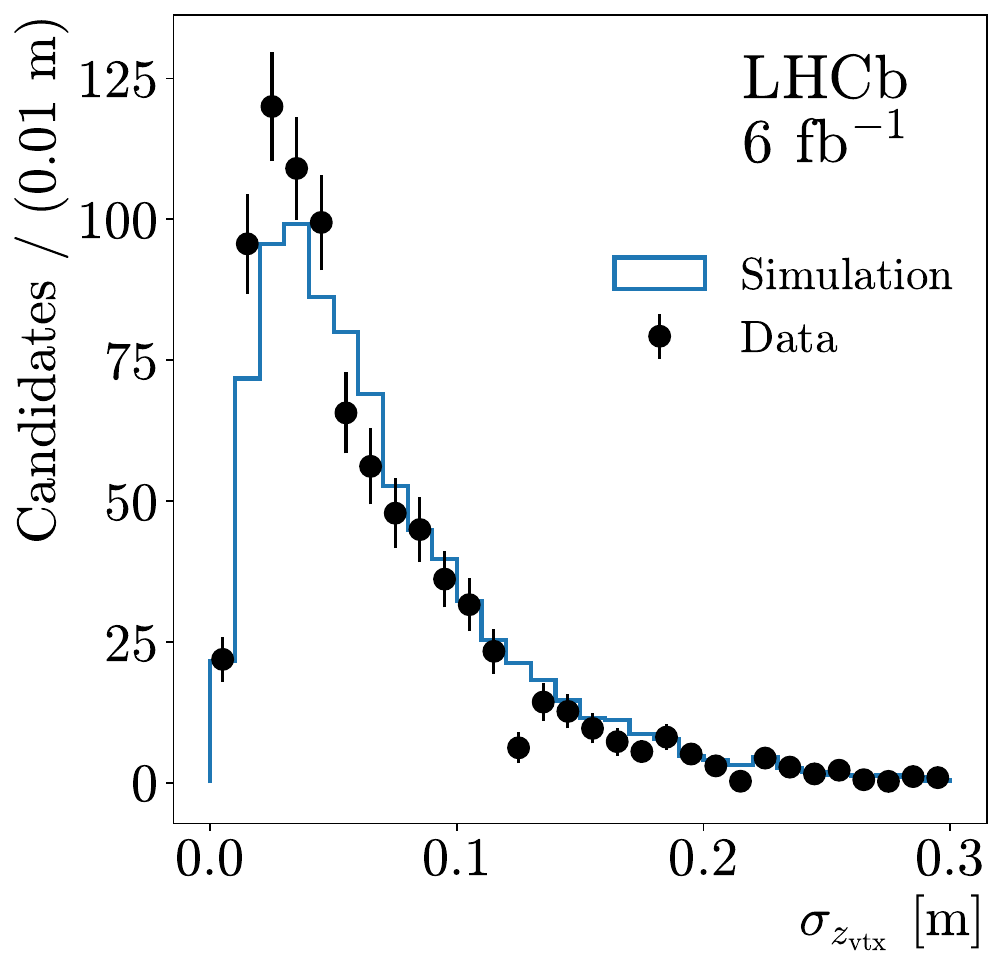}}
\caption{(Left) most probable event-by-event uncertainty divided by the mean value of the reconstructed $\varLambda$ 
vertex position along the $z$ axis, for \mbox{$\varLambda_b^0 \rightarrow J/\psi \varLambda$}  
signal candidates in simulation (orange squares) and data (black points), and in different $z_{\rm vtx}$ regions. 
The resolutions in simulation estimated from residuals for {\em Good} 
events are also shown (blue triangles). The binning scheme is the same as in Fig.~\ref{fig:vertex_resolution}.
(Right) distributions of the $\sigma_{z_{\rm vtx}}$ uncertainties for \mbox{$\varLambda_b^0 \rightarrow J/\psi \varLambda$}  
simulation (histogram) and data (points) candidates falling into the second bin of the reconstructed $z$ position.
}
\label{fig:PeE_MC_data}
\end{figure}

The measurement of the \Lz vertex resolution provided by the vertex fitter offers a data procedure to estimate on a event-by-event basis the vertex resolution. Figure~\ref{fig:PeE_MC_data} (left) illustrates the most probable values of these estimates along the $z$ axis, 
$\sigma_{z_{\rm vtx}}$, 
for both simulation and data, and different regions (bins) of $z_{\rm vtx}$. Error bars are evaluated adding in quadrature the 68.3\% CL region of the per-event distribution divided by the root square of the yields, and the half difference between the most probable and the median values.
Figure~\ref{fig:PeE_MC_data} (right) shows the distributions of the $\sigma_{z_{\rm vtx}}$ estimates for candidates falling in the second bin of the (left), for simulation and data. The distributions can be described using a Johnson ${\rm S_U}$ function~\cite{JohnsonSU}. 
There is a good agreement simulation and data.
The global offset with respect to the resolution in simulation estimated from residuals, also shown, is a consequence of the non-Gaussian effects in the reconstruction and the irreducible offset of the resolution function. 
The $\sigma_{z_{\rm vtx}}$ distributions are unaffected by the presence of \ghost events.

\begin{figure}[htb]
\centering
\includegraphics[height=7.1cm]{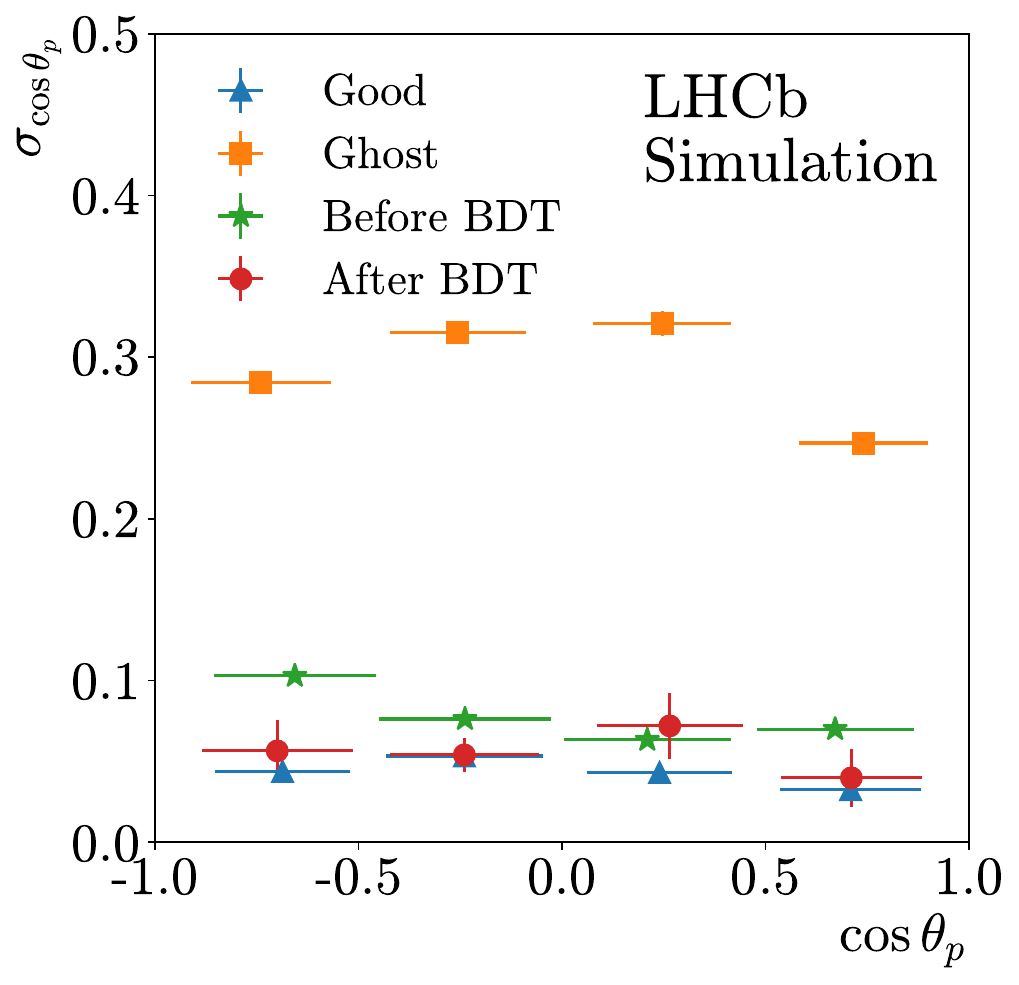}
\includegraphics[height=7.1cm]{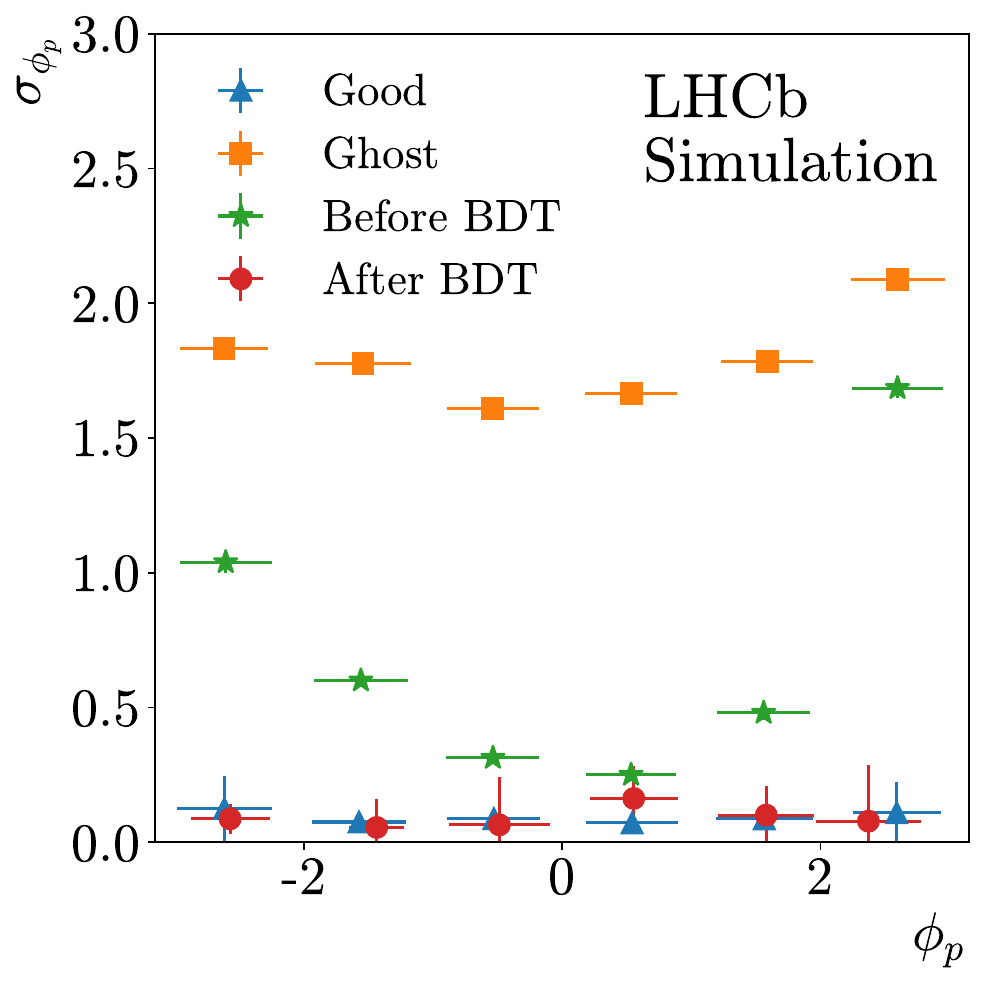}
\caption{Resolution on the (left) $\cos\theta_p$ and (right) $\phi_p$ helicity angles for \mbox{$\varLambda_b^0 \rightarrow J/\psi \varLambda$} 
signal candidates in simulation tagged as {\em Good} 
(blue triangles) and {\em Ghost} 
(orange squares), as well as for candidates before (green stars) and after (red points) the BDT selection. 
The resolution achieved after the BDT selection approaches closely that obtained for {\em Good} 
events.
}
\label{fig:helicity_angles_Lambdab}
\end{figure}

The polar and azimuthal angles, $\theta_p$ and $\phi_p$, of the proton direction in the \Lz rest frame, reached from the \Lb rest frame and rotated by the spherical angles of the \Lz momentum in the \Lb frame, are essential ingredients of the \Lz polarisation determination required for the 
measurements of the magnetic and electric dipole moments~\cite{Botella:2016ksl,LHCb-PAPER-2020-005}. The resolution on these helicity angles as obtained using the decay chain fitter and residuals in simulation is shown in Fig.~\ref{fig:helicity_angles_Lambdab}. It is strongly affected by the presence of \ghost events. Candidates passing the BDT selection have a resolution close to that of \good events.

\section{Reconstruction of \texorpdfstring{\bfBdToJpsiKS}{B0} decays}
\label{sec:reconstructionB0}

A sample of long-lived \KS mesons from $\Bz\rightarrow\jpsi\KS$ decays is also reconstructed using a similar procedure to that described in Sec.~\ref{sec:reconstructionLb} for \LbToJpsiLz decays.
The loose selection criteria to identify \Bz candidates are summarised in  Table~\ref{tab:B0_preselections} of Appendix~\ref{appendix:LbKS}.

\begin{figure}[tb]
\centering
\includegraphics[height=6cm]{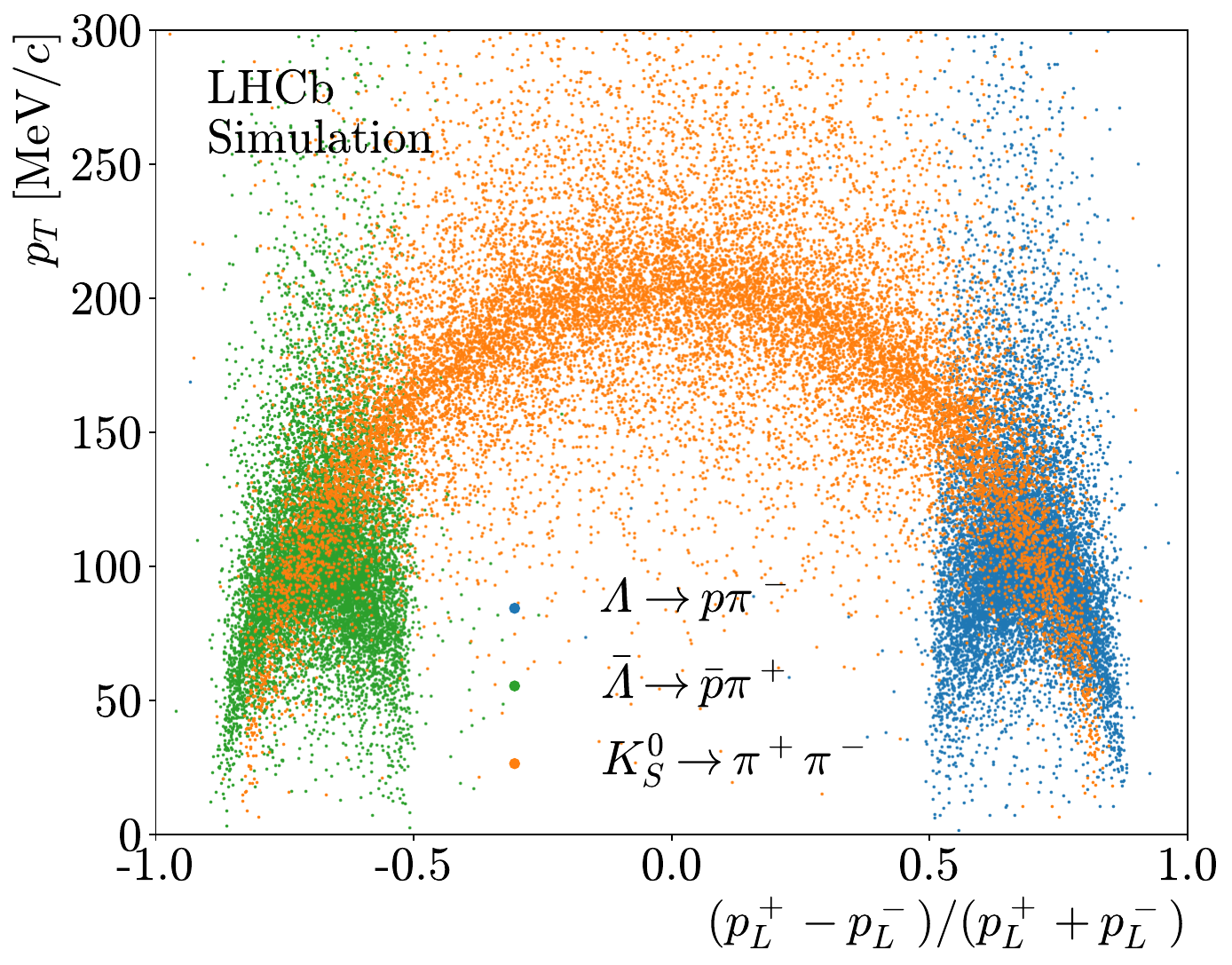}
\includegraphics[height=6cm]{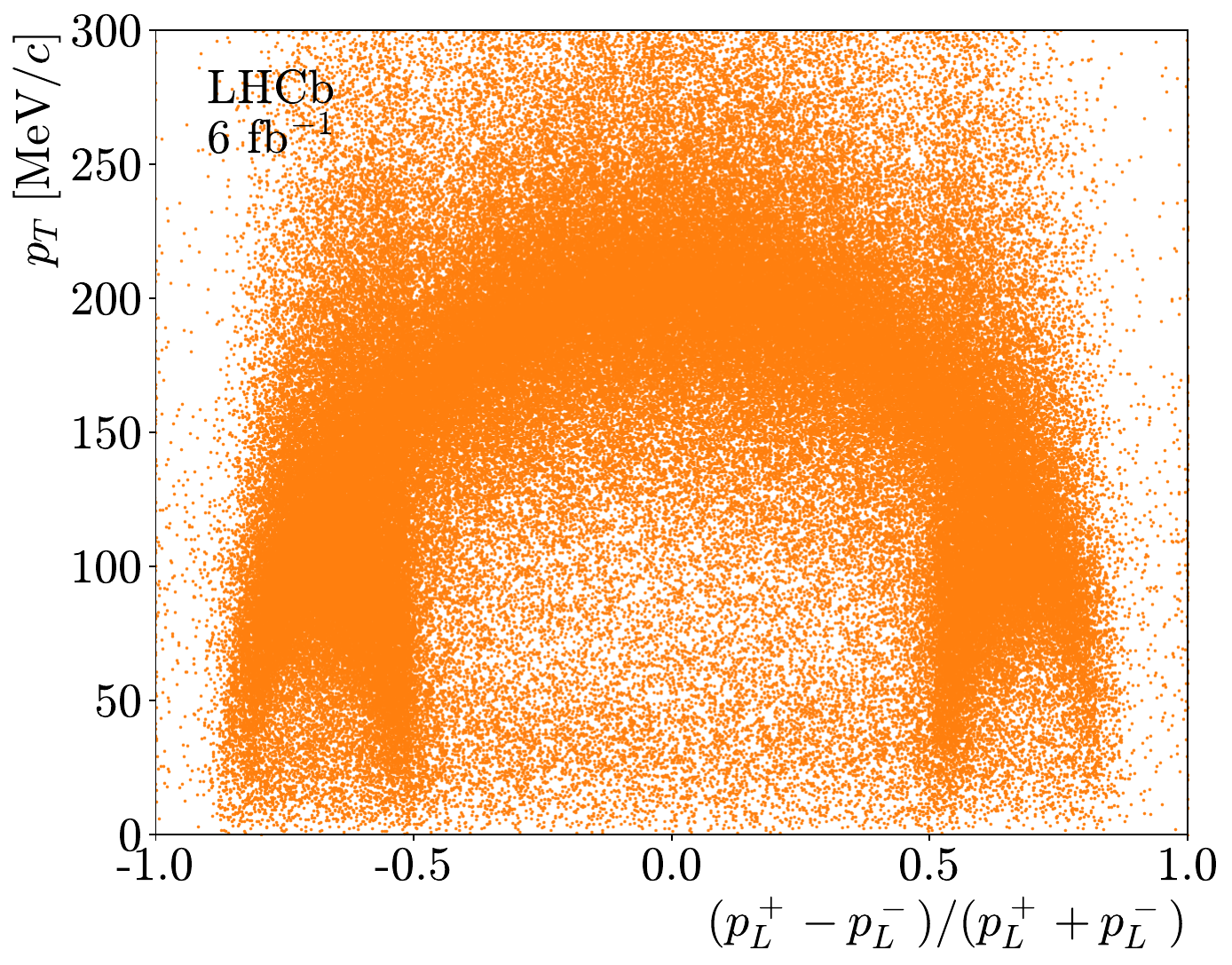}
\caption{The AP plot for \mbox{$B^0 \rightarrow J/\psi K_S^0$} 
candidates after the loose, HBDT and $\varLambda_b^0$ 
veto selection criteria, for (left) simulated \mbox{$\varLambda_b^0 \rightarrow J/\psi \varLambda$} 
and \mbox{$B^0 \rightarrow J/\psi K_S^0$} 
signal and (right) \mbox{Run 2} 
data. 
}
\label{fig:armenteros_podolanski_B0}
\end{figure}

Due to the similar topologies and the lack of PID information for T tracks in this analysis,
the \LbToJpsiLz decays represent a 
background source for the reconstruction of \BdToJpsiKS candidates. 
A veto is applied to reject candidates with $m(\jpsi\Lz)$ in the range $\pm150$\mevcc around the known \Lb mass~\cite{PDG2022}, where the positive pion from the \KS candidate is reconstructed with the proton mass hypothesis. This results in a $96$\% selection efficiency for the \Bz signal and $37$\% for the \Lb background.  
Similarly to the reconstruction of \LbToJpsiLz decays, an HBDT classifier is trained using about 22\,000 simulated \Bz signal decays and about 
$6\times10^6$
background candidates reconstructed in data
and selected from the lower and upper sidebands of the $m(\jpsi\KS)$ distribution.
The optimal HBDT operating point results in a signal efficiency of 91\% and a figure-of-merit ratio of 89.
Furthermore, by requiring the longitudinal momentum asymmetry in the \mbox{\BdToJpsiKS} AP plot,
shown in Fig.~\ref{fig:armenteros_podolanski_B0} for simulation and data, to be within the range $-0.5$ to $0.5$, 
99\% of the \mbox{\LbToJpsiLz} background is rejected while 66\% of the signal is retained.

\begin{figure}[tb]
\centering
\includegraphics[height=6.5cm]{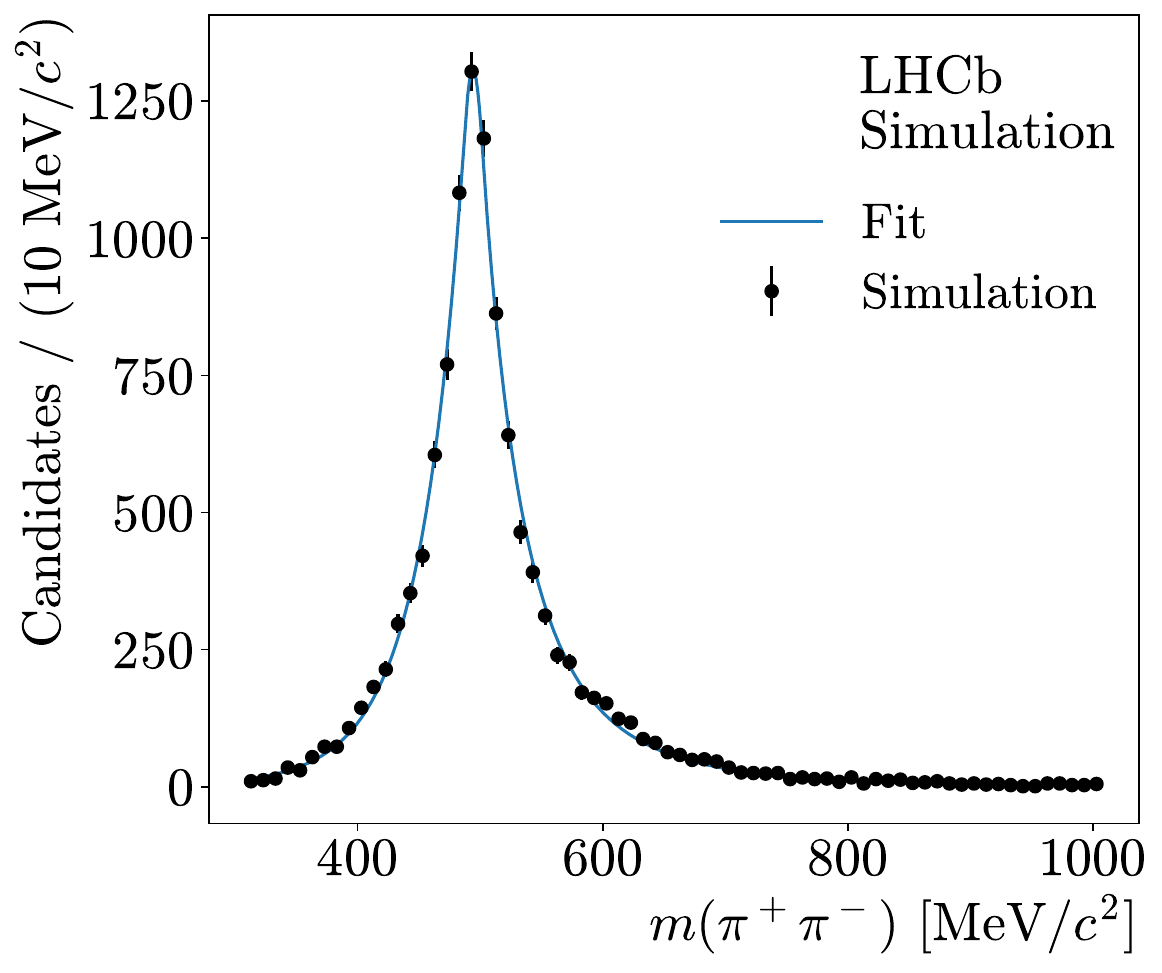}
\includegraphics[height=6.5cm]{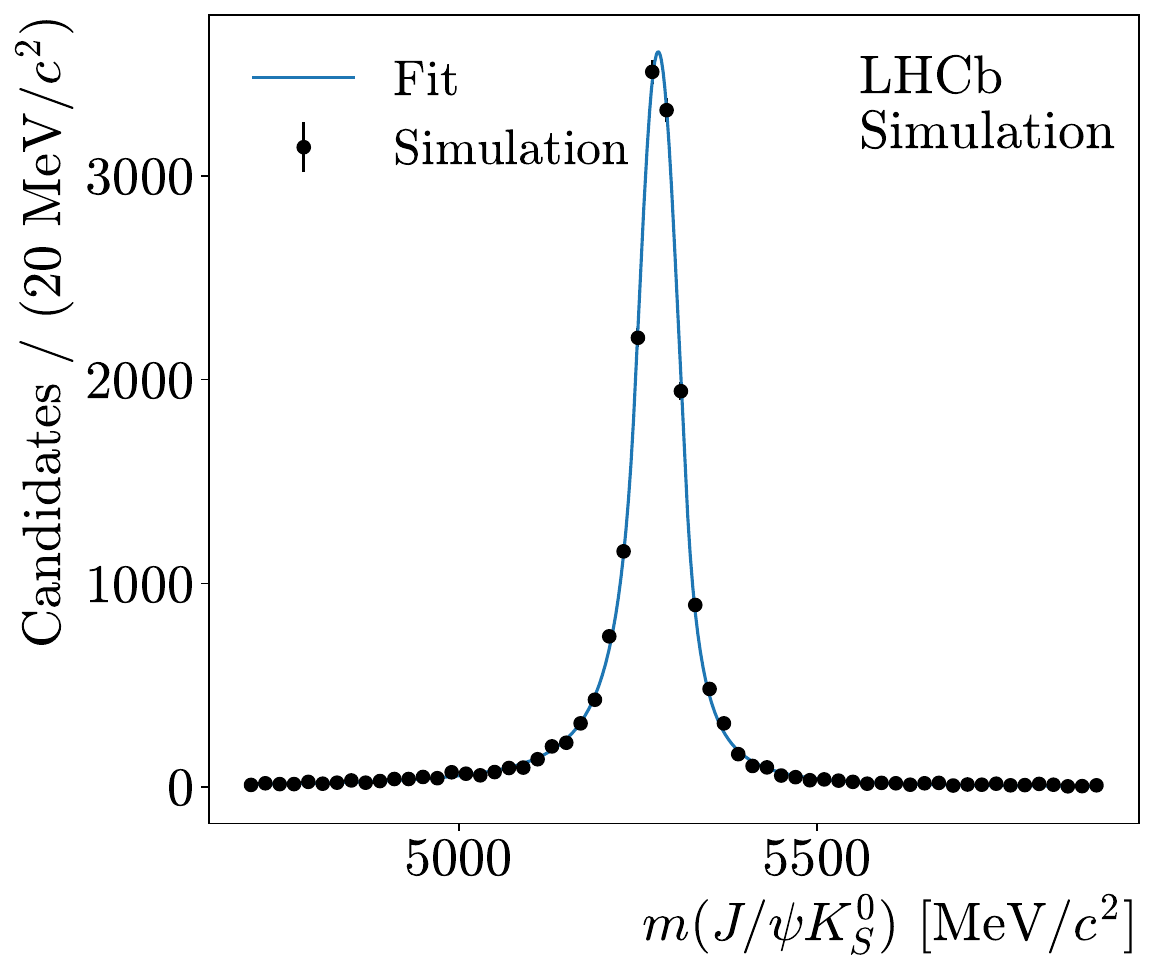}
\caption{Invariant-mass distribution (left) $m(\pi^+\pi^-)$ 
and (right) $m(J/\psi K^0_S)$ 
for simulated  \mbox{$B^0 \rightarrow J/\psi K_S^0$} 
signal decays. The mass fit results are overlaid.
}
\label{fig:inv_mass_MC_B0}
\end{figure}

\begin{figure}[tb]
\centering
\includegraphics[height=6.5cm]{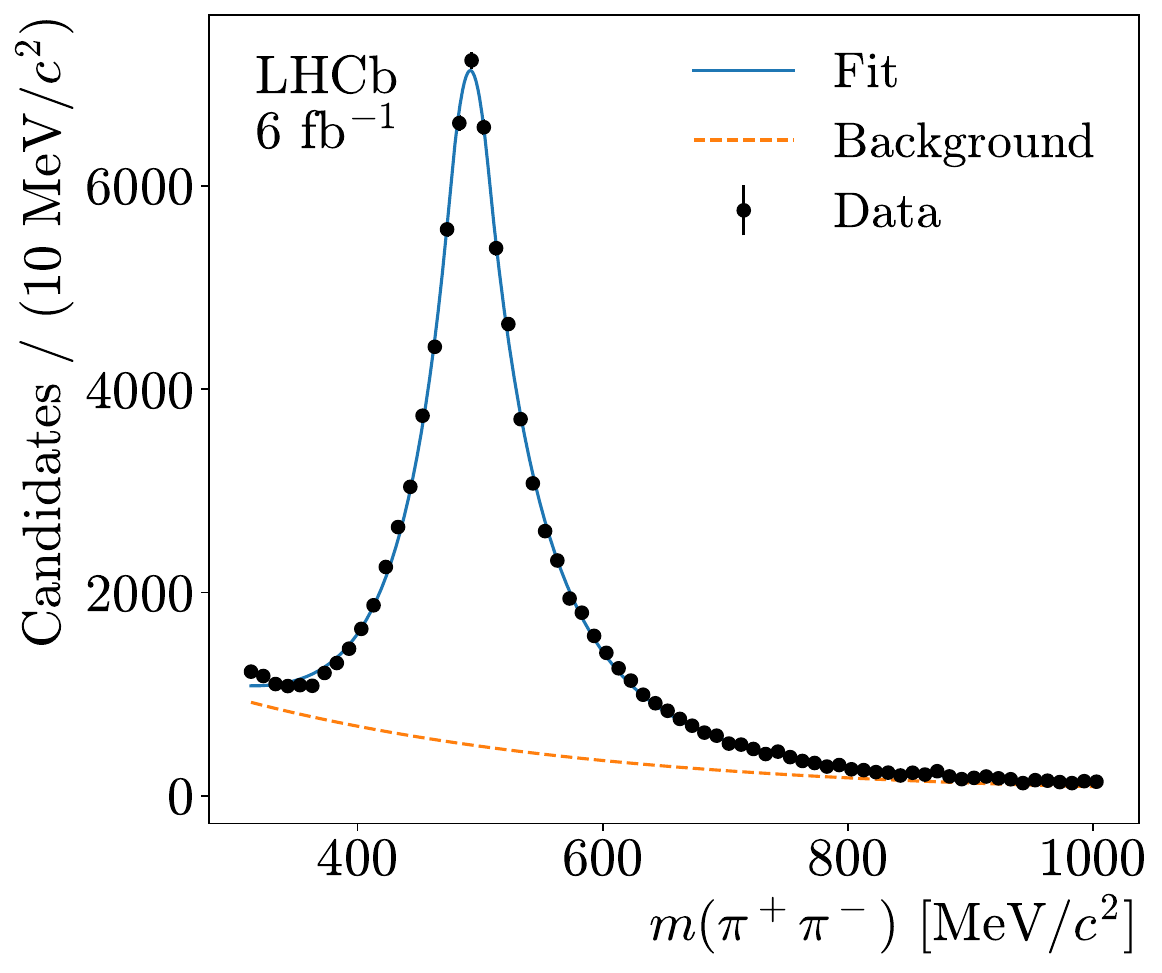}
\includegraphics[height=6.5cm]{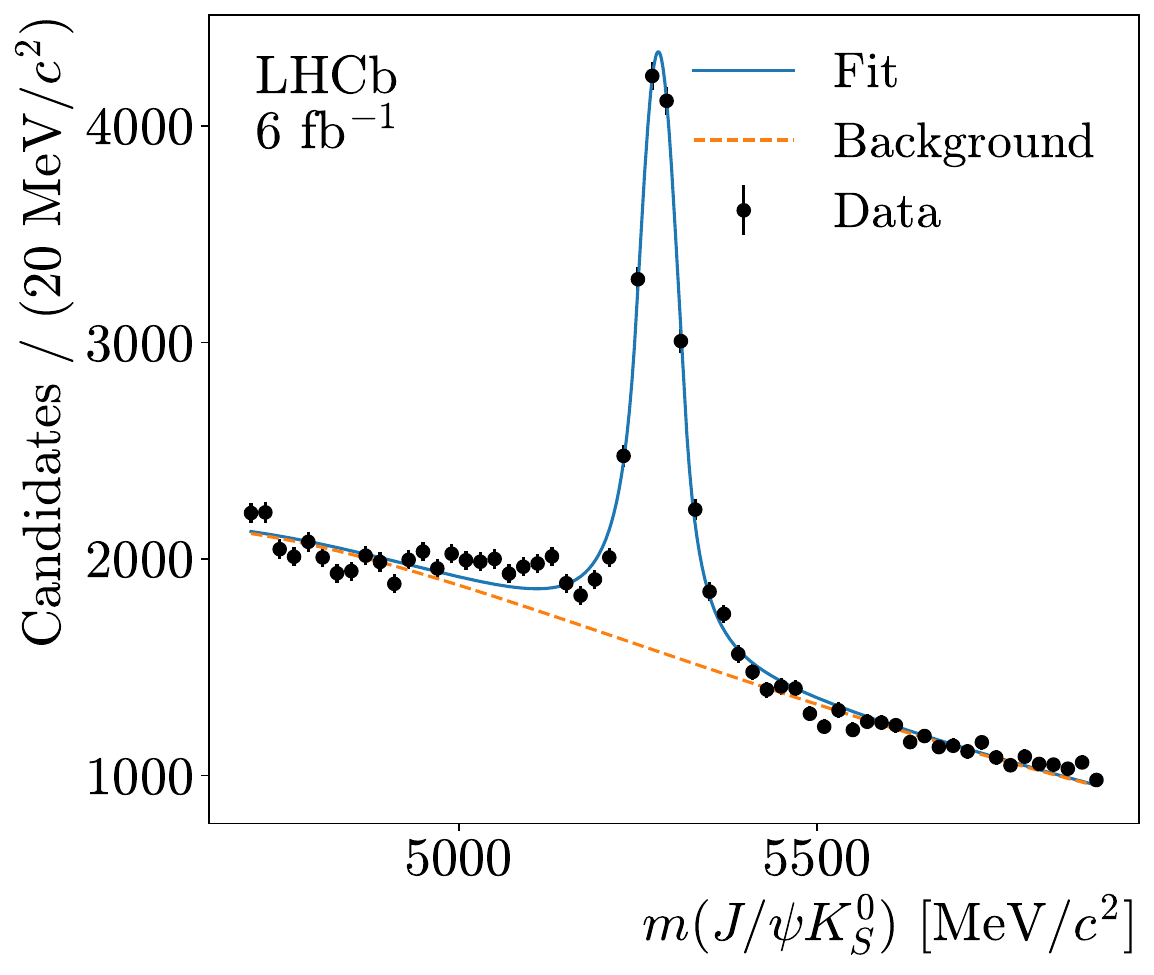}
\hspace{0.5cm}
\caption{Invariant-mass distribution (left) $m(\pi^+\pi^-)$ 
and (right) $m(J/\psi K^0_S)$ 
for \mbox{$B^0 \rightarrow J/\psi K_S^0$} 
candidates using \mbox{Run 2} 
data after all selection criteria. The mass fit results are overlaid.
} 
\label{fig:inv_mass_data_B0}
\end{figure}

Figure~\ref{fig:inv_mass_MC_B0} shows the $m(\pip\pim)$ and $m(\jpsi\KS)$ distributions of the reconstructed and selected \BdToJpsiKS candidates from simulation, about $13\,000$ in total,
along with the mass fits. Figure~\ref{fig:inv_mass_data_B0} shows the corresponding distributions for \Bz candidates reconstructed in data after selection requirements.
The signal shapes are modelled using an asymmetric double-tail Crystal Ball function~\cite{Skwarnicki:1986xj}, while the background contribution to the $m(\pip\pim)$ and
$m(\jpsi \KS)$ distributions, 
of combinatorial nature,
are described using 
an exponential function.
A total of $120\,000$ \KS and $13\,300$ $\Bz\to\jpsi\KS$ signal candidates are reconstructed in data, with mass resolutions of \mbox{$51.0\pm0.3$\mevcc} and \mbox{$47.1\pm0.5$\mevcc}, respectively, which can be compared to \mbox{$51.0\pm0.5$\mevcc} and \mbox{$48.5\pm0.5$\mevcc} in simulation. 
The background yield in the $m(\jpsi\KS$) region, defined by three times the invariant-mass resolution, amounts to $13\,200$.

\begin{figure}[tb]
\centering
\includegraphics[width=\textwidth]{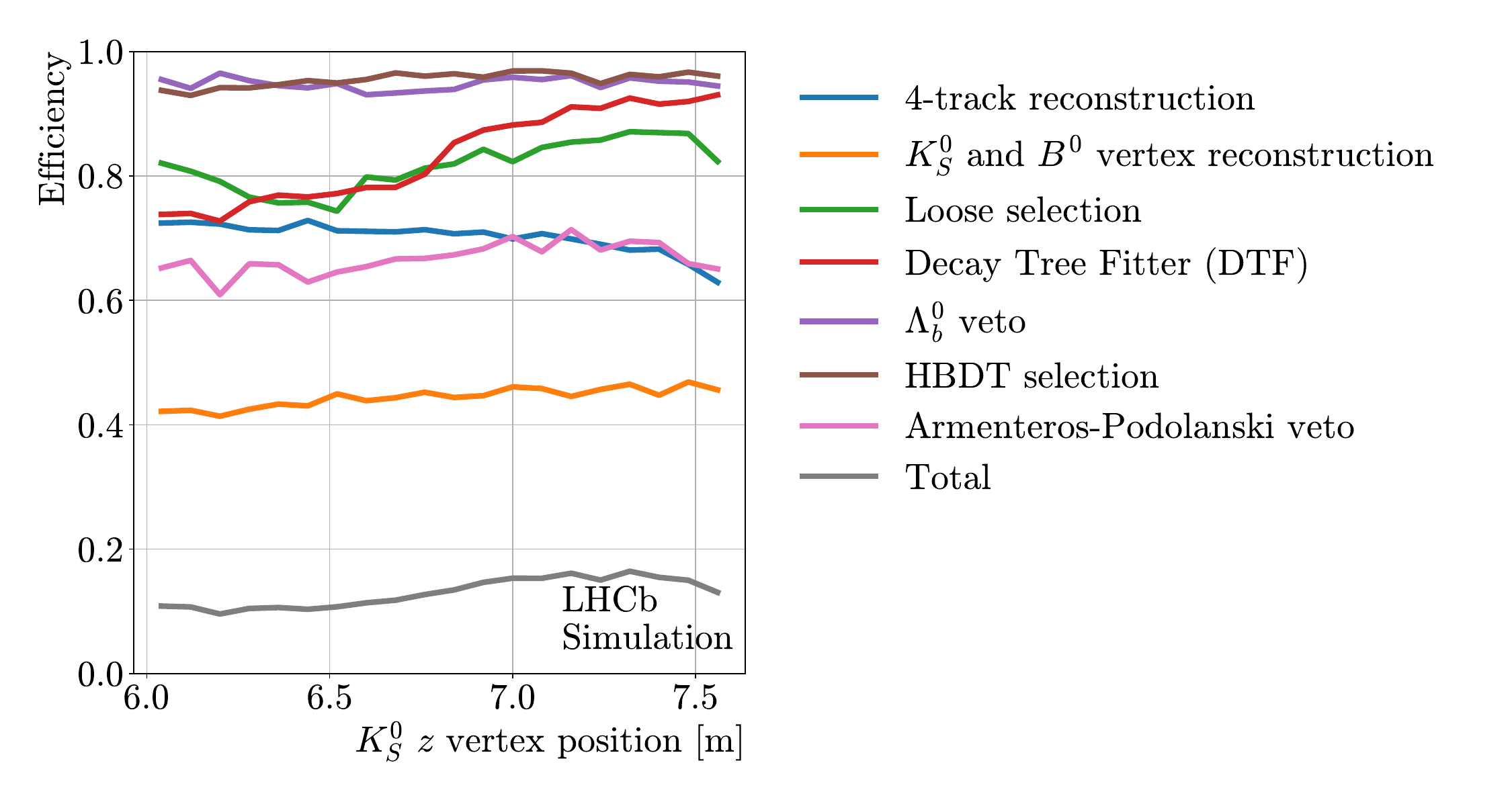}
\caption{Reconstruction and selection efficiency as a function of the true $K^0_S$ 
decay vertex position along the beam axis for simulated \mbox{$B^0 \rightarrow J/\psi K_S^0$}  
signal events (grey) and its breakdown for different reconstruction and selection steps: the reconstruction and quality selections of the four tracks of the final state (blue); the reconstruction of the $K^0_S$ 
and $B^0$ 
decay vertices (orange); the initial selection applied to the final-state and decaying particles as described in Table~\ref{tab:B0_preselections} (Appendix~\ref{appendix:LbKS}, green); the fraction of successful 
decay chain vertex fits (DTF, red); the \Lb mass veto (violet); the selection efficiency of the HBDT (maroon); and the AP veto (pink). The inefficiencies of the vertex reconstruction and the decay chain fit are dominated by the reconstruction of the $\KS$ candidate.}
\label{fig:B0_efficiency}
\end{figure}

The total reconstruction and selection efficiency,
defined previously in Sec.~\ref{subsec:signal_yields_mass_resol},
ranges from about 
$10$\% 
for \KS decays taking place at $6.0$\,m from the nominal IP to about 
$16$\% 
when they occur closest to the T station, as illustrated in Fig.~\ref{fig:B0_efficiency}, along with the breakdown of the steps in the reconstruction and selection sequence. We observe that the tracking and vertexing efficiencies are very similar to those obtained for the \Lb case (Fig.~\ref{fig:efficiency}), while the decay chain vertex fit performance is slightly better. The lower efficiency of the loose selection is partially compensated by the higher efficiency of the HBDT selection. The total efficiency is lower than for \Lb decays due to the lower efficiency of the AP veto.

\begin{figure}[tb]
\centering
\includegraphics[height=6.7cm]{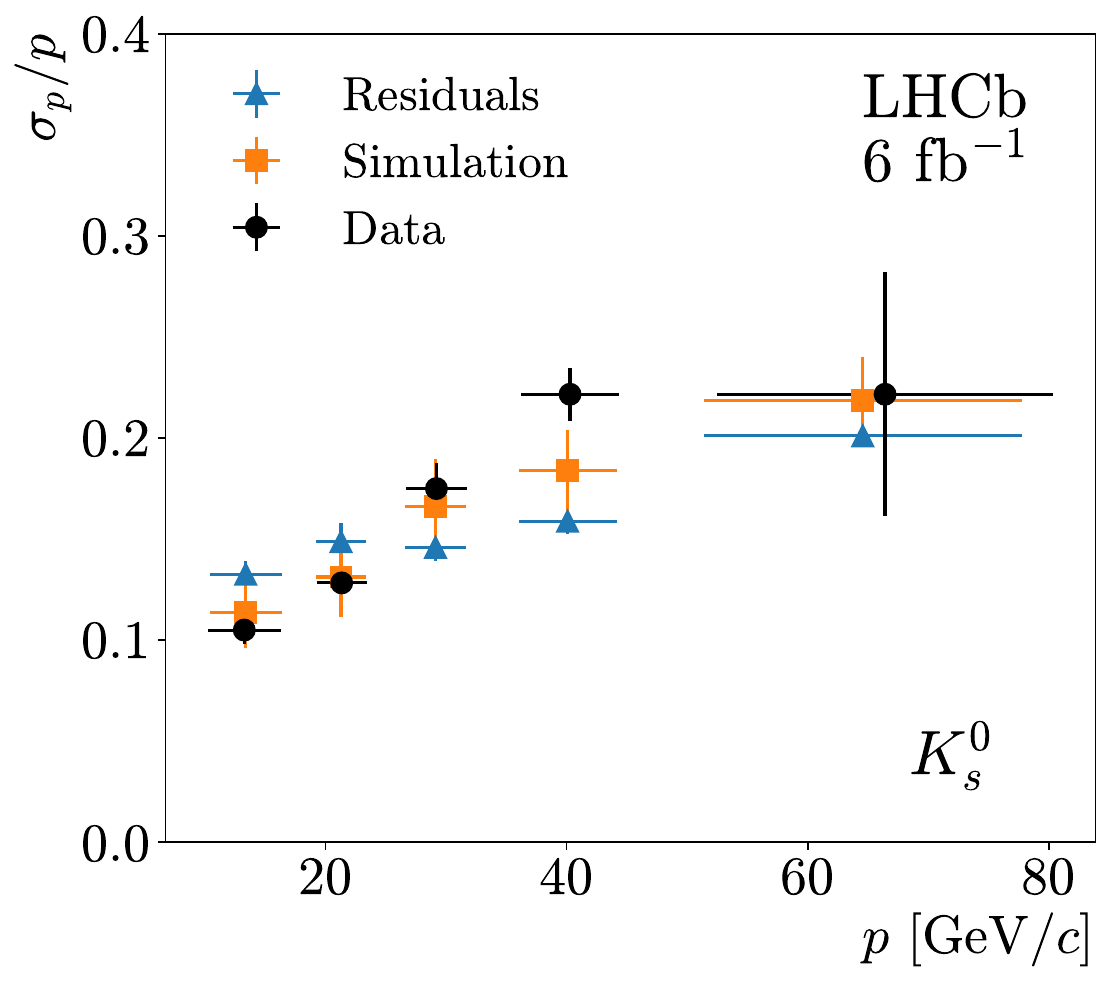}
\includegraphics[height=6.7cm]{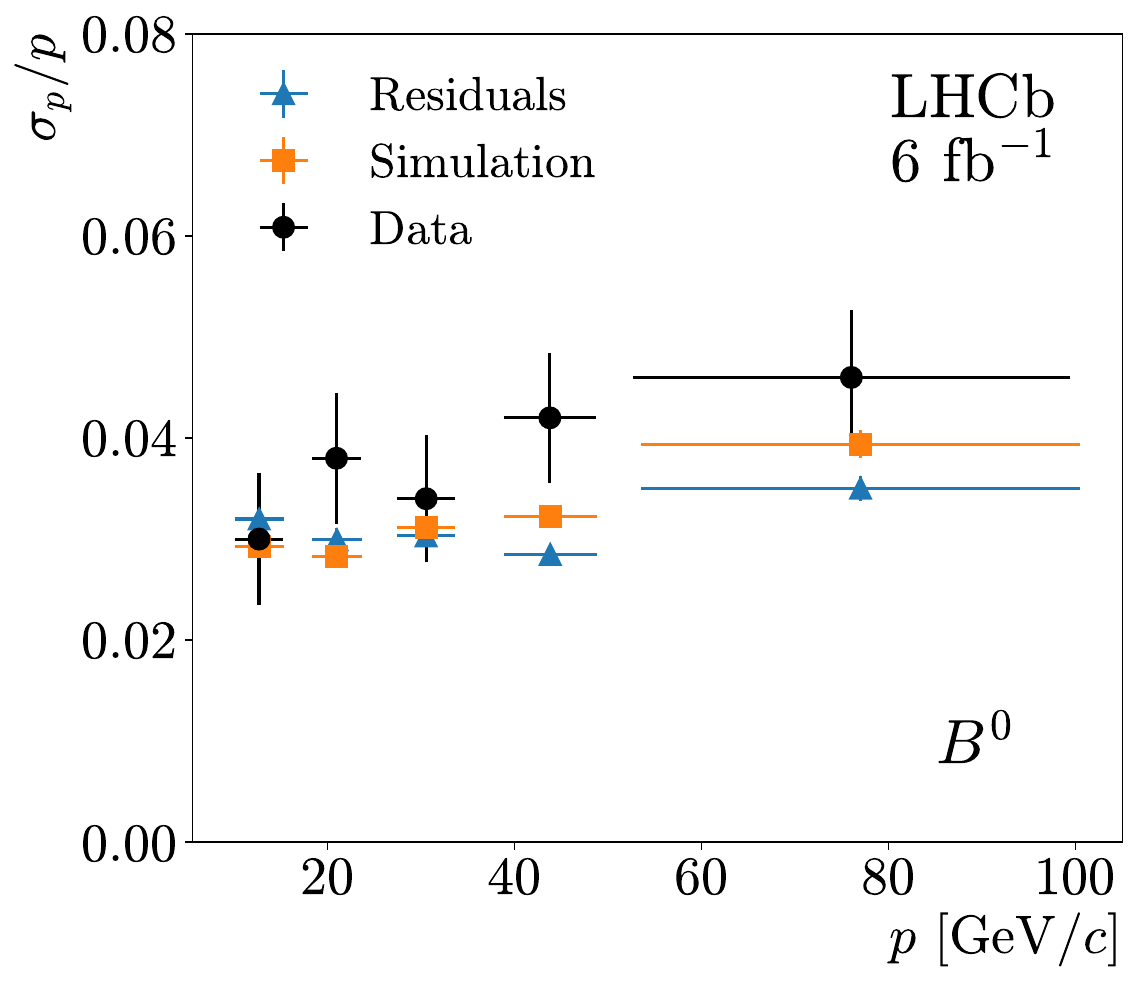}
\caption{Relative momentum resolution as a function of momentum for pions reconstructed as final state of the \mbox{$B^0 \rightarrow J/\psi K_S^0$} 
decay, obtained with the (left) vertex and (right) decay chain fitters. The resolution from residuals in simulation (blue triangles) is compared to that obtained from the data procedure 
applied to both simulation (orange squares) and data (black points). The improved momentum resolution on the right compared to the left is due to the geometric and kinematic constraints imposed when fitting the whole decay chain.}
\label{fig:B0_mom_res}
\end{figure}

\begin{figure}[tb]
\centering
{\includegraphics[height=7cm]{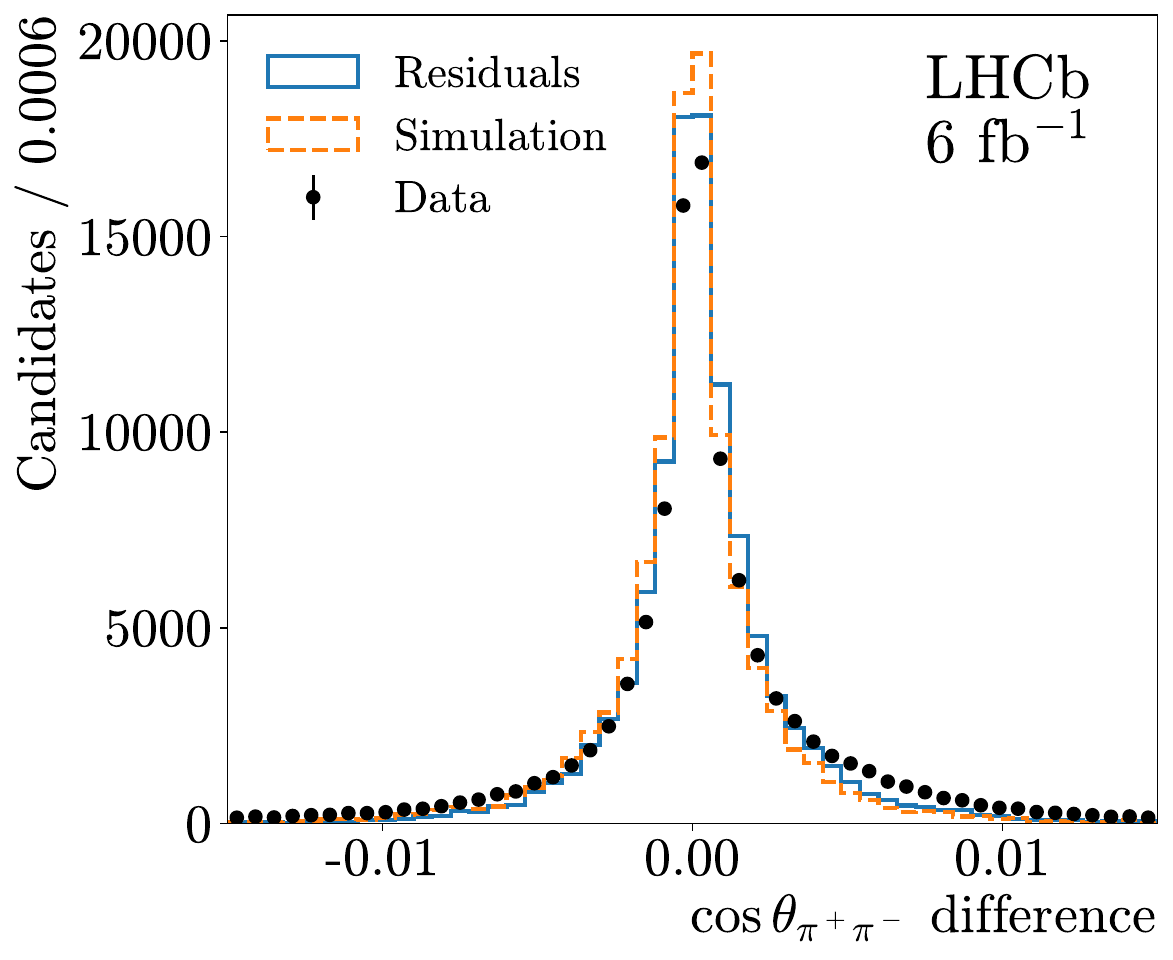}}
\hspace{0.5cm}
\caption{Distributions of the event-by-event differences between the cosine of the two-pion opening angle at the vertex position reconstructed using the vertex and the 
decay chain fitters, for \mbox{$B^0 \rightarrow J/\psi K_S^0$} 
signal candidates in simulation (orange histogram) and data (black points). For simulation, the residual distribution using the vertex fitter is overlaid (blue histogram).
}
\label{fig:B0_angular_resolution}
\end{figure}

\begin{figure}[tb]
\centering
\includegraphics[width=.7\textwidth]{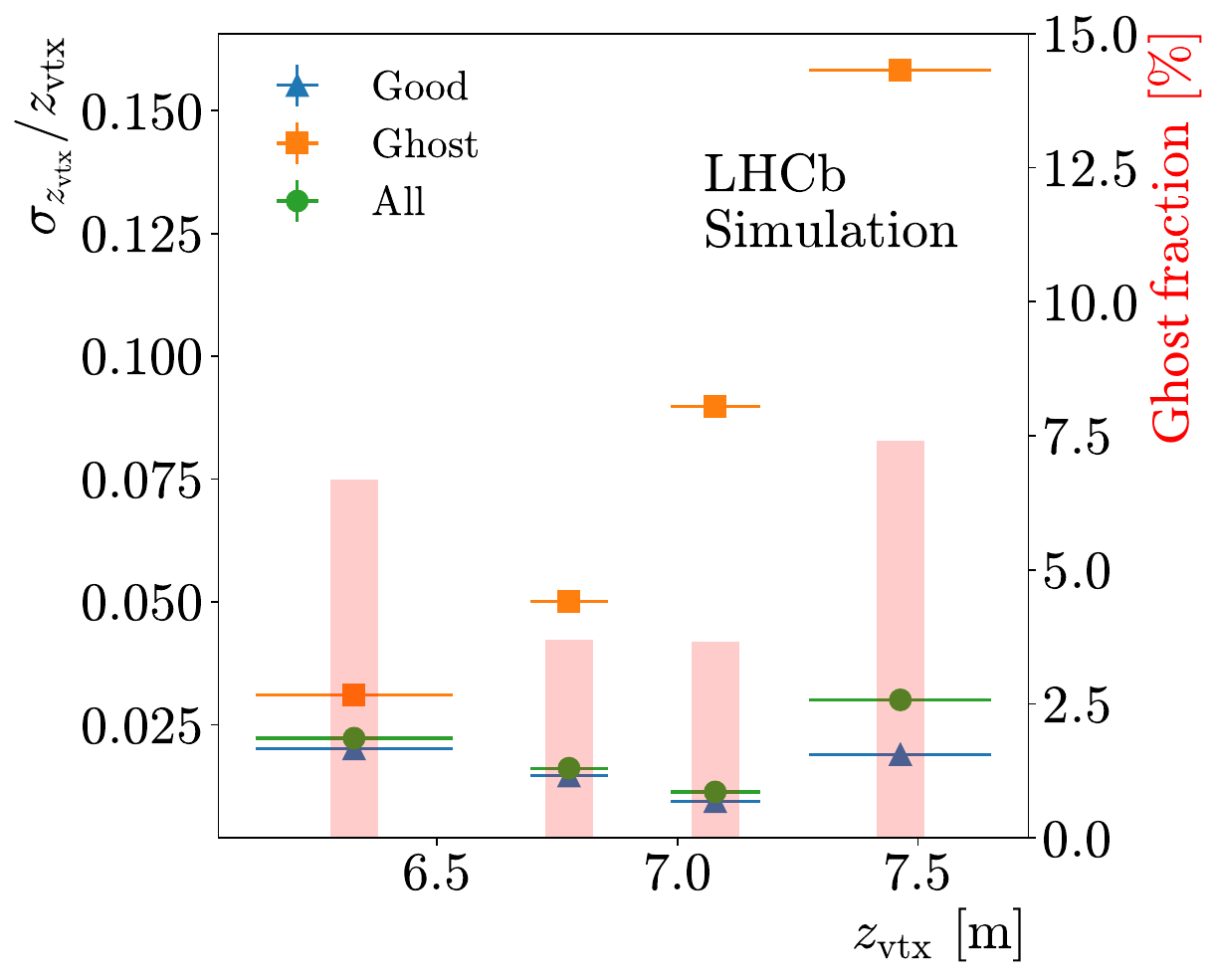}
\caption{Relative precision on the $K^0_S$ 
decay vertex position along the beam axis for \mbox{$B^0 \rightarrow J/\psi K_S^0$} 
signal candidates in simulation tagged as {\em Good} 
(blue triangles) and {\em Ghost} 
(orange squares), as well as for all events (green stars). 
The fraction of {\em Ghost} 
events is also indicated (pink shadow). 
The resolution is evaluated as the central 68.3\% CL region of the underlying residual distribution, adding in quadrature the offset, and approaches that obtained for {\em Good} 
events due to the small fraction of {\em Ghost} 
events.
}
\label{fig:vertex_resolution_B0}
\end{figure}

\begin{figure}[tb]
\centering
\includegraphics[height=7cm]{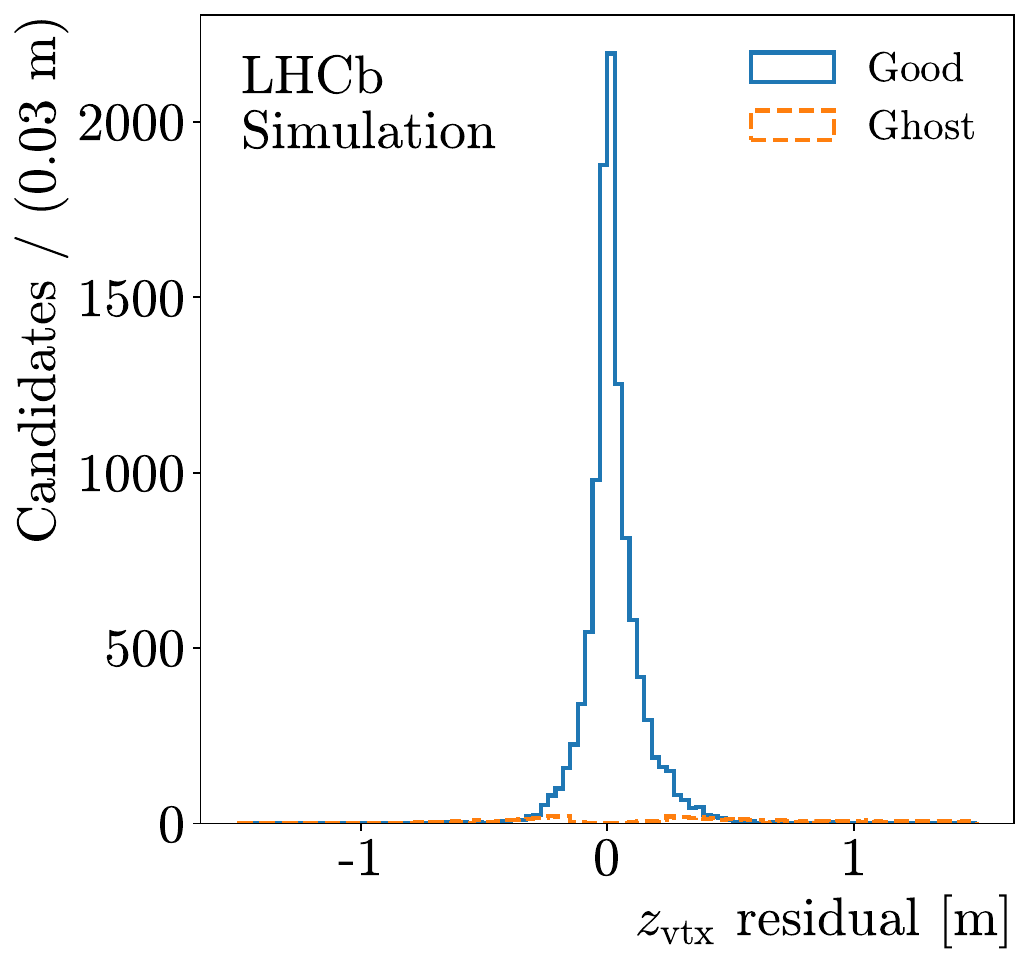}
\caption{Residual distribution of the reconstructed $K_S^0$ 
vertex $z$ position for \mbox{$B^0 \rightarrow J/\psi K_S^0$} 
signal candidates in simulation tagged as {\em Good} 
and {\em Ghost}, 
with true $K_S^0$ 
vertex along the $z$ axis above 6.0\,m.
}
\label{fig:select_real_ghosts_B0}
\end{figure}

\begin{figure}[tb]
\centering
{\includegraphics[height=7cm]{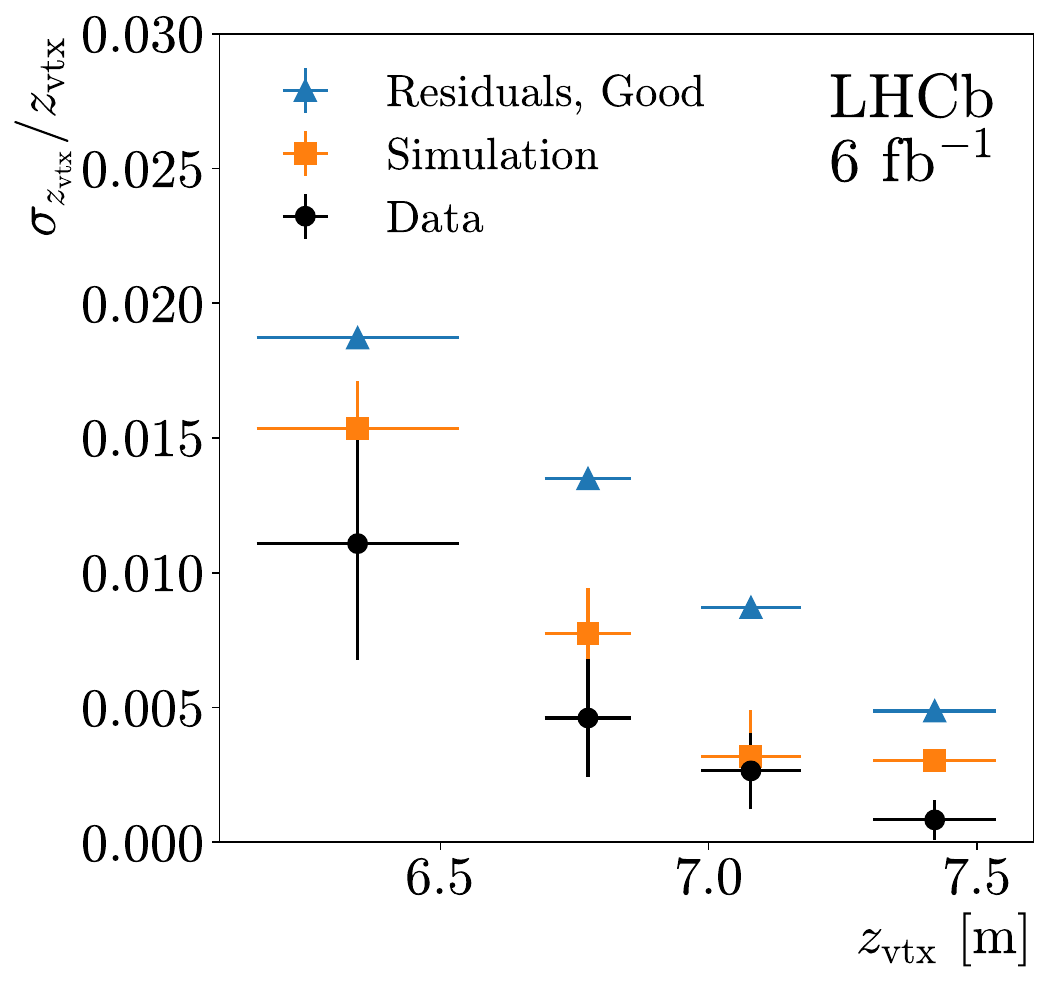}}
{\includegraphics[height=6.9cm]{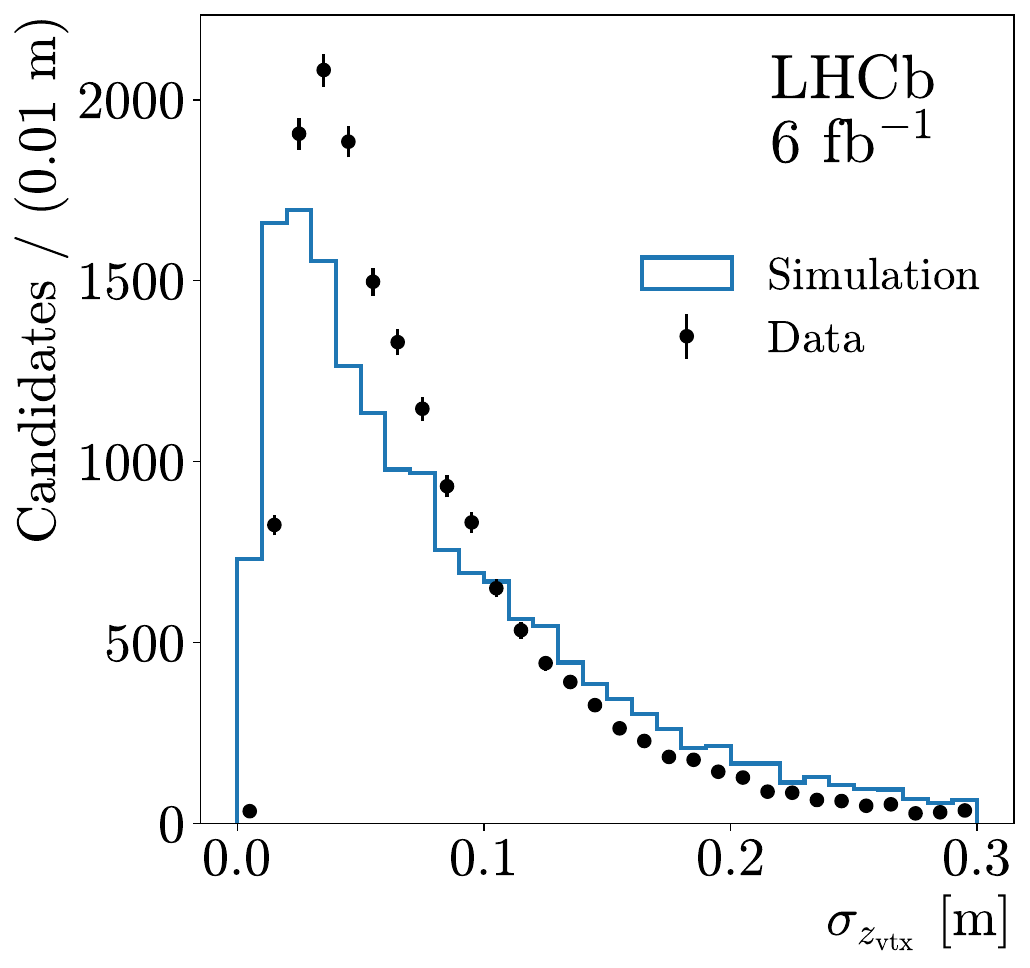}}
\caption{(Left) most probable event-by-event uncertainty divided by the mean value of the reconstructed $K_S^0$ 
vertex position along the $z$ axis, for \mbox{$B^0 \rightarrow J/\psi K_S^0$} 
signal candidates in simulation (orange squares) and data (black points), and in different $z_{\rm vtx}$ regions. The resolutions estimated from residuals for {\em Good} 
events in simulation are also shown (blue triangles).
The binning scheme is the same as in Fig.~\ref{fig:vertex_resolution_B0}.
(Right) distributions of the $\sigma_{z_{\rm vtx}}$ uncertainties for \mbox{$B^0 \rightarrow J/\psi K_S^0$} 
signal candidates in simulation (histogram) and data (points) falling into the second bin of the reconstructed $z$ position.
}
\label{fig:PeE_MC_data_B0}
\end{figure}

All the momentum, vertex and angular resolution studies and comparisons between simulation and data, and between data procedures and residuals in simulation developed for \mbox{\LbToJpsiLz} decays
can directly be applied to \mbox{\BdToJpsiKS} decays. 
The track momentum resolution, shown in Fig.~\ref{fig:B0_mom_res}, lies in the range 15--20\%, and improves to about 4\% when the decay chain constraints are applied.  
Error bars are now smaller since the per-event distributions of the relative momentum uncertainty are more symmetric than for \Lz decays, as it can be observed in Fig.~\ref{fig:sigmapoverp_bin3} of 
Appendix~\ref{appendix:pfromdata}.
%
The resolution on the cosine of the opening angle between the two pions is illustrated in Fig.~\ref{fig:B0_angular_resolution}, reaching in simulation
$0.002$ with the vertex fitter, and about a factor of five smaller with the decay tree fitter.
The former in data is about 40\% larger. Compared to \Lz decays, the better angular resolution of \KS decays is due to the larger opening angle of its decay products, itself a consequence of the larger $Q$ value of the decay, determined by the mass difference between the initial and final-state particles.
%
%
The relative precision of the \KS vertex position along the $z$ axis, illustrated in Fig.~\ref{fig:vertex_resolution_B0}, shows similar features to that of \Lz decays. A relevant difference is the smaller fraction of \ghost vertices present in the sample after all selection criteria, as illustrated in Fig.~\ref{fig:select_real_ghosts_B0}, now less than 6\%.
In this case,
the larger $Q$ value,
together with the equal masses of the final-state particles results in the majority of the decays having an opening-track topology, similar to that sketched in \mbox{Fig.~\ref{fig:event_displays} (right)}. 
Therefore, for \KS decays the benefits of a dedicated BDT to remove \ghost events is largely reduced in comparison to \Lz decays.
The \KS vertex resolution along the $z$ axis provided event-by-event by the vertex fitter, shown in Fig.~\ref{fig:PeE_MC_data_B0}, behaves almost identically as for \Lz decays.

{The resolution on the $\cos\theta_\pip$ and $\phi_\pip$ helicity angles for all candidates, as obtained using the decay chain fitter and residuals in simulation, is illustrated in Fig.~\ref{fig:helicity_angles_B0}. The resolution is slightly better than for \LbToJpsiLz decays, again due to the larger $Q$ value. 
The $\KS\to\pip\pim$ decay offers a data procedure to estimate the $\cos\theta_\pip$ resolution, as follows.
Candidates with the two pions having a longitudinal momentum along the \KS line-of-flight in the laboratory frame differing by less than 5\gevc are selected. This requirement collects \KS decays with the pions being preferentially aligned in the \KS rest frame at $\pm \pi/2$~rad with respect to the \KS direction, thus satisfying $\cos\theta_\pip \approx 0$. The $\cos\theta_\pip$ distribution for candidates passing this selection is adequately described by a Gaussian function. The width of this distribution accounts for momentum and angular resolutions in the evaluation of the helicity angle, as well as the variation of the true $\cos\theta_\pip$ value due to the finite range and resolution of the longitudinal momentum selection criteria. The latter can be estimated from the truth $\cos\theta_\pip$ distribution in simulation, itself described correctly by a Gaussian function as well. The quadratic difference between the two widths represents the $\cos\theta_\pip$ resolution measured as root-mean-square (rms).
Similarly, candidates with $\cos\theta_\pip \approx 0.5~(-0.5) $ can be isolated by selecting \KS candidates with momentum ranging between 30 and 70\gevc and a cosine of the angle between the \KS and the \pim (\pip) in the laboratory frame less than 0.005. 
The rms values obtained with this procedure in simulation and data are shown in Fig.~\ref{fig:helicity_angles_B0} (left), and are compared to the corresponding values obtained directly from residuals in simulation. Note that the differences between the rms and the 68.3\% CL resolutions are due to the non-Gaussian behaviour of the  residual distributions in simulation.

\begin{figure}[tb]
\centering
\includegraphics[height=7cm]{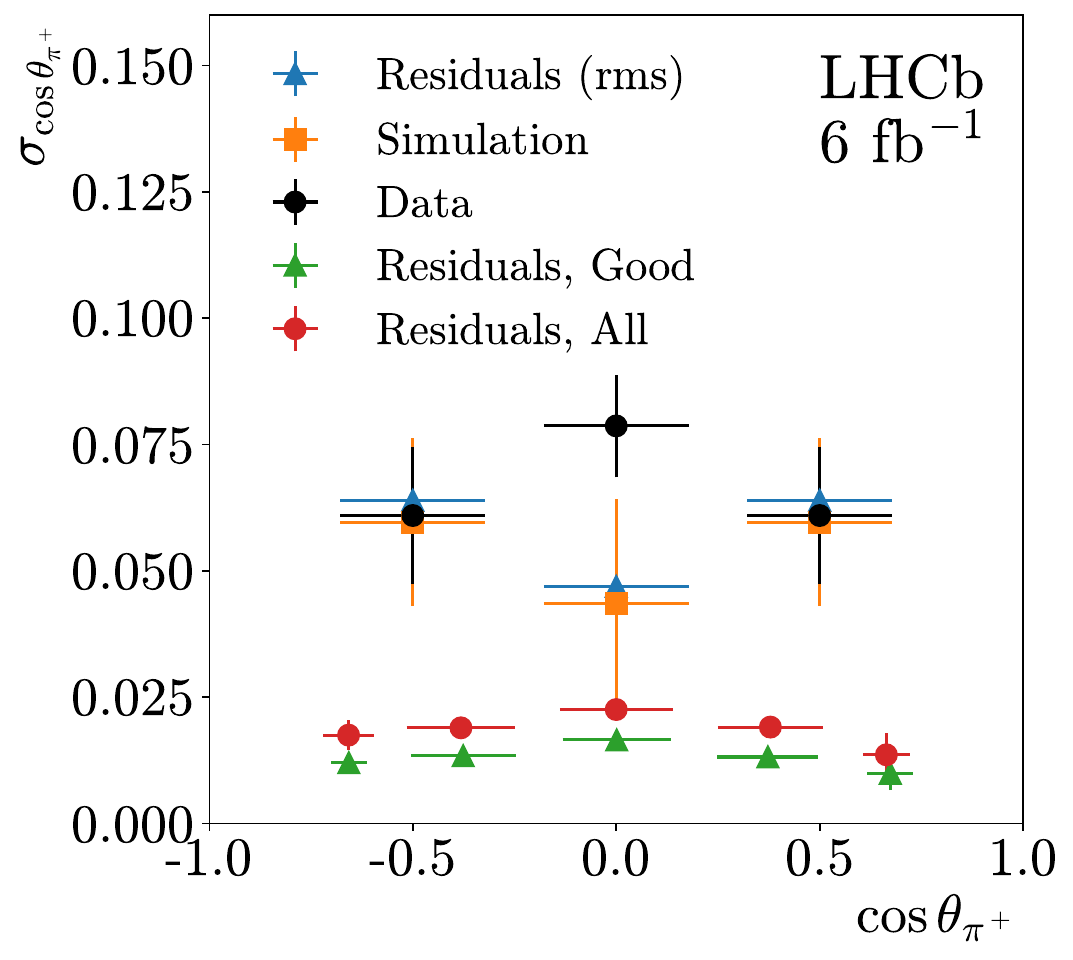}
\includegraphics[height=7.1cm]{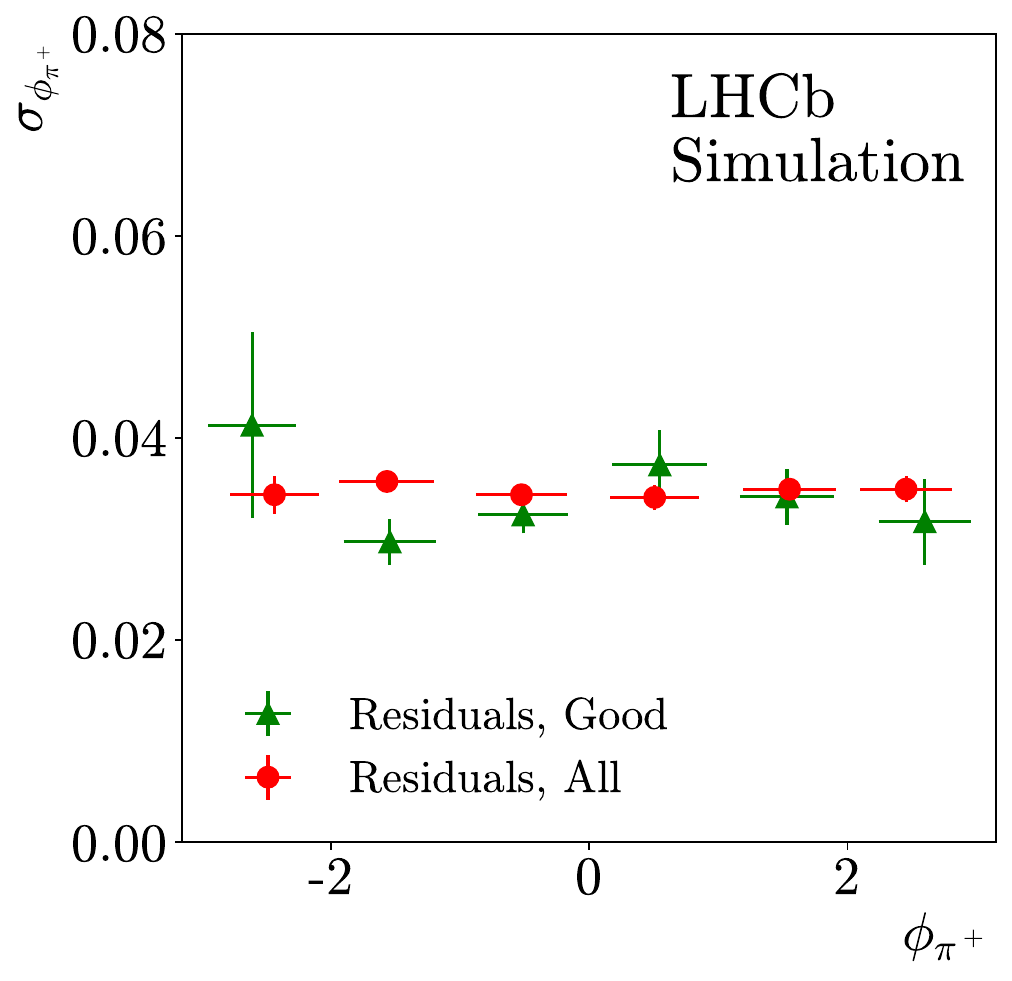}
\caption{Resolution on the (left) $\cos\theta_{\pi^+}$ and (right) $\phi_\pip$ helicity angles for \mbox{$B^0 \rightarrow J/\psi K_S^0$} 
signal candidates in simulation (red points) and those tagged as {\em Good} 
(green triangles), evaluated as the central 68.3\% CL region of the  residuals.
The rms resolution in three different regions of the $\cos\theta_{\pi^+}$ distribution is also evaluated using a data procedure (see text) applied to the simulation (orange squares) and the data (black points), and is compared to that obtained from  residuals in simulation (blue triangles). 
}
\label{fig:helicity_angles_B0}
\end{figure}

\section{Conclusions and prospects}
\label{sec:conclusions}

The feasibility of the reconstruction of long-lived particles decaying inside the \lhcb dipole magnet region is demonstrated and the performance is evaluated in both simulation and data.
Samples of \Lz and \KS hadrons, with the vertex position between 6.0 and 7.6\,m away from the IP, have been reconstructed 
inclusively and exclusively from \mbox{\LbToJpsiLz} and \mbox{\BdToJpsiKS} decays,
using data recorded by \lhcb during the \lhc \runII, corresponding to an integrated luminosity of 6\invfb. 
Strange hadrons decaying in this region have traversed most of the magnetic field
and experience maximal spin precession, 
thereby providing maximal sensitivity for measurements of magnetic and electric dipole moments.
%
The reconstruction can directly be extended upstream to about $2.5\m$, thus expanding the decay volume of the detector reachable using Long and Downstream tracks.
Due to the long extrapolation distances
across the magnet region, the vertex reconstruction using tracks with hits only in the tracking stations downstream of the dipole magnet requires an accurate track transporter based on the Runge--Kutta method instead of the usual polynomial approximation. 
The experimental resolution significantly benefits from the geometric and kinematic constraints of the decay chain when the long-lived particles are produced from exclusive \bquark-hadron decays. 
The combined use of a multivariate classifier,
invariant-mass veto with particle mass hypothesis substitutions, and
the Armenteros-Podolanski technique, maximises the selection performance and mitigates cross-feed from other long-lived decays. 
In addition, a second multivariate classifier specifically designed to remove wrongly reconstructed \ghost decays with closing-track topology, occurring especially in \mbox{$\Lz\to\proton\pim$} decays, has significantly improved the vertex and helicity angle resolutions.

Several improvements to the reconstruction and selection of LLPs using T tracks are 
in progress.
These include the use of PID information based on measurements from the RICH2, calorimeter and muon systems located downstream of the T stations,
and 
the adaptation of the vertexing algorithms to enhance the reconstruction efficiency and further extend the 
decay
volume inside the magnet region.
Furthermore, the implementation of dedicated trigger lines are in progress for \runIII data and beyond, taking advantage of the 
flexible software-based trigger of the \lhcb upgraded detector. 
Dedicated vertexing algorithms for T tracks are also under development.
The trigger, reconstruction and selection of LLPs using T tracks will offer new opportunities to extend the physics reach of the \lhcb experiment through measurements of the magnetic and electric dipole moments of the \Lz baryon and searches for LLPs predicted by various BSM theories. Physics projections and analyses exploiting these new possibilities are in progress.

\section*{Acknowledgements}
%
%
\noindent We express our gratitude to our colleagues in the CERN
accelerator departments for the excellent performance of the LHC. We
thank the technical and administrative staff at the LHCb
institutes.
We acknowledge support from CERN and from the national agencies:
CAPES, CNPq, FAPERJ and FINEP (Brazil); 
MOST and NSFC (China); 
CNRS/IN2P3 (France); 
BMBF, DFG and MPG (Germany); 
INFN (Italy); 
NWO (Netherlands); 
MNiSW and NCN (Poland); 
MCID/IFA (Romania); 
MICIU and AEI (Spain);
SNSF and SER (Switzerland); 
NASU (Ukraine); 
STFC (United Kingdom); 
DOE NP and NSF (USA).
We acknowledge the computing resources that are provided by CERN, IN2P3
(France), KIT and DESY (Germany), INFN (Italy), SURF (Netherlands),
PIC (Spain), GridPP (United Kingdom), 
CSCS (Switzerland), IFIN-HH (Romania), CBPF (Brazil),
and Polish WLCG (Poland).
We are indebted to the communities behind the multiple open-source
software packages on which we depend.
Individual groups or members have received support from
ARC and ARDC (Australia);
Key Research Program of Frontier Sciences of CAS, CAS PIFI, CAS CCEPP, 
Fundamental Research Funds for the Central Universities, 
and Sci. \& Tech. Program of Guangzhou (China);
Minciencias (Colombia);
EPLANET, Marie Sk\l{}odowska-Curie Actions, ERC and NextGenerationEU (European Union);
A*MIDEX, ANR, IPhU and Labex P2IO, and R\'{e}gion Auvergne-Rh\^{o}ne-Alpes (France);
AvH Foundation (Germany);
ICSC (Italy); 
Severo Ochoa and Mar\'ia de Maeztu Units of Excellence, GVA, XuntaGal, GENCAT, InTalent-Inditex and Prog.~Atracci\'on Talento CM (Spain);
SRC (Sweden);
the Leverhulme Trust, the Royal Society
 and UKRI (United Kingdom).

\newpage
\appendix
\section{Loose selection criteria}
\label{appendix:LbKS}

Tables~\ref{tab:preselections} and~\ref{tab:B0_preselections} report the loose selection requirements applied to the \LbToJpsiLz and \BdToJpsiKS with \mbox{$\jpsi\rightarrow \mup\mun$}, \mbox{$\Lz \rightarrow \proton\pim$} and \mbox{$\KS \rightarrow \pip\pim$} candidates passing the online event selection.
\begin{table}[htb!]
\centering
\caption{Loose selection requirements applied to $\Lb\to\jpsi\Lz$ candidates passing the online event selection. The invariant masses $m(p\pim)$ and $m(\jpsi\Lz)$ are obtained from the vertex fitter.
}
\begin{tabular}{llrr}
Variable                      & Units & Minimum           & Maximum \\
\hline\\
$p\,(\pim)$                    & \mevc  & $2\,000$            & $500\,000$\\
$p\,(p)$                      & \mevc  & $10\,000$          & $500\,000$\\
$\pt\,(p)$                  & \mevc  & $400$             & $-$ \\
$m(p\pim)$                    & \mevcc & $600$             & $1\,500$ \\
$z_{\rm vtx}^{\Lz}$          & m  & $5.5$  & $8.5$ \\
$\cos\xi_p\,(\Lz)$      &        & $0.9999$          & $-$ \\
$\chi^2_{\rm IP}\,(\Lz)$&        & $-$               & $200$ \\
$\chi^2_{\rm vtx}\,(\Lz)$     &        & $-$               & $750$ \\
$|m(\mup\mun) - m(\jpsi)|$ & \mevcc & $-$ & $90$ \\
$\pt(\Lz)$                 & \mevc  & $450$             & $-$ \\
$m(\jpsi\Lz)$                & \mevcc   & $4\,700$     & $8\,500$ \\
$\cos\xi_p\,(\Lb)$      &          & $0.99$          & $-$ \\$\chi^2_{\rm IP}\,(\Lb)$&          & $-$             & $1\,750$ \\
$\chi^2_{\rm vtx}\,(\Lb)$     &          & $-$             & $150$ \\
\\[-2.0ex] \hline
\end{tabular}
\label{tab:preselections}
\end{table}


\begin{table}[tb]
\centering
\caption{Loose selection requirements applied to $\Bz\to\jpsi\KS$ candidates passing the online event selection. 
The invariant masses $m(\pip\pim)$ and $m(\jpsi\KS)$ are obtained from the vertex fitter. 
}
\begin{tabular}{llrr}
Variable                    & Units & Minimum           & Maximum \\
\hline\\
$p(\pip,\pim)$  & \mevc   & $2\,500$              & $500\,000$\\
$m(\pip\pim)$ & \mevcc &     $100$        & $1\,300$ \\
$z_{\rm vtx}^{\KS}$ & m &   $5.0$     & $9.5$ \\
$\cos\xi_p({\KS})$ &  & $0.9995$          & $-$ \\
$\chi^2_{\rm IP}(\KS)$ & &  $-$               & $200$ \\
$\chi^2_{\rm vtx}(\KS)$ &         & $-$               & $100$ \\
$|m(\mup\mun) - m(\jpsi)|$  & \mevcc  & $-$ & $90$ \\
$p_{T}(\KS)$  & \mevc        & $600$             & $-$ \\
$p(\jpsi)$  & \mevc          & $18\,000$           & $-$\\
$\pt(\jpsi)$  & \mevc       & $250$             & $-$\\
$m(\jpsi\KS)$ & \mevcc       & $4\,700$             & $6\,500$\\
$\cos\xi_p({\Bz})$    &  &   $0.995$           & $-$ \\$\chi^2_{\rm IP}(\Bz)$ &  &   $-$               & $60$ \\
$\chi^2_{\rm vtx}(\Bz)$      &   &   $-$               & $35$ \\
\hline
\end{tabular}
\label{tab:B0_preselections}
\end{table}

\section{Data-based momentum resolution procedure}
\label{appendix:pfromdata}

Let us consider a particle 
decaying into two particles with masses $m_1$ and $m_2$ and four-momenta $(E_1,{\bf p}_1)$ and $(E_2,{\bf p}_2)$. The invariant mass of the decaying particle follows as
\begin{eqnarray} 
\label{eq:m}
m^2 & = &
m_1^2 + m_2^2 + 2 \left( E_1 E_2 - p_1 p_2 \cos\theta_{12} \right) \, ,
\end{eqnarray}
where $\theta_{12}$ is the angle between ${\bf p}_1$ and ${\bf p}_2$, $p_1$ and $p_2$ are their magnitudes, and \mbox{$E_1=\sqrt{p_1^2+m_1^2}$} and
\mbox{$E_2=\sqrt{p_2^2+m_2^2}$} the energies. 
Error propagation on Eq.~(\ref{eq:m}) and approximating $\sin\theta_{12} \approx \theta_{12}$ 
lead to
\begin{eqnarray} 
\label{eq:sigmap}
\left( \frac{\sigma_{p}}{p} \right)^2 & = & g_m \left( \frac{\sigma_m}{m}\right)^2  - g_\theta  \left(\frac{\sigma_{\theta_{12}}}{\theta_{12}}\right)^2 \, ,
\end{eqnarray}
where
\begin{eqnarray} 
\label{eq:gmgtheta}
g_m & = & g \left( \frac{m}{p} \right)^4 \, , \nonumber \\
g_\theta & = & g \xi^2\theta_{12}^4 \, , 
\end{eqnarray}
with
\begin{eqnarray} 
\label{eq:g}
\frac{1}{g} & = & \xi^2 \left( \frac{\xi_E}{\xi} - \cos\theta_{\theta_{12}}\right)^2 + \xi^2 f^2(p,\xi,\eta_1,\eta_2) \left( \frac{\xi}{\xi_E} - \cos\theta_{12} \right)^2 \, . 
\end{eqnarray}
Here, we have introduced the kinematic ratios $\xi=p_1/p_2$ and $\xi_E = E_1/E_2$, and the ratio of relative momentum uncertainties 
\begin{eqnarray} 
\label{eq:feta}
f(p,\xi,\eta_1,\eta_2) & = &  
\frac{1}{\xi} \frac{\sigma_p(\xi p, \eta_1)}{\sigma_p(p,\eta_2)} \, ,
\end{eqnarray}
where $\sigma_p(p,\eta)/p$ stands for the relative momentum uncertainty of the daughter particle 1 (2) evaluated as a function of its momentum $\xi p$ ($p$) and pseudorapidity $\eta_1$ ($\eta_2$).
The mass and angular factors, $g_m$ and $g_\theta$, respectively,
encapsulate the kinematic dependence when propagating the relative mass and angle uncertainties to the relative momentum uncertainty. 

In the relativistic limit, $E_1\approx p_1$ and $E_2 \approx p$, and  small angle approximation, \mbox{$1-\cos\theta_{12} \approx \theta_{12}^2/2$}, 
Eq.~(\ref{eq:sigmap}) holds whereas~(\ref{eq:gmgtheta}) and~(\ref{eq:g})
reduce to
\begin{eqnarray} 
\label{eq:gmthetafactors_simplified1}
g_m & \approx & \frac{4}{1+f^2(p,\xi,\eta_1,\eta_2)} \frac{1}{\xi^2\theta_{12}^4} \left(\frac{m}{p}\right)^4 \, , \nonumber \\
g_\theta & \approx & \frac{4}{1+f^2(p,\xi,\eta_1,\eta_2)}\, .
\end{eqnarray}
Further assuming $m\gg m_1,m_2$, it follows that
\begin{eqnarray} 
\label{eq:gmthetafactors_simplified2}
g_m \approx g_\theta & \approx & \frac{4}{1+f^2(p,\xi,\eta_1,\eta_2)} \, , 
\end{eqnarray}
since in this case $m^2 \approx \xi p^2 \theta_{12}^2$. For $m_1 \approx m_2$ and assuming candidates satisfying $\xi \approx 1$ and $\eta_1 \approx \eta_2$, leads to $f(p,\xi,\eta_1,\eta_2) \approx 1$ and $g_m\approx g_\theta \approx 2$.
Equation~(\ref{eq:sigmap}) can then be written as 
\begin{eqnarray} 
\label{eq:sigmap_LHCbperformance}
\left( \frac{\sigma_p}{p} \right)^2 & \approx & 2 \left( \frac{\sigma_m}{m}\right)^2  - 2 \left(\frac{\sigma_{\theta_{12}}}{\theta_{12}}\right)^2 \, ,
\end{eqnarray}
in agreement with Eq.~(1) of Ref.~\cite{LHCb-DP-2014-002},\footnote{There is a misprint in the factor $p/mc$ of the angular term, which should not be present.} where this approximate expression has been been applied to $\jpsi \to \mu^+ \mu^-$ decays to measure the momentum resolution of Long tracks. 

Equations~(\ref{eq:sigmap}) to~(\ref{eq:g}) can be exploited to measure the relative track momentum resolution in regions (bins) of momentum for T tracks from $\Lz\to p \pim$ and $\KS \to \pip \pim$ decays. It should be noted that the approximate Eqs.~(\ref{eq:gmthetafactors_simplified1}) to~(\ref{eq:sigmap_LHCbperformance}) do not hold in either of these cases.
After applying all selection criteria, signal candidates in data are binned according to the reconstructed momentum of the proton and the two pions, for $\Lz\to p \pim$ and $\KS \to \pip \pim$ decays, respectively. The \sPlot technique~\cite{Pivk:2004ty} with $m(p\pim)$ and $m(\pip\pim)$ as the discriminating variable is adopted to statistically subtract the background contribution. 
In order to assess its validity, the procedure is also applied to simulation, in addition to the evaluation of the resolution from the distributions of differences between reconstructed and true quantities (residuals).
In the following, we identify particle 2 with the positively charged decay product, \ie the proton (\pip) for \Lz (\KS) decays.

For each momentum bin, the mass resolution $\sigma_m$ is determined from the 68.3\% CL region of the $m(p\pim)$ and $m(\pip\pim)$ invariant-mass distributions. 
The world-average \Lz and \KS masses~\cite{PDG2022} are used for $m$, although there is no sizeable impact on the resolution when the mean or most probable values of the distributions are used instead.
The angular resolution $\sigma_{\theta_{12}}$ per bin is determined analogously from the distribution of differences between the reconstructed $\theta_{12}$ angle and the same angle reconstructed with the decay chain constraints, since the latter has an uncertainty ranging between a factor of four and five times smaller.
Figures~\ref{fig:angular_resolution} and~\ref{fig:B0_angular_resolution} compare
the distributions of these residuals for signal events in simulation and data, integrated over momenta, for \Lz and \KS decays, respectively. 

The kinematic factor $\xi$, along with $\xi_E$, can be determined directly on an event-by-event basis. Its average, $\langle \xi \rangle$, amounts to $\approx 6$ for \Lz decays, ranging from about $4.7$ to $6.8$ with increasing momentum bin, whereas for \KS decays it is close to unity with marginal variation across the bins, for both simulation and data. 
Approximate average values can also be estimated analytically, given the momentum of particle 1 along the (longitudinal) direction of motion of the decaying particle,
\begin{eqnarray} 
\label{eq:p1}
p_{1,L}(\cos\theta^*) & = & \gamma p^* \left( \cos\theta^* + \beta \sqrt{1+\frac{m_1^2}{p^{*2}}} \right) \, , 
\end{eqnarray}
(and analogously for particle 2, reversing the sign of $\cos\theta^*$ and replacing $m_1$ by $m_2$),
where $p^*$ and $\theta^*$ are the momentum and angle of particle 1 (and 2) in the centre-of-mass frame of the decaying particle, and $\beta\gamma \gg 1$ its Lorentz boost.    
The average value is obtained integrating over $\cos\theta^*$,
\begin{eqnarray} 
\label{eq:eta_analytical}
\langle \xi \rangle & \approx & \frac{ \bigint p_{1,L}^*(\cos\theta^*) \Gamma(\cos\theta^*) {\rm d}\cos\theta^* }{ \bigint p_{2,L}^*(\cos\theta^*) \Gamma(\cos\theta^*) {\rm d}\cos\theta^* } \, ,
\end{eqnarray}
where $\Gamma(\cos\theta^*)$ represents the angular distribution of the $m \to m_1 \ m_2$ decay.
For \mbox{$\Lz\to p \pim$} decays, taking $\Gamma(\cos\theta^*)=\frac 12 (1 + \alpha \cos\theta^*)$ with $\alpha \approx 0.75$~\cite{PDG2022},
yields $\langle \xi \rangle \approx 6.5$. 
For \mbox{$\KS\to\pip\pim$} decays, with $\Gamma(\cos\theta^*)=1/2$ and $m_1=m_2$, one obtains $\langle \xi \rangle \approx 1$.

The $f(p,\xi,\eta_1,\eta_2)$ factor is estimated from the ratio of relative momentum uncertainties of the two daughter particles evaluated at $(\xi p,\eta_1)$ and $(p,\eta_2)$, following Eq.~(\ref{eq:feta}). The dependence of the relative momentum uncertainty is itself obtained from simulation, binning in momentum and pseudorapidity.
Since the relative momentum resolution does not depend on the particle mass, protons and pions from both \Lz and \KS decays are combined. The two-dimensional binning along with the 68.3\% CL region of the residual distribution in each bin are illustrated in 
Fig.~\ref{fig:sigmapop_vspeta} (left).
\begin{figure}[tb]
\centering
{\includegraphics[height=5.7cm]{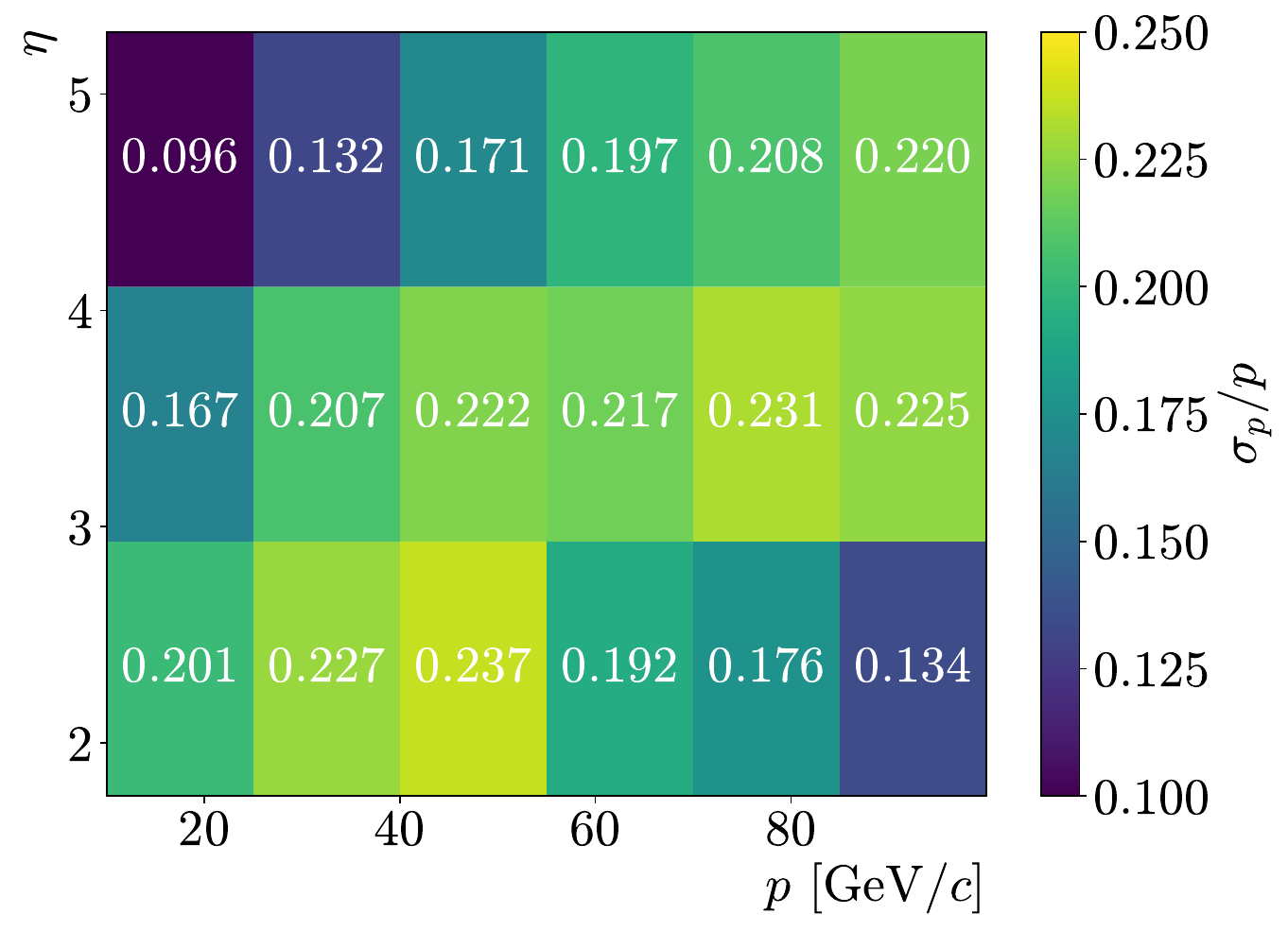}}
{\includegraphics[height=5.7cm]{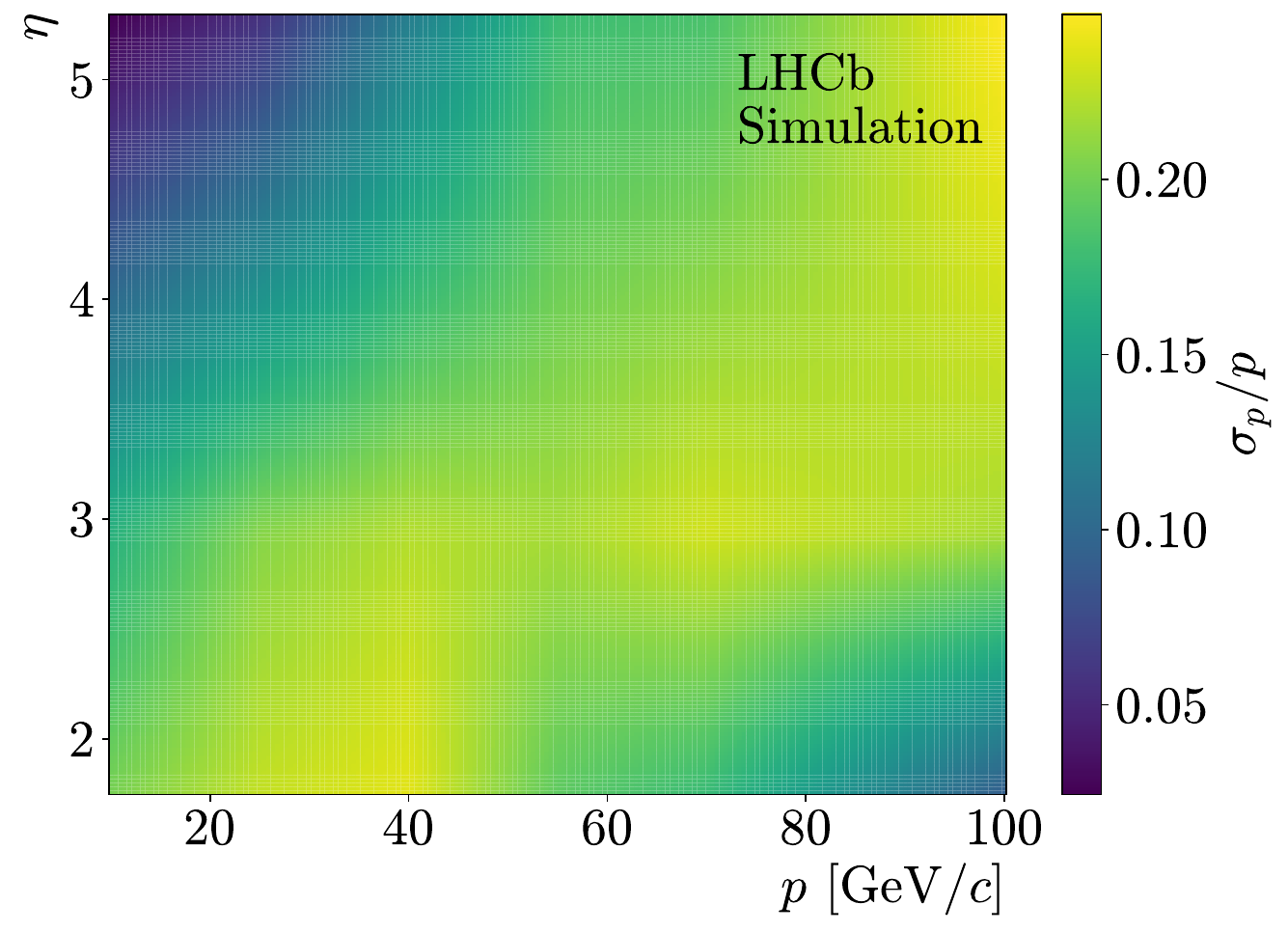}}
\hspace{0.5cm}
\caption{(Left) Relative momentum resolution for protons and pions from $\varLambda$ 
and $K_S^0$ 
decays (left) in bins of momentum and pseudorapidity,
evaluated with residual distributions (68.3\% CL region) from simulated signal events passing all selection criteria, and (right) after spline interpolation.
The statistical uncertainties are stable across the bins and amount to about 0.003. 
}
\label{fig:sigmapop_vspeta}
\end{figure}
For $\KS\to\pip\pim$ decays, where the two pions are selected in the same momentum bin, \ie $\xi\approx 1$, it follows $f(p,\xi,\eta_1,\eta_2) \approx 1$ when $\eta_1 \approx \eta_2$.

The relative resolution $\sigma_{p}/p$ can then be evaluated on an event-by-event basis using the per-bin mass and angular resolutions, the angle $\theta_{12}$, the momentum $p$,  the kinematic factors $\xi$ and $\xi_E$, the pseudorapidities $\eta_1$ and $\eta_2$, and the $f(p,\xi,\eta_1,\eta_2)$ ratio evaluated using Eq.~(\ref{eq:feta}) and Fig.~\ref{fig:sigmapop_vspeta} (right). Figure~\ref{fig:sigmapoverp_bin3} (left) illustrates the $\sigma_{p}/p$ distributions for the central bins [see Figs.~\ref{fig:mom_res} and~\ref{fig:B0_mom_res} (left)].
The wider shape and positive tail of the distribution for \Lz decays reflects the larger event-by-event variation of its kinematics with respect to \KS decays.
An average relative momentum resolution per bin is evaluated using the most probable value of the distribution, with uncertainty determined from the central 68.3\% CL region divided by the square root of the bin yields, added in quadrature to the half difference between the most probable value and the mean of the distribution. The broader and more asymmetric distribution for \Lz decays result in larger uncertainties as compared to \KS decays.
Other sources of uncertainty originated by the propagation of uncertainties on $\sigma_m$ and $\sigma_{\theta_{12}}$, estimated through the method of moments~\cite{rao73},
and the $f(p,\xi,\eta_1,\eta_2)$ statistical uncertainties, are also taken into account and found to be subdominant.
The final results of the procedure are shown in Figs.~\ref{fig:mom_res} and~\ref{fig:B0_mom_res} (left).
In these figures, the central values and error bars along the horizontal axis are evaluated as the mean and 68\% CL interval of the momentum distribution in each bin, respectively\footnote{The same procedure is adopted for Figs.~\ref{fig:PeE_MC_data} (left), \ref{fig:helicity_angles_Lambdab}, \ref{fig:PeE_MC_data_B0} (left), and~\ref{fig:helicity_angles_B0}.}.

\begin{figure}[tb]
\centering
{\includegraphics[height=6.2cm]{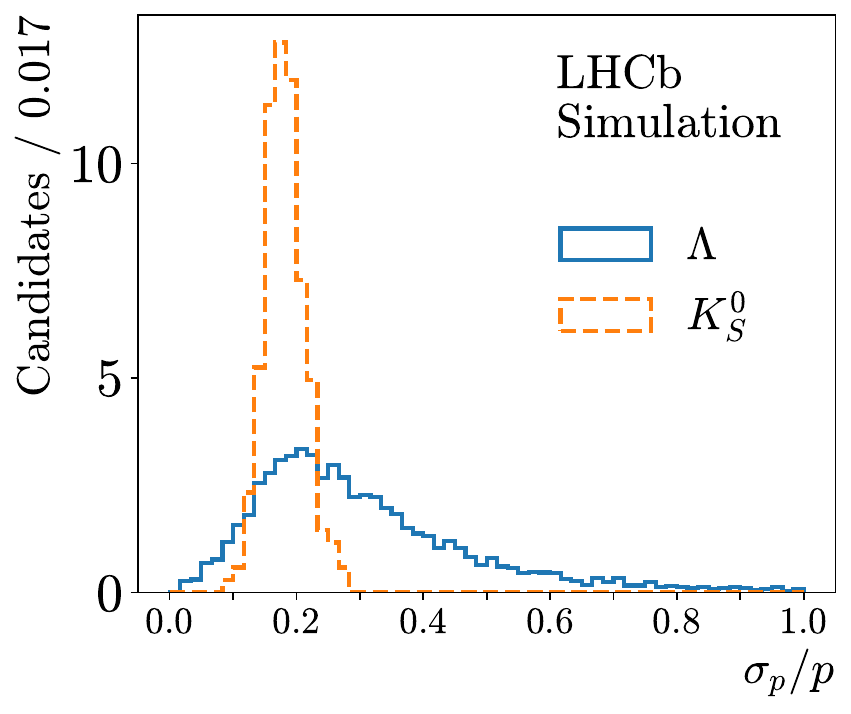}}
{\includegraphics[height=6.2cm]{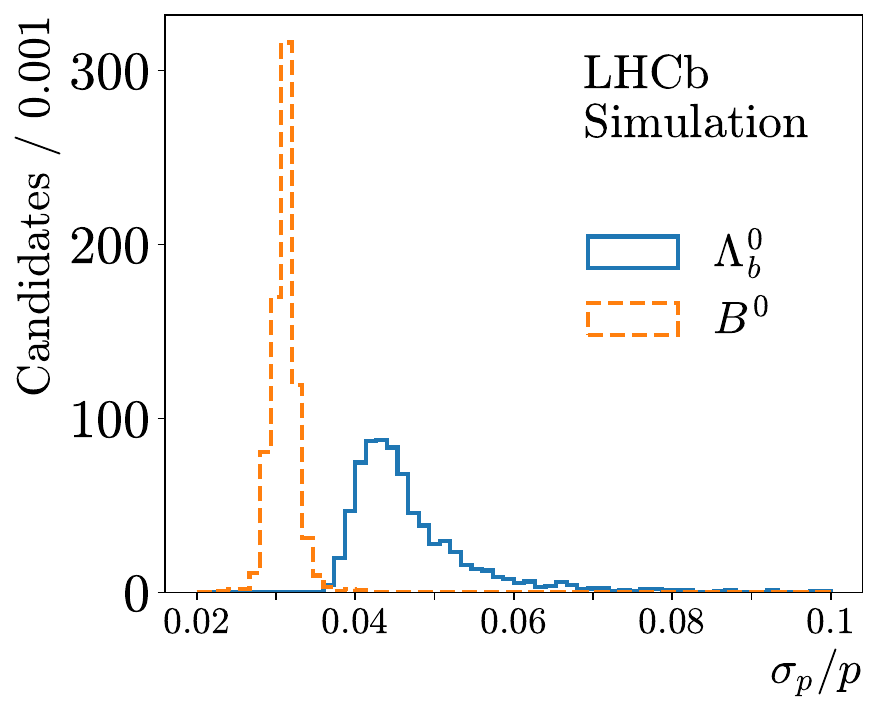}}
\hspace{0.5cm}
\caption{Distributions of the relative momentum resolution for T tracks from $\varLambda\rightarrow p \pi^-$  
and $K^0_S \rightarrow \pi^+ \pi^-$ 
decays for the central momentum bin, (left) from track reconstruction and (right) after decay chain fitting.}
\label{fig:sigmapoverp_bin3}
\end{figure}

The method can be elaborated further to consider a decay chain 
\mbox{$m \to m_3 \ m_4$}, \mbox{$m_4 \to m_1 \ m_2$}.
The invariant mass of the first decaying particle reads now
\begin{eqnarray} 
\label{eq:mDTF}
m^2 & = &
m_3^2 + m_4^2 + 2 E_3 (E_1 + E_2) - 2 p_3 \sqrt{ (E_1+E_2)^2 - m_4^2 } \cos\theta_{34} \, ,
\end{eqnarray}
where $p_3$ and \mbox{$E_3=\sqrt{p_3^2+m_3^2}$} are the momentum and energy of particle 3, and $\theta_{34}$ is the angle between the three-momenta of particles 3 and 4.
Error propagation assuming that $m_4$ is kinematically (mass) constrained, \ie its value is well defined and has no associated uncertainty, yields
\begin{eqnarray} 
\label{eq:sigmaq}
\left( \frac{\sigma_{p}}{p} \right)^2 & = & 
g_m \left( \frac{\sigma_m}{m}\right)^2  - 
g_\theta  \left(\frac{\sigma_{\theta_{34}}}{\theta_{34}}\right)^2 -
g_{p_3} \left( \frac{\sigma_{p_3}}{p_3}\right)^2  \, .
\end{eqnarray}
The factors encapsulating the kinematic dependence now read 
\begin{eqnarray} 
\label{eq:gmgthetaDTF}
g_m & = & g \left( \frac{m}{p} \right)^4 \, , \nonumber \\
g_\theta & = & g \left( \frac{p_4}{p} \right)^4 \kappa^2\theta_{34}^4 \, , \nonumber \\
g_{p_3} & = & g \left( \frac{p_4}{p} \right)^4 \kappa^2 \left( \frac{\kappa}{\kappa_E} - \cos\theta_{34} \right)^2 \, ,
\end{eqnarray}
with
\begin{eqnarray} 
\label{eq:gDTF}
\frac{1}{g} & = &  \kappa^2 \left( \frac{\kappa_E}{\kappa} - \cos\theta_{34}\right)^2 \left( 1 + \xi_E \right)^2  \left[ 1 + \frac{\xi^4}{\xi_E^2} f^2(p,\xi,\eta_1,\eta_2)\right]
\, , 
\end{eqnarray}
where $\kappa$ and $\kappa_E$ are the corresponding kinematic factors for particles 3 and 4,
\ie \mbox{$\kappa=p_3/p_4$}, \mbox{$\kappa_E = E_3/E_4$}, with $p_4$ and \mbox{$E_4=\sqrt{p_4^2+m_4^2}$} being the momentum and energy of particle 4. 

In the relativistic limit, $E_i\approx p_i$ with $i=1$ to $4$,
and small angle approximation, 
$1-\cos\theta_{34} \approx \theta_{34}^2/2$,
the factors $g_{p_3}$ and $g$ simplify to
\begin{eqnarray} 
\label{eq:gmthetafactorsDTF_simplified}
g_{p_3} & \approx & \frac{g}{4} \left( \frac{p_4}{p} \right)^4 \kappa^2  \theta_{34}^4 \, , 
\end{eqnarray}
and
\begin{eqnarray} 
\label{eq:gDTF_simplified}
\frac{1}{g} & = &  \frac 14 \kappa^2 \theta_{34}^4 \left( 1 + \xi \right)^2  \left[ 1 + \xi^2 f^2(p,\xi,\eta_1,\eta_2)\right]
\, . 
\end{eqnarray}

Equations~(\ref{eq:sigmaq}) to~(\ref{eq:gDTF}) can be used to measure the relative track momentum resolution for T tracks from $\Lz\to p \pim$ and $\KS \to \pip \pim$ decays when the whole decay chains \mbox{$\Lb\to\jpsi\Lz[\to p \pim]$} and 
\mbox{$\Bz\to\jpsi\KS[\to\pip\pim]$} are reconstructed and fitted simultaneously with geometric and kinematic constraints (see Sec.~\ref{subsec:reconstructionLb} for details).
%
The procedure follows as previously,
with four main differences.
First, the $\theta_{34}$ (\ie the angle between the \jpsi and the \Lz or \KS hadrons) resolution is evaluated from the difference between the true and reconstructed angles in simulation, and then is used in the procedure with simulation and data. In contrast to the previous case, this angular resolution term is now largely subdominant with respect to the mass contribution.
Second, the simulation shows that the relative momentum uncertainty is largely independent of momentum and pseudorapidity, as a consequence of the constraints applied to the decay chains, resulting in a ratio $f(p,\xi,\eta_1,\eta_2) \approx 1$.
Third, the last term in Eq.~(\ref{eq:sigmaq}) can be neglected since the \jpsi momentum resolution, $\sigma_{p_3}/p_3$, is about 0.5\%~\cite{LHCb-DP-2014-002}, an order of magnitude less than the other two contributing terms.
Last, the $m(\jpsi\Lz)$ and $m(\jpsi\KS)$ invariant masses are used as discriminating variables to statistically subtract the background contribution in the data.
The average relative momentum resolution per bin along with its uncertainty are shown in Figs.~\ref{fig:mom_res} and~\ref{fig:B0_mom_res} (right).
Figure~\ref{fig:sigmapoverp_bin3} (right) illustrates the $\sigma_{p}/p$ distributions for the central momentum bin.


\newpage
\addcontentsline{toc}{section}{References}
\bibliographystyle{LHCb}
\bibliography{main,standard,LHCb-PAPER,LHCb-CONF,LHCb-DP,LHCb-TDR}

\newpage
\centerline
{\large\bf LHCb collaboration}
\begin
{flushleft}
\small
R.~Aaij$^{37}$\lhcborcid{0000-0003-0533-1952},
A.S.W.~Abdelmotteleb$^{56}$\lhcborcid{0000-0001-7905-0542},
C.~Abellan~Beteta$^{50}$,
F.~Abudin{\'e}n$^{56}$\lhcborcid{0000-0002-6737-3528},
T.~Ackernley$^{60}$\lhcborcid{0000-0002-5951-3498},
A. A. ~Adefisoye$^{68}$\lhcborcid{0000-0003-2448-1550},
B.~Adeva$^{46}$\lhcborcid{0000-0001-9756-3712},
M.~Adinolfi$^{54}$\lhcborcid{0000-0002-1326-1264},
P.~Adlarson$^{81}$\lhcborcid{0000-0001-6280-3851},
C.~Agapopoulou$^{14}$\lhcborcid{0000-0002-2368-0147}
% ...

{\footnotesize \it

% ...
$^{14}$Universit{\'e} Paris-Saclay, CNRS/IN2P3, IJCLab, Orsay, France\\
% ...
$^{37}$Nikhef National Institute for Subatomic Physics, Amsterdam, Netherlands\\
% ...
$^{46}$Instituto Galego de F{\'\i}sica de Altas Enerx{\'\i}as (IGFAE), Universidade de Santiago de Compostela, Santiago de Compostela, Spain\\
% ...
$^{50}$Physik-Institut, Universit{\"a}t Z{\"u}rich, Z{\"u}rich, Switzerland\\
% ...
$^{54}$H.H. Wills Physics Laboratory, University of Bristol, Bristol, United Kingdom\\
% ...
$^{56}$Department of Physics, University of Warwick, Coventry, United Kingdom\\
% ...
$^{59}$School of Physics and Astronomy, University of Glasgow, Glasgow, United Kingdom\\
$^{60}$Oliver Lodge Laboratory, University of Liverpool, Liverpool, United Kingdom\\
% ...
$^{68}$Syracuse University, Syracuse, NY, United States\\
% ...
$^{81}$Department of Physics and Astronomy, Uppsala University, Uppsala, Sweden, associated to $^{59}$\\
% ...
}
\end{flushleft}

\end{document}